\documentclass[twocolumn,showpacs,preprintnumbers,amsmath,amssymb]{revtex4}
\usepackage{subfigure,dcolumn,jabbrv}
\usepackage{graphicx}% Include figure files
\usepackage{dcolumn}% Align table columns on decimal point
\usepackage{bm}% bold math
\usepackage{CJK}
\usepackage{url}
\usepackage{color}
\usepackage{booktabs}
\usepackage{natbib}
\usepackage{hyperref}
\usepackage{cleveref}

\begin{document}

% Use the \preprint command to place your local institutional report
% number in the upper righthand corner of the title page in preprint mode.
% Multiple \preprint commands are allowed.
% Use the 'preprintnumbers' class option to override journal defaults
% to display numbers if necessary
%\preprint{}

%Title of paper
\title{Bayesian Inference of Fine-Features of Nuclear Equation of State from Future Neutron Star Radius Measurements to 0.1km Accuracy}

% repeat the \author .. \affiliation  etc. as needed
% \email, \thanks, \homepage, \altaffiliation all apply to the current
% author. Explanatory text should go in the []'s, actual e-mail
% address or url should go in the {}'s for \email and \homepage.
% Please use the appropriate macro foreach each type of information

% \affiliation command applies to all authors since the last
% \affiliation command. The \affiliation command should follow the
% other information
% \affiliation can be followed by \email, \homepage, \thanks as well.
%$\footnote{Corresponding author. fszhang@bnu.edu.cn}

\author{Bao-An Li$^{1}$$\footnote{Corresponding Author: Bao-An.Li@Tamuc.edu}$, Xavier Grundler$^{1}$$\footnote{xgrundler@leomail.tamuc.edu}$, Wen-Jie Xie$^{1,2,3}$$\footnote{wenjiexie@yeah.net}$,
and Nai-Bo Zhang$^{1,4}$$\footnote{naibozhang@seu.edu.cn}$}
\affiliation{$^1$Department of Physics and Astronomy, Texas A$\&$M University-Commerce, Commerce, TX 75429-3011, USA}
\affiliation{$^{2}$Department of Physics, Yuncheng University, Yuncheng 044000, China}
\affiliation{$^3$Guangxi Key Laboratory of Nuclear Physics and Nuclear Technology, Guangxi Normal University, Guilin 541004, China}
\affiliation{$^4$School of Physics, Southeast University, Nanjing 211189, China}
\date{\today}

\begin{abstract}
To more precisely constrain the Equation of State (EOS) of supradense neutron-rich nuclear matter, future high-precision X-ray and gravitational wave observatories are proposed to measure the radii of neutron stars (NSs) with an accuracy better than about 0.1 km. However, it remains unclear what particular aspects (other than the stiffness generally spoken of in the literature) of the EOS and to what precision they will be better constrained. In this work, within a Bayesian framework using a meta-model EOS for NSs, we infer the posterior probability distribution functions (PDFs) of incompressibility $K_{0}$ and skewness $J_{0}$ of symmetric nuclear matter (SNM) as well as the slope $L$, curvature $K_{\rm{sym}}$, and skewness $J_{\rm{sym}}$ characterizing the density dependence of nuclear symmetry energy $E_{\rm{sym}}(\rho)$, respectively, from mean values of NS radii consistent with existing observations and an expected accuracy $\Delta R$ ranging from about 1.0 km to 0.1 km. We found that (1) the $\Delta R$ has little effect on inferring the stiffness of SNM at suprasaturation densities, (2) smaller $\Delta R$ reveals more accurately not only the PDFs but also pairwise correlations among parameters characterizing high-density $E_{\rm{sym}}(\rho)$, (3) a double-peak feature of the PDF($K_{\rm{sym}}$) corresponding to the strong $K_{\rm{sym}}-J_{\rm{sym}}$ and $K_{\rm{sym}}-L$ anti-correlations is revealed when $\Delta R$ is less than about 0.2 km, and the locations of the two peaks are sensitive to the maximum value of $J_{\rm{sym}}$ reflecting the stiffness of $E_{\rm{sym}}(\rho)$ above about 3 times the saturation density $\rho_0$ of SNM, (4) the high-precision radius measurement for canonical NSs is more useful than that for massive ones for constraining the EOS of nucleonic matter around $(2-3)\rho_0$. 
\end{abstract}

\maketitle

%\newpage
%\noindent{\bf Introduction:}
\section{Introduction}
Neutron stars (NSs) are natural testing grounds for our knowledge about the nature and Equation of State (EOS) of supradense neutron-rich matter, see, e.g., Refs.\cite{Wat,sathyaprakash2019,Bogdanov:2019owz,Baiotti2019,Li:2021thg,Sorensen:2023zkk,Lovato:2022vgq,MUSES:2023hyz}. In particular, the radii of canonical NSs of masses around 1.4 M$_{\odot}$ carry important information about the EOS, especially its symmetry energy term, at densities around twice the saturation density $\rho_0$ of symmetric nuclear matter (SNM) \cite{Lattimer:2000nx}. While much progress has been made, especially since GW170817, measuring precisely the radii of NSs has been extremely challenging. Some surveys of extensive analyses of X-rays and gravitational waves in the recent literature indicate that the mean radius of a canonical NS is about $R_{1.4}=12.0\pm 1.13$ km at 68\% credibility assuming all reports are equally reliable \cite{LiEPJA}, and empirically $R_{1.4}\approx R_{1.8}\approx R_{2.0}$ within about 1.0 km precision \cite{MR-Russia}. While the available NS observational data have improved our understanding of the EOS and nucleon effective mass in dense neutron-rich matter, see, e.g., Ref.\cite{Prad}, more precise radius measurements are considered a critical task for making further progress. Indeed, high-precision NS radius measurements have been identified by the astrophysics community as a major science driver of the next-generation X-ray pulse profile observatories, e.g., The enhanced X-ray Timing and Polarimetry mission (eXTP) \cite{eXTP:2018anb}, Spectroscopic Time-Resolving Observatory for Broadband Energy X-rays (STROBE-X) \cite{STROBE-XScienceWorkingGroup:2019cyd}, 
and third-generation gravitational-wave detectors \cite{Hild:2009ns,LIGOScientific:2020zkf}, e.g., Einstein Telescope \cite{Sathyaprakash:2012jk} and Cosmic Explorer \cite{Evans:2021gyd}.  
Based on several recent analyses and simulations, see, e.g., Refs. \cite{Chatziioannou:2021tdi,Pacilio:2021jmq,Bandopadhyay:2024zrr,Finstad:2022oni,Walker:2024loo}, the planned new facilities can measure the $R_{1.4}$ to a precision better than 2.0\%. For example, considering only the 75 loudest events of the expected more than $3\times 10^5$  binary NS mergers in the one-year operation of a network consisting of one Cosmic Explorer and the Einstein Telescope, the radii of NSs in the mass range (1-1.97) M$_{\odot}$ will be constrained to at least $\Delta R\leq 0.2$ km at 90\% credibility \cite{Walker:2024loo}. It was also pointed out that
a single 40 km Cosmic Explorer detector can pin down the NS radius to an accuracy of 10 meters within a decade, whereas the current generation of ground-based detectors like the Advanced LIGO-Virgo network would take $\mathcal{O}(10^5)$ years to do so \cite{Bandopadhyay:2024zrr}.

Certainly, there is no doubt that measuring precisely the radii of NSs is scientifically invaluable, especially for resolving many mysteries associated with NSs and the EOS of the densest visible matter in the universe. On the other hand, given the extreme difficulties and the required super-expensive investments in several measures, thoroughly investigating the scientific returns in multiple areas from such measurement is also invaluable and probably a prerequisite. Acknowledging {\it a priori} that probably many important new physics that can be discovered from the proposed high-precision NS radius measurements is beyond the current knowledge of the authors, here we focus on investigating what the proposed high-precision radius measurements will teach us about the EOS of supradense neutron-rich nucleonic matter within the minimum model of NSs consisting of neutrons, protons, electrons, and muons ($npe\mu$) at $\beta-$equilibrium. Such NS EOS is the most fundamental basis for constructing more complicated EOS models considering more degrees of freedom and possible phase transitions. Since going beyond the minimum NS EOS model inevitably involves more assumptions and uncertainties, our study here can thus be considered the most conservative one. Of course, we can not exclude any physics associated with new particles beyond $npe\mu$ and/or possible phase transitions in NSs. 

In the literature, there are a lot of general speculations on how the high-precision radius measurements may constrain the stiffness of supradense NS matter EOS. However, it is presently unclear which particular features of the EOS and to what extent they can be more precisely constrained compared to our current best knowledge. To fill this knowledge gap and reveal more clearly the most important EOS features determining NS radii, in this work, we infer the PDFs of parameters characterizing the density dependence of both SNM EOS and nuclear symmetry energy using imagined high-precision NS radius data within a Bayesian framework. 
In many impactful papers in the literature, one simply refers an EOS giving a larger (smaller) radius and/or tidal deformation as being stiff (soft). As we shall demonstrate in several ways in this work, such description is ambiguous if not misleading about the fundamental physics underlying the NS radius/deformation. As discussed in detail in Ref. \cite{Lattimer:2000nx}, because the NS radius $R$ is determined by the radial coordinate where the pressure vanishes, $R$ scales with the pressure around $(1-2)\rho_0$ \cite{Lattimer:2000nx} and is thus most sensitive to the EOS of neutron-rich matter around $2\rho_0$. Contributions to the pressure around this density from SNM EOS and nuclear symmetry energy compete strongly depending on the values of $L$, $K_{\rm{sym}}$, and $J_0$ \cite{Li:2005sr,LCK}. More specifically, using machine learning techniques to large sets of posterior EOS parameters from Bayesian analyses of existing NS radius data,
it was found earlier that the most important EOS parameters determining $R_{1.4}$ are (in order of decreasing importance): curvature $K_{\rm{sym}}$, slope $L$, skewness $J_{\rm{sym}}$ of nuclear symmetry energy, skewness $J_{0}$, incompressibility $K_{0}$ of symmetric nuclear matter, and the magnitude $E_{\rm{sym}} (\rho_0)$ of symmetry energy at the saturation density $\rho_0$ of nuclear matter \cite{Jake}. Thus, the radii of canonical NSs are expected to be most useful for constraining the EOS of neutron-rich matter around $2\rho_0$ instead of the much higher core densities, and we have no reason to believe the radii (but the masses themselves) of massive NSs are more useful than those of canonical ones for this purpose. In this work, we shall study how the PDFs and pairwise correlations of these EOS parameters may be altered by using mean values of NS radii consistent with existing observations and varying the accuracy $\Delta R$ from about 1.0 km to 0.1 km. These results are useful for further developing and constraining nuclear many-body theories for supradense neutron-rich nuclear matter \cite{Sorensen:2023zkk,Lovato:2022vgq}. They also provide useful guidance and reference for laboratory tests of these EOS parameters using heavy-ion reactions induced by high-energy radioactive beams \cite{FRIB400,Almaalol:2022xwv,Xu:2022mqn}.

The rest of the paper is organized as follows. In the next section, we shall first outline the meta-model EOS for NS matter, the Bayesian statistical framework we employ and the existing NS radius data that we use as a basis for our study with expected future precision. In section \ref{Direct}, some expectations on what we can learn about the EOS of supradense neutron-rich matter are discussed based on direct inversions of NS radii, the maximum mass as well as causality in a 3-dimensional high-density EOS parameter space.
In section \ref{Results}, we present and discuss our results. Finally, a summary is given. In three appendices, typical results testing MCMC convergence and the level of autocorrelations in the inferred PDFs of EOS parameters are presented.

\section{Theoretical framework: meta-model EOS and Bayesian approach}%\label{Theo}
For completeness and ease of our discussions, we recall in the following several main aspects of our approach. More technical details and examples of applications can be found in our earlier publications (e.g.
\cite{zhang2018combined,Zhang:2019fog,Zhang:2021xdt,xie2019bayesian,xie2020bayesian,xie2021bayesian}) and reviews \cite{LiEPJA,Li:2021thg}.

\subsection{Explicitly isospin-dependent meta-model EOS for neutron stars}
\label{nseos}
A meta-model is a template for building models. In data science or software engineering, a meta-model describes the structure, rules, or behavior of other models. In short, it is a model of models. We construct an NS meta-model that can mimic all existing EOSs for $npe\mu$ matter at $\beta-$equilibrium. It is based on parameterizing the binding energy per nucleon $E(\rho,\delta)$ in neutron-rich matter at nucleon density $\rho=\rho_n+\rho_p$ and isospin asymmetry $\delta\equiv (\rho_n-\rho_p)/\rho$ \cite{bombaci1991asymmetric}
\begin{equation}\label{eos}
E(\rho,\delta)=E_0(\rho)+E_{\rm{sym}}(\rho)\cdot \delta ^{2} +\mathcal{O}(\delta^4)
\end{equation}
where $E_0(\rho)$ is the SNM EOS and $E_{\rm{sym}}(\rho)$ is nuclear symmetry energy at density $\rho$. 
The corresponding pressure can be written as
\begin{eqnarray}\label{pressure1}
  P(\rho, \delta)=\rho^2\frac{dE(\rho,\delta)}{d\rho}&=&\rho^2[\frac{dE_{\rm{0}}(\rho)}{d\rho}+\frac{dE_{\rm{sym}}(\rho)}{d\rho}\delta^2]\nonumber\\
  &=&P_{\rm{SNM}}(\rho)+\rho^2\frac{dE_{\rm{sym}}(\rho)}{d\rho}\delta^2
\end{eqnarray}
where $P_{\rm{SNM}}(\rho)\equiv \rho^2\frac{dE_{\rm{0}}(\rho)}{d\rho}$ is the pressure in SNM. Including contributions from leptons, the pressure in NSs is obtained from
\begin{equation}\label{pressure}
  P(\rho, \delta) = \rho^2 \frac{d\epsilon(\rho,\delta)/\rho}{d\rho},
\end{equation}
where $\epsilon(\rho, \delta) = \epsilon_n(\rho, \delta) + \epsilon_l(\rho, \delta)$ is the energy density of NS matter. Here, $\epsilon_n(\rho, \delta)$ and $\epsilon_l(\rho, \delta)$ are the energy densities of nucleons and leptons, respectively. The energy density of leptons, $\epsilon_l(\rho, \delta)$, is determined using the noninteracting Fermi gas model \cite{oppenheimer1939massive}. On the other hand, the energy density of nucleons, $\epsilon_n(\rho, \delta)$, is linked to the $E(\rho, \delta)$, and the average mass $M_N$ of nucleons via
\begin{equation}\label{lepton-density}
  \epsilon_n(\rho, \delta)=\rho [E(\rho,\delta)+M_N].
\end{equation}
The density profile of isospin asymmetry $\delta(\rho)$ is obtained by using the $\beta$-equilibrium condition $\mu_n-\mu_p=\mu_e=\mu_\mu\approx4\delta E_{\rm{sym}}(\rho)$ and the charge neutrality requirement $\rho_p=\rho_e+\rho_\mu$. Here the chemical potential $\mu_i$ for a particle $i$ is calculated from the energy density via
$\mu_i=\partial\epsilon(\rho,\delta)/\partial\rho_i.
$
Before muons start appearing, the proton fraction $x_p$ (or isospin asymmetry $\delta=(1-2x_p)$)
in $npe$ matter can be written down analytically as \cite{Lattimer:2000nx}
%\begin{eqnarray}\label{xp}
$
x_p(\rho)= 0.048 \left[E_{\rm sym}(\rho)/E_{\rm sym}(\rho_0)\right]^3
(\rho/\rho_0)(1-2x_p(\rho))^3.$
%\end{eqnarray}
It is seen that $x_p$ is uniquely determined by the density dependence of symmetry energy $E_{\rm sym}(\rho)$. At suprasaturation densities (see numerical examples in Ref. \cite{Zhang-EPJA23}), one can easily show that when the $E_{\rm sym}(\rho)$ is very stiff (with either very large $K_{\rm{sym}}$ and/or $J_{\rm{sym}}$ the $\delta$ approaches zero (symmetric nuclear matter). On the other hand, when the $E_{\rm sym}(\rho)$ is very soft (with either very large but negative $K_{\rm{sym}}$ and/or $J_{\rm{sym}}$ the $\delta$ approaches 1.0 (pure neutron matter).

With the above $\delta(\rho)$, both the pressure $P(\rho,\delta(\rho))$ and energy density $\epsilon(\rho, \delta(\rho))$ become barotropic, i.e., depending on density only. 
The core EOS described above is then connected with the crust EOS at a transition density $\rho_t$ consistently determined by examining when the core EOS becomes thermodynamically unstable 
against spinodal decomposition by forming clusters
\cite{Lattimer:2006xb,kubis2007nuclear,Xu:2009vi}.

Specifically, the incompressibility $K_{\mu}$ of uniform NS core at $\beta$-equilibrium can be written as
\begin{eqnarray}\label{kmu1}
&&K_{\mu}= \rho^2 \frac{d^2 E_0}{d \rho^2} + 2 \rho \frac{dE_0}{d \rho}\\
&+& \delta^2
\left[ \rho^2 \frac{d^2 E_{\rm sym}}{d \rho^2}
+2 \rho \frac{d E_{\rm sym}}{d \rho} - 2 E^{-1}_{\rm sym}(\rho)
\left(\rho \frac{d E_{\rm sym}}{d \rho}\right)^2\right]. \nonumber
\end{eqnarray}
In terms of the EOS parameters, the transition density $\rho_t$ is determined by the condition \cite{LiMagno}
\begin{eqnarray}\label{kmu2}
&&K_{\mu}=\frac{1}{9} (\frac{\rho}{\rho_0})^2 K_0+ 2 \rho \frac{dE_0}{d \rho}\\&+& \delta^2
\left[\frac{1}{9} (\frac{\rho}{\rho_0})^2 K_{\rm{sym}}
+\frac{2}{3}\frac{\rho}{\rho_0}L
-2E^{-1}_{\rm sym}(\rho)(\frac{1}{3}\frac{\rho}{\rho_0}L)^2\right]=0. \nonumber
\end{eqnarray}
As shown in great details in Refs. \cite{Zhang-JPG,LiMagno},
while $K_0$ and $L$ also play significant roles,
the transition density $\rho_t$ is most sensitive to the vriation of $K_{\rm{sym}}$. Going beyond the limits of our present meta-EOS model, there are other well-known uncertainties regarding the crust-core transition density, see e.g., Refs. \cite{Pro14,Newton14} for reviews.

We adopt the Negele-Vautherin (NV) EOS \cite{negele1973neutron} for the inner crust and the Baym-Pethick-Sutherland (BPS) EOS \cite{baym1971ground} for the outer crust. Finally, the complete NS EOS in the form of pressure versus energy density $P(\epsilon)$ is used in solving the Tolman-Oppenheimer-Volkoff (TOV) equations 
\cite{tolman1934effect,oppenheimer1939massive}. 

To this end, it is important to note that while the thickness of NS crust is only about (10-15)\% of its radius \cite{Xu:2009vi}, there are many interesting physical processes happening inside and around the crust, see, e.g., Refs. \cite{Newton21,LiN23,Sotani24,Sh24}. Employing different crust models can introduce an 
uncertainty in the radius calculations.
This relates to the application of either a unified crust model (pasta phase) or multiple distinct crust models as addressed in the literature, see, e.g. Ref. \cite{Davis24} for a very recent review. The quantitative estimate of the associated uncertainty in the predicted NS radius is model dependent ranging from about 0.1 to 0.7 km (Table I of Ref. \cite{Fortin}). Thus, while the thickness of NSs is compatible to the current uncertainty of radius measurements, its own uncertainty is relatively large. This is mostly because the physics going on in the crust is very poorly known, and there is little if any confirmed observational evidence for the various proposed physics going on inside NS crusts. Certainly, future high-precision radius measurements will provide much needed data and probably evidence of new phenomena associated uniquely with NS crusts. As stated repeatedly, we focus on the physics at suprasaturation densities far above the crust-core transition densities predicted. 

Similarly, there are still some uncertainties even about the saturation density $\rho_0$ of SNM and its binding energy there. Some of them may have some appreciable effects on some properties of NSs, see, e.g. Refs. \cite{Mondal22,Car24}.
We concur that they are important for NS physics. 
In our opinion, however, the related physics can be better addressed with nuclear laboratory experiments measuring masses, charge radii, neutron-skins, etc of especially those extremely neutron-rich nuclei made possible recently by the advanced rare isotope beam facilities in several countries, see, e.g., \cite{FRIB}, probably long before NS radius is actually measured with a precision $\Delta R$ around 0.1 km. In this work, we thus use the currently known most probable values for quantities describing such saturation properties.

We adopt the empirical binding energy $E_0(\rho_0)=-16$ MeV at $\rho_0=0.16/\rm{fm}^3$ used in most nuclear physics textbooks. For the other EOS parameters, as we shall discuss in detail next, we generate them randomly within their currently known uncertainty ranges consistent with existing constraints from nuclear experiments and theories especially predictions of the various Energy Density Functionals (EDFs). With the risk of going far away from the main topic of this work, we notice briefly that there is a historical problem of predicting different $E_0(\rho_0)$ and $\rho_0$ values by different nuclear many-body theories using various interactions, depending especially on the form and strength of the tensor and three-body forces used and whether short-range correlations induced by them are considered or not. Such problem still exists today especially with the EDFs where effects of the tensor forces are often ignored and many-body forces are modeled differently. Consequently, the saturation properties predicted by these models can diverge widely. For instance, as summarized in Refs. \cite{Dutra:2012mb,Dutra:2014qga,Mar17}, with the Relativistic Mean Field (RMF) models and Hartree-Fock Approch with Skyrme, Gogny or other forces, even the resulting $E_0(\rho_0)$ and $\rho_0$ values can be appreciably different from their widely used empirical values given above. 

In both the forward-modeling and Bayesian inference, the variation of the $E_0(\rho_0)$ and $\rho_0$ values has been found in some models to affect significantly the predicted NS properties or the inferred PDFs of nuclear matter saturation properties , see, e.g., Ref. \cite{Pro-RMF} for quantitative examples using three different RMF models. As we said repeatedly, for achieving the goals of this work we purposely fix the $E_0(\rho_0)$ and $\rho_0$ at their empirical values although their variations may have been found to significantly affect NS properies in some models. It is a common practice in most studies in the literature. For example, in studying properties of NSs in Ref. \cite{Mar18} after the authors have surveyed the predictions on saturation properties of nuclear matter by various EDFs in Ref. \cite{Mar17},  they stated {\it ``for the simplicity of the discussion as well as to keep computing time within a reasonable range, we have decided to fix the value of these parameters in this work to be: $E_0(\rho_0)=-15.8$ MeV at $\rho_0=0.155/\rm{fm}^3$".} The justifications they gave are {\it ``we have evaluated in paper \cite{Mar17} that these empirical parameters are sufficiently well known and/or have a very weak impact on the dense matter EoS"}. It is understandable that their choice and assement based on their calculations may be different from those by some other people using other theories/models. Nevertheless, the inconsistent findings and different opinions in the field present us a dilemma in considering effects due to the possible variations of $E_0(\rho_0)$ and $\rho_0$ in studying properties of NSs, if one trusts all publications equally and respects different opinions all based on sound physics with minimum biases.

We fully acknowledge the fact that some studies have found that even the $E_0(\rho_0)$ and $\rho_0$ (we do not mention here other EOS properties especially the high-density ones as the main purpose of this work is to explore how the high-precision NS radius measurement may help constrain them) can affect NS properties especilly considering the correlations among all the EOS parameters. For example, it was found in Ref. \cite{Pro-RMF} with the RMF models that even under the same constraints within a Bayesian inference framework, different types of RMF models produce varying PDFs of nuclear saturation properties, and these differences are significant and cannot be ignored. We emphasize here again that we are NOT ignoring any effects of varying the EOS parameters that are approaching asymptotically to saturation properties as $\rho\rightarrow \rho_0$, except those associated with varying the $E_0(\rho_0)$ and $\rho_0$. Then, why do we not consider effects of varying the later in our study here? Besides agreeing with the justficiations given in Ref. \cite{Mar18} for such choice, we have the following consideration that the readers are welcome to disagree with:
\begin{enumerate}
\item The physics origins of the well-known ``Coester band" \cite{Coester} regarding the varying predictions for the $E_0(\rho_0)$ and $\rho_0$ are relatively well understood, see, e.g., Ref. \cite{Baldo} and references therein, by the nuclear physics community.  What physics ingredients are missing in some of the theories and/or models are largely known and progress is being made in improving them based on various existing data from terrestrial laboratories. To our best knowledge, there is no sound and convincing physics reason to rely on very limited NS data currently availavle to further constrain $E_0(\rho_0)$ and $\rho_0$ especially considering practically the gain/cost ratio and the extreme difficulties of getting accurate data in astrophysics observations compared to terrestrial nuclear experiments.
\item It is well known that the advantage of NS data is to provide access to supra-dense neutron-rich matter that can not be obtained from terrestrial experiments. While it has been shown in several studies that the uncertainties of $E_0(\rho_0)$ and $\rho_0$ affect the extraction of
other saturation properties characterizing supra-dense matter EOS and it may be possible to indeed constrain all saturation properties of nuclear matter if really thousands or more NS-NS mergers had been detected, in our opinion, the currently available data from only one confirmed NS-NS merger and several mass-radius mesurements having still very large errors can not constrain accurately $E_0(\rho_0)$ and $\rho_0$ in a productive manner. On the contrary, because of the correlations among the EOS parameters, considering the variations of $E_0(\rho_0)$ and $\rho_0$ in NS models reduces the chance or masks the clarity of seeing effects of the high-density EOS parameters that are currently much more poorly known but are most interesting and important for NS physics, besides requiring more computing time and enlarging the necessary parameter space.
\item The predicted sensitivites of NS proprties to variations of $E_0(\rho_0)$ and $\rho_0$ certainly provide a strong science motivation for the nuclear physics community to further refine them using terrestrial experiments and more advanced nuclear many-body theories. However, before plenty of high-precision NS observational data at approximately the same relative accuracy level as current terrestrial nuclear experiments in measuring atomic masses and radii become available, we do not advocate any suggestion to constrain the $E_0(\rho_0)$ and $\rho_0$ using NS data. 
\end{enumerate}

\begin{table}[htbp]
\centering
\caption{Prior ranges of the six EOS parameters (MeV).}\label{tab-prior}
 \begin{tabular}{lccccccc}
  \hline\hline
%  \colhead{Parameters} & \colhead{Lower limit} & \colhead{Upper limit (MeV)}
   Parameters&~~~~Lower limit  &~~~~Upper limit \\
    \hline\\
  \vspace{0.2cm}
$K_0$ & 220 & 260 \\
$J_0$ & -400 & 400 \\
$K_{\mathrm{sym}}$  & -400 & 100 \\
$J_{\mathrm{sym}}$ & -200 & 800 \\
$L$ & 30 & 90 \\
$E_{\mathrm{sym}}(\rho_0)$ & 28.5 & 34.9 \\
 \hline
 \end{tabular}
\end{table}
Returning to the construction of our parameterized meta-model, we recall that the $E_0(\rho)$ and $E_{\rm{sym}}(\rho)$ can be parameterized as
\begin{eqnarray}\label{E0para}
  E_{0}(\rho)&=&E_0(\rho_0)+\frac{K_0}{2}(\frac{\rho-\rho_0}{3\rho_0})^2+\frac{J_0}{6}(\frac{\rho-\rho_0}{3\rho_0})^3,\\
  E_{\rm{sym}}(\rho)&=&E_{\rm{sym}}(\rho_0)+L(\frac{\rho-\rho_0}{3\rho_0})+\frac{K_{\rm{sym}}}{2}(\frac{\rho-\rho_0}{3\rho_0})^2\nonumber\\
  &+&\frac{J_{\rm{sym}}}{6}(\frac{\rho-\rho_0}{3\rho_0})^3\label{Esympara}.
\end{eqnarray}
The coefficients $K_0$ and $J_0$ defined as
\begin{eqnarray}
    K_0&=&9\rho_0^2[\partial^2 E_0(\rho)/\partial\rho^2]|_{\rho=\rho_0},\\\
    J_0&=&27\rho_0^3[\partial^3 E_0(\rho)/\partial\rho^3]|_{\rho=\rho_0} 
\end{eqnarray}
are the SNM incompressibility and skewness, respectively. The $E_{\rm{sym}}(\rho_0)$, 
\begin{eqnarray}
    L&=&3\rho_0[\partial E_{\rm{sym}}(\rho)/\partial\rho]|_{\rho=\rho_0},\\\ K_{\rm{sym}}&=&9\rho_0^2[\partial^2 E_{\rm{sym}}(\rho)/\partial\rho^2]|_{\rho=\rho_0},\\\
    J_{\rm{sym}}&=&27\rho_0^3[\partial^3 E_{\rm{sym}}(\rho)/\partial\rho^3]|_{\rho=\rho_0} 
\end{eqnarray}
are the magnitude, slope, curvature and skewness of nuclear symmetry energy at $\rho_0$, respectively. 
While these coefficients are all defined at $\rho_0$, the high-order derivatives characterize the high-density behaviors of  
$E_0(\rho)$ and $E_{\rm{sym}}(\rho)$, respectively. In particular, the skewness parameters $J_0$ and $J_{\rm{sym}}$ characterize the stiffness of 
$E_0(\rho)$ and $E_{\rm{sym}}(\rho)$ around $(3-4)\rho_0$ \cite{Xie:2020kta}, and $K_{\rm{sym}}$ characterizes the stiffness of $E_{\rm{sym}}(\rho)$ around $(2-3)\rho_0$. In our meta-model EOS used in the Bayesian framework, these coefficients are randomly generated in their prior uncertainty ranges listed in Table \ref{tab-prior} \cite{LiEPJA} in each Markov-Chain Monte Carlo (MCMC) step. These uncertainty ranges are based on the knowledge the community has accumulated over the last 40 years from analyzing terrestrial experiments, astrophysical observations, and extensive theoretical studies. In particular,
$K_0$ has been relatively well constrained to $K_0=240\pm20$ MeV \cite{Garg18,Shlomo06}. Similarly, $E_{\rm sym}(\rho_0)$ and $L$ are relatively well constrained to $E_{\rm sym}(\rho_0)=31.7\pm3.2$ MeV and $L=58.7\pm28.1$ MeV at 68\% confidence level \cite{Li13,Oertel17,Li:2017nna}, respectively. While the parameters characterizing the isospin-asymmetric EOS at suprasaturation densities, i.e., $K_{\rm sym}=-100\pm100$ MeV \cite{Mondal17,Mar17,Mar18,Som21,Gra22,Li:2021thg}, $J_0=-190\pm100$ MeV \cite{Zhang:2019fog,Xie:2020kta,xie2019bayesian,xie2020bayesian,Cai17}, and $-200<J_{\rm sym}<800$ MeV \cite{Dutra:2012mb,Dutra:2014qga,Zhang17,Tews17,Li-PPNP} still have rather large errors.

In comparison with many NS EOS models in the literature, some key features of the meta-model EOS described above deserve emphasizing:
\begin{enumerate}
\item It is well known that the TOV equations are degenerate about the compositions of NSs (i.e., composition blind) in the sense that once the EOS $P(\epsilon)$ is given, the mass-radius sequence is uniquely determined regardless of how the $P(\epsilon)$ is constructed 
or what particles/phases are considered. At supersaturation densities, since there is presently no completely reliable theory, one often parameterizes directly the $P(\epsilon)$ using piecewise polytropes or spectrum functions \cite{Lee2018}. These types of parameterizations, however, carry no non-degenerate information about NS composition. They can not be used to reveal the underlying nuclear symmetry energy which is presently the most uncertain term in the EOS of supradense neutron-rich nucleonic matter but most important for determining the radii of NSs and/or the cooling mechanisms of protoneutron stars \cite{Esym-review}. Our meta-model EOS is built from a more fundamental level with NS composition determined self-consistently by the density dependence of nuclear symmetry energy. It has the same flexibility as the piecewise polytropes by varying the EOS parameters within their prior ranges constrained by existing nuclear laboratory data and theoretical predictions. On the contrary, there is little direct experimental or prior information about parameters describing especially the high-density pieces of the polytropical EOSs. 

\item On the other hand, compared to the microscopic and/or phenomenological nuclear many-body theories for NS matter, our meta-model EOS is not so fundamental or microscopic enough for some purposes. Nevertheless, it can mimic most EOSs predicted by these many-body theories. Most importantly, it provides the flexibility and diversity necessary for the purposes of this work in a Bayesian statistical framework. 

\item The parameterizations for $E_0(\rho)$ and $E_{\rm{sym}}(\rho)$ look like Taylor expansions for known energy density functionals. 
Indeed, the EOS parameters asymptotically become the coefficients of the Taylor expansions as $\rho\rightarrow \rho_0$.
Here, they are just parameters to be inferred from the observational/experimental data instead of expanding some known energy density functionals \cite{Cai:2021ucx}. The issue of convergence associated with Taylor expansions does not exist in our Bayesian analyses. Moreover, the last parameters, i.e. $J_0$ and $J_{\rm{sym}}$, carry all information about the high-density EOS of neutron-rich matter. 
In Bayesian analyses, once the EOS parameters are randomly generated within their prior ranges in each MCMC step, the NS EOS can be constructed as described above. 
\end{enumerate}

\subsection{Bayesian inference approach}\label{bayes}
For completeness, we recall that the Bayesian theorem reads
\begin{equation}\label{Bay1}
P({\cal M}|D) = \frac{P(D|{\cal M}) P({\cal M})}{\int P(D|{\cal M}) P({\cal M})d\cal M}.
\end{equation}
Here $P({\cal M}|D)$ represents the posterior probability of the model $\cal M$ (described here by the 6 meta-model EOS parameters) given the dataset $D$. Meanwhile, $P(D|{\cal M})$ is the likelihood function that a given theoretical model $\cal M$ predicts the data $D$, and $P({\cal M})$ is the prior probability of the model $\cal M$ before encountering its prediction with the given data. The denominator in Eq. (\ref{Bay1}) is a normalization constant. 

\begin{table}[htbp]
\centering
\caption{Neutron star data used in the present work. For the radii of the four millisecond pulsars (MPSs), the masses are from Refs. \cite{Nice,Webb:2019tkw,Guo:2021bqa,NANOGrav:2017wvv} while the radii and their uncertainties correspond to approximately (5-10)\% constraints on R with the extent in M corresponding to the $1\sigma$ bounds from the radio measurements are from simulations using the AP3 EOS \cite{AP3} reported in Fig. 8 of Ref. \cite{Wat}.}\label{tab-data}
 \begin{tabular}{lccccccc}
  \hline\hline
   Mass($\mathrm{M}_{\odot}$)&Radius $R$ (km)  &~~~~Source and Reference \\
    \hline\hline\\
  \vspace{0.2cm}
 1.4 & 11.9$^{+1.4}_{-1.4}$(90\% CFL)&GW170817\cite{abbott2018gw170817} \\
1.4 &10.8$^{+2.1}_{-1.6}$ (90\% CFL)&GW170817 \cite{de2018tidal} \\
1.4  & 11.7$^{+1.1}_{-1.1}$ (90\% CFL)&QLMXBs \cite{lattimer2014constraints} \\
$1.34_{-0.16}^{+0.15}$ &$12.71_{-1.19}^{+1.14}$ (68\% CFL)&PSR J0030+0451 \cite{riley2019nicer} \\
$1.44_{-0.14}^{+0.15}$ &$13.0_{-1.0}^{+1.2}$ (68\% CFL)&PSR J0030+0451 \cite{Miller:2019cac} \\
$2.08_{-0.07}^{+0.07}$ &$13.7_{-1.5}^{+2.6}$ (68\% CFL)&PSR J0740+6620 \cite{fonseca2021refined} \\
\hline
$1.26_{-0.14}^{+0.14}$ &$12.28_{-0.45}^{+0.45}$ &MSP J0751+1807 \cite{Nice,Wat}\\
$1.54_{-0.03}^{+0.03}$ &$11.57_{-0.56}^{+0.56}$ &MSP J1909+3744 \cite{Webb:2019tkw,Wat}\\
$1.831_{-0.01}^{+0.01}$ &$12.53_{-0.57}^{+0.57}$ &MSP J2222+0137 \cite{Guo:2021bqa,Wat}\\
$1.908_{-0.016}^{+0.016}$ &$11.93_{-0.50}^{+0.50}$ &MSP J1614+2230 \cite{NANOGrav:2017wvv,Wat}\\
  \hline\hline
 \end{tabular}
\end{table}

We randomly sample uniformly the six EOS parameters within their prior ranges specified by their minimum and maximum values given in Table \ref{tab-prior}. For each set of the EOS parameters, we construct the NS EOS at $\beta-$equilibrium as described earlier. The corresponding NS mass-radius sequence is determined by solving the TOV equations. The resultant theoretical radius $R_{\mathrm{th},j}$ is then utilized to assess the likelihood of the selected EOS parameter set to reproduce the observed radius $R_{\mathrm{obs},j}$, where $j$ ranges from 1 to the total number N of radius data used in the particular analysis. This radius likelihood is calculated from
\begin{eqnarray}\label{Likelihood-R}
 && P_\mathrm{R}[D(R_{1,2,\cdots N})|{\cal M}(p_{1,2,\cdots N})]\nonumber\\
 &&=\prod_{j=1}^{N}\frac{1}{\sqrt{2\pi}\sigma_{\mathrm{obs},j}}\exp[-\frac{(R_{\mathrm{th},j}-R_{\mathrm{obs},j})^{2}}{2\sigma_{\mathrm{obs},j}^{2}}],
\end{eqnarray}
where $\sigma_{\mathrm{obs},j}$ represents the $1\sigma$ error bar associated with the result from analyzing the observation $j$. If there is more than one analysis for the same NS, e.g., the independent analyses using somewhat different approaches of the same observational data by NICER or LIGO/VIRGO Collaboration, we treat them as equally reliable within the error bars published. Namely, they are considered as independent. Technically, without double counting, in preparing $P_\mathrm{R}$ one can either uses the product of two Gaussian functions from the two independent analyses of the same NS observational data as indicated by the above equation in the whole MCMC chain, or in each MCMC step one first compares 1/2 with a random number generated between 0 and 1 to choose randomly with equal weight one Gaussian distribution function for the radius from the two independent analyses. Numerically, we found in our previous Bayesian analyses that these two methods give almost identical posterior PDFs. This is mainly because they have statistically the same effects in filtering the model predicted radius values.

Depending on the purpose of each analysis, the total likelihood function can be constructed differently by selecting and multiplying the individual likelihood components. Our base/default total likelihood function can be formally written as
\begin{equation}\label{Likelihood}
  P(D|{\cal M})_{\rm base} = P_{\rm{filter}} \times P_{\rm{mass,max}} \times P_\mathrm{R}. 
\end{equation}
Here, $P_{\rm{filter}}$ and $P_{\rm{mass,max}}$ indicate that the generated EOSs must satisfy the following conditions: (i) The crust-core transition pressure remains positive; (ii) The thermodynamic stability condition, $dP/d\epsilon\geq0$, holds at all densities; (iii) The causality condition is upheld at all densities; (iv) The generated NS EOS should be sufficiently stiff to support NSs at least as massive as 1.97 M$_{\odot}$ (i.e., the minimum M$_{\rm TOV}$ which is the maximum mass a given EOS can support) as in the original analysis of GW170817 by the LIGO/VIRGO Collaborations \cite{abbott2017gw170817}. To avoid causing confusion, we emphasize that while the conditions represented by $P_{\rm{filter}}$ and $ P_{\rm{mass,max}}$ are consistently implemented in calculating $P_\mathrm{R}$ in solving the TOV equations, the expression in Eq. (\ref{Likelihood}) is still valid. 
The individual effects of varying the minimum M$_{\rm TOV}$ and its uncertainty as well as the way to implement them (sharp-cut off or as a Gaussian), and turning on/off the causality condition on inferring the PDFs of EOS parameters for a given set of NS radius data have been studied in our previous publication \cite{xie2019bayesian}. Indeed, enforcing differently additional conditions, i.e., the $P_{\rm{filter}}$ and $ P_{\rm{mass,max}}$, give rise to non-trivial modifications to the PDFs of some EOS parameters obtained only with the $P_\mathrm{R}$ (note that we can not do the other way around, namely using only the $P_{\rm{filter}}$ and $ P_{\rm{mass,max}}$ without a specific $P_\mathrm{R}$). The main findings from Ref. \cite{xie2019bayesian} will be summarized briefly in the next section together with discussions on the intertwining effects of radius measurements, causality and the minimum M$_{\rm TOV}$ as well as the possibility to disentangle them in inferring the PDFs of EOS parameters.

For this work, we use the mean or most probable radii of several NSs listed in Table \ref{tab-data} but vary their precision $\Delta R=\sigma_{\mathrm{obs},j}$. In particular, we consider the following cases:
\begin{enumerate}
    \item A single canonical NS with $R_{1.4}=11.9$ km as inferred by the LIGO/VIRGO Collaborations from GW170817 \cite{abbott2017gw170817} with 
    $\Delta R=1.0, 0.5, 0.2$, and $0.1$ km, respectively.
     \item Three canonical NSs with $R_{1.4}=11.8$, 11.9 and 12.0 km, respectively, all with $\Delta R=0.1$ km.
     \item A single NS with a mass of 2.0M$_{\odot}$ and $R_{2.0}=11.9$ km with 
    $\Delta R=1.0, 0.5, 0.2$, and $0.1$ km, respectively.
     \item Four NSs with a mass of 1.4, 1.6, 1.8 and 2.0M$_{\odot}$, respectively, all having the same radius of 11.9 km 
      with $\Delta R=1.0, 0.5, 0.2$, and $0.1$ km, respectively.
     \item An analysis using all data listed in the first six rows in Table \ref{tab-data} from LIGO/VIRGO, Chandra+XMM-Newton, and NICER as a representation of the best data available. The resulting PDFs are used as references in studying the effects of improving the accuracy of measuring the radii of the four millisecond pulsars (MSPs) listed in the last four rows of Table \ref{tab-data}. Their masses are well-determined observationally but their radii are only roughly known based an EOS model prediction \cite{AP3} as reported in Fig. 8 of Ref. \cite{Wat}. We use $\Delta R=1.0, 0.5, 0.2$, and $0.1$ km, respectively, for all four MSPs.
\end{enumerate}

\begin{figure}[h]
\begin{center}
\vspace{-0.5cm}
%\centering
\hspace{-1.cm}
  \resizebox{0.55\textwidth}{!}{
  \includegraphics[width=5cm,height=4cm]{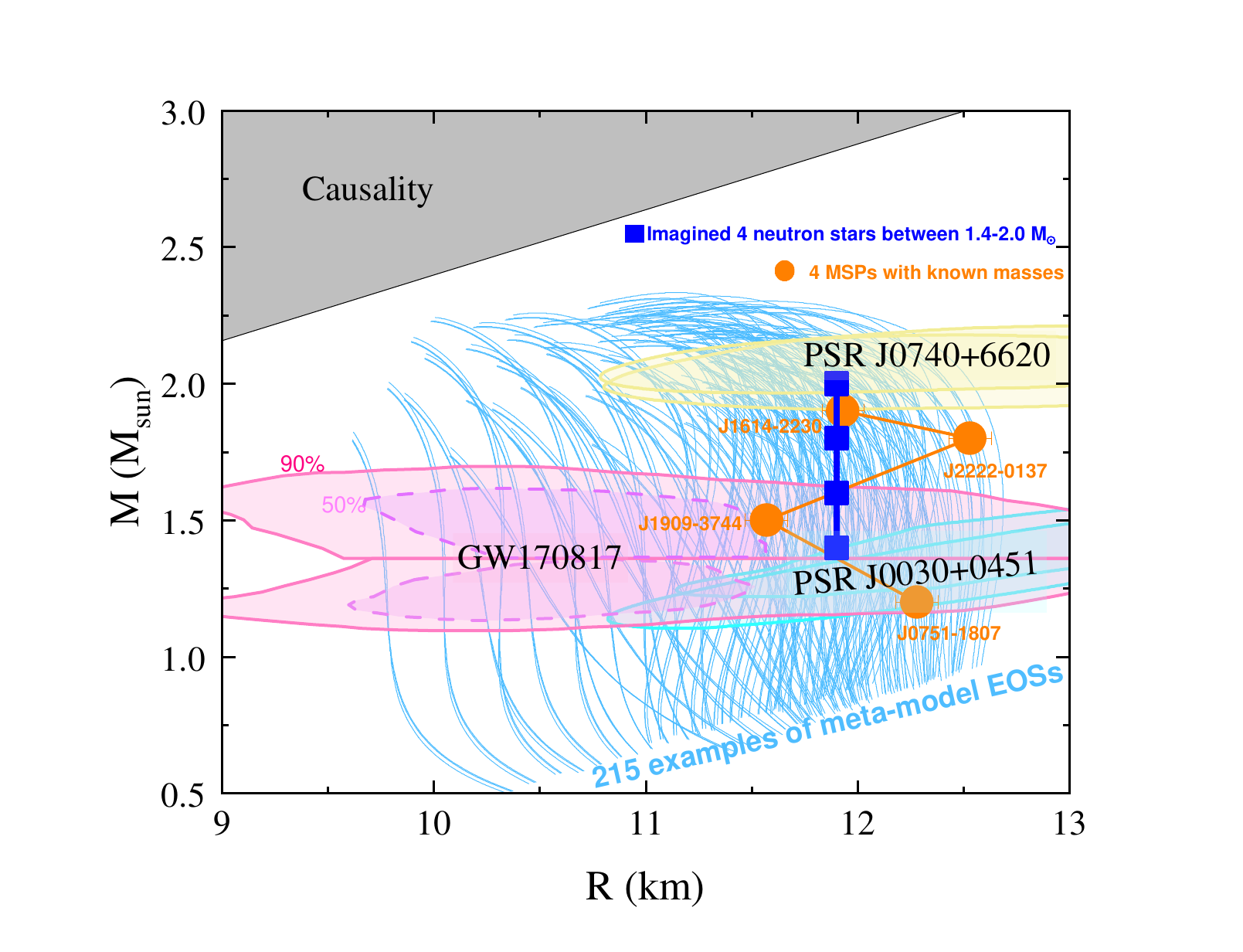}
}
  \caption{(color online) Mass-radius sequences from using 215 examples of meta-model EOSs in comparison with the actual data from observations listed in Table \ref{tab-data} as well as some imagined NSs for the analyses in this work.}\label{MR1plot}
\end{center}
\end{figure}
The real and imagined data used in the analyses listed above are illustrated in Fig. \ref{MR1plot} on the background of mass-radius M(R) sequences predicted by 215 examples of meta-model EOSs with parameters covering their prior ranges listed in Table \ref{tab-prior}. 
It is seen that even with this small set of representative meta-model EOSs, the predicted M(R) curves are rather diverse with the global slope dM/dR from positive to negative. Moreover, for NSs heavier than 1.4M$_{\odot}$ but below the maximum mass (i.e., M$_{{\rm TOV}}$) on a given M(R) curve, the local slope dM/dR can changes its sign. These results from using the meta-model EOS of purely $npe\mu$ matter are similar to the results shown in Fig. 1 of Ref. \cite{Ferreira:2024hxc} from using 40435 EOSs constructed by using piecewise polytropes for the high-density EOSs (which does not know the composition or nature of the matter as discussed earlier) satisfying all existing theoretical and observational constraints. 

It is necessary to note here that similarly diverse predictions for the M(R) curve have been made by many groups over many years. Often, a changing sign of dM/dR in a large range of mass or radius is associated with the onset of new degrees of freedom (e.g., strange/quark stars with hyperons or quark deconfinement), while a consistently positive one is for nucleonic or hadronic NSs. For a very brief summary of the predictions, see, e.g., the introduction of Ref. \cite{Ferreira:2024hxc} and/or Fig.1 of Ref. \cite{Han22}. Thus, both global and local dM/dR slopes carry very rich information about the physics underlying the M(R) curve and the nature of NSs. Surely, knowing them from observations would place strong constraints on the EOS models. 
In particular, information about the local dM/dR is important for investigating the possible existence of strange stars \cite{Ferreira:2024hxc} and/or twin stars having approximately the same mass but different radii \cite{Alford:2013aca,Han22} as well as fine features of quarkyonic matter \cite{Zhao20}. Actually, one possible reason for the non-detection of twin stars so far is that their radius separation predicted with most EOSs are mostly less than the best precision $\Delta R$ of present NS radius measurements \cite{Zhang:2024npg}.

Indeed, stimulated by the promised high-precision radius measurements especially for massive NSs, significant efforts have been made recently by several groups to investigate implications of knowing observationally the slope dM/dR and the fundamental physics underlying it. Naturally, with only few NS radius measurements available so far and they all have relatively large error bars compared to their masses, theoretical studies going after differential properties of M(R) curves can have a lot of freedom for imagination. With possibly some biases, we briefly mention below three examples most relevant to the present work. 
\begin{enumerate}
    \item 
Example-1: Very recently, the authors of Ref.\ \cite{Ferreira:2024hxc} classified their 40435 M(R) curves into three groups with the global dM/dR negative, infinity and positive. They then explored how the global and/or local dM/dR slopes at 1.4M$_{\odot}$ and two heavier mass points can inform us about M$_{{\rm TOV}}$ and the corresponding radius R$_{{\rm TOV}}$\cite{Ferreira:2024hxc}. In turn, observational constraints on the latter are expected to constrain the sign of dM/dR globally or locally and thus the underlying dense matter EOS. 
\item
Example-2: Earlier, in Ref.\ \cite{Han22}, stimulated by the prospects of establishing the radii of massive neutron stars in PSR J1614-2230 and PSR J0740+6620 from NICER and Chandra observatories, the authors studied how the radius measurement of the most massive NSs together with existing information about the radii of canonical neutron stars can be used to determine the global dM/dR. They further studied how properties of both hadronic and quark matter EOSs affect the dM/dR in a hybrid model coupling a hadronic EOS through a first-order phase transition to a quark matter EOS characterized by a constant speed of sound. Among several interesting findings, it was pointed out that a common radius exists for two different masses in the case of dM/dR$>0$ for hadronic NSs similar to some of the examples shown in Fig. \ref{MR1plot}. Moreover, it was noticed that such behavior is also possible in quarkyonic models as demonstrated in Fig. 8 of Ref. \cite{Zhao20}. 
\item
Example-3: Earlier, in Ref.\ \cite{xie2020bayesian}, using three imagined M(R) functions with a constant global dM/dR positive, negative and infinity, respectively, within the same model framework as in the present work, the inferred PDFs of EOS parameters in the three cases were compared by two of us. Specifically, we used $R(M)=R_{1.4}\pm 4.2(M/M_{1.4}-1)$ as well as a constant of
$R(M)=R_{1.4}=11.9\pm 1.4$ km as shown in Fig.2 of Ref. \cite{xie2020bayesian}.
The PDFs of $L$ and $K_{\rm sym}$ parameters were found to have the most significant dependence on the dM/dR of the input M(R) curve. Moreover, the inferred SNM EOS $E_0(\rho)$ and nuclear symmetry energy $E_{\rm sym}(\rho)$ from the three imagined M(R) curves are significantly different at suprasaturation densities. However, only a single precision of $\Delta R=1.4$ km at 90\% confidence level was used. The present work is a natural extension of that in Ref. \cite{xie2020bayesian}.
\end{enumerate}

Realizing the physics goals in the examples listed above relies on the high-precision radius measurements. For example, to determine the local slope dM/dR may require high-precision observations of several NSs clustered around a given mass \cite{Ferreira:2024hxc}. However, such observations are extremely challenging practically. Results of our following calculations using imagined radius data with high-precision hopefully will help further motivate astrophysicists to carry out such observations. Moreover, we will feel being greatly honored if our results based on mostly imagined data have some values in helping plan future NS observations and/or understand the physics underlying the real data hopefully coming soon.

\begin{figure*}[ht]
%\vspace{-0.6cm}
\begin{center}
 \resizebox{0.49\textwidth}{!}{
  \includegraphics[width=12cm,height=10cm]{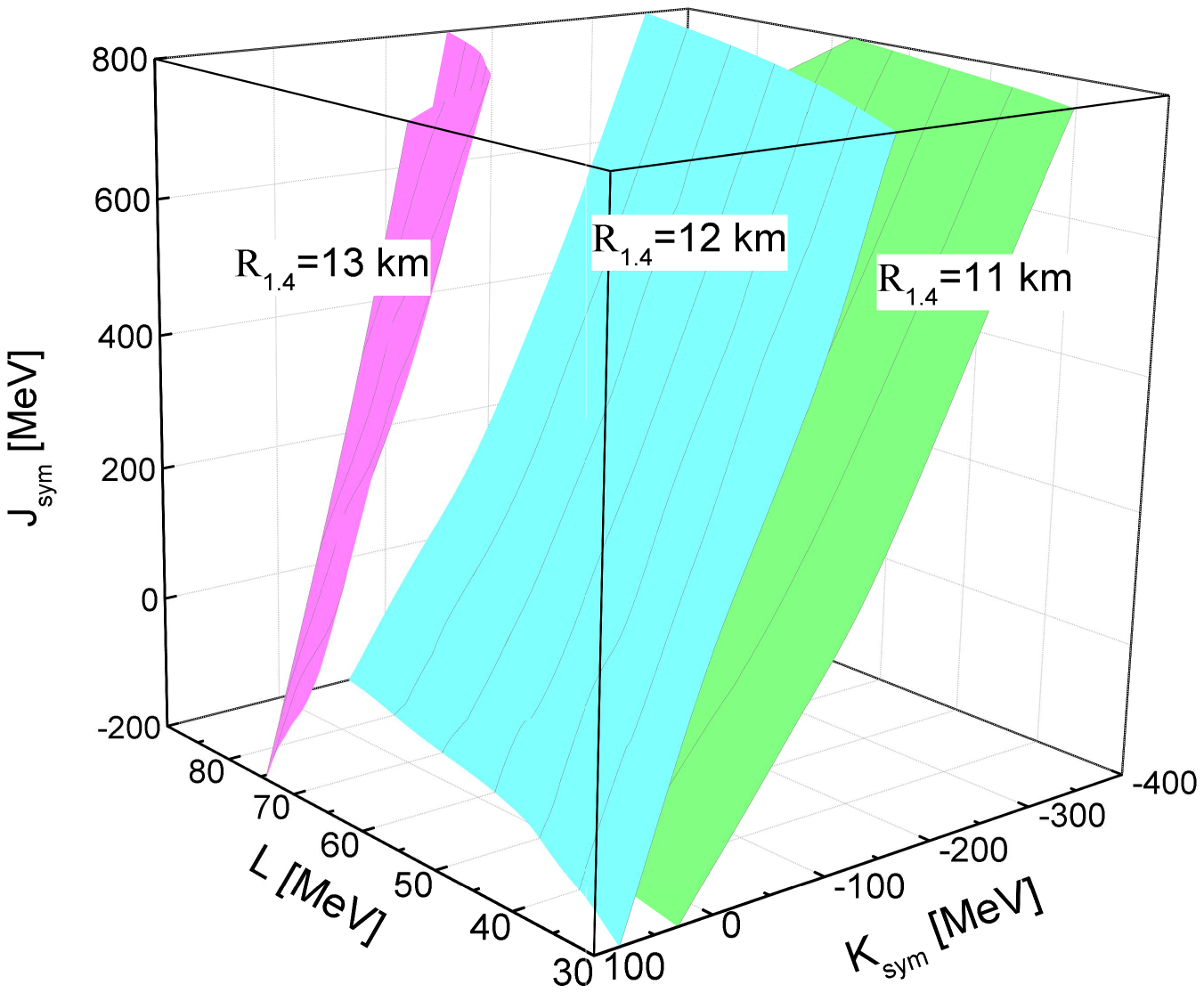}
  }
 \resizebox{0.49\textwidth}{!}{
  \includegraphics[width=12cm,height=10cm]{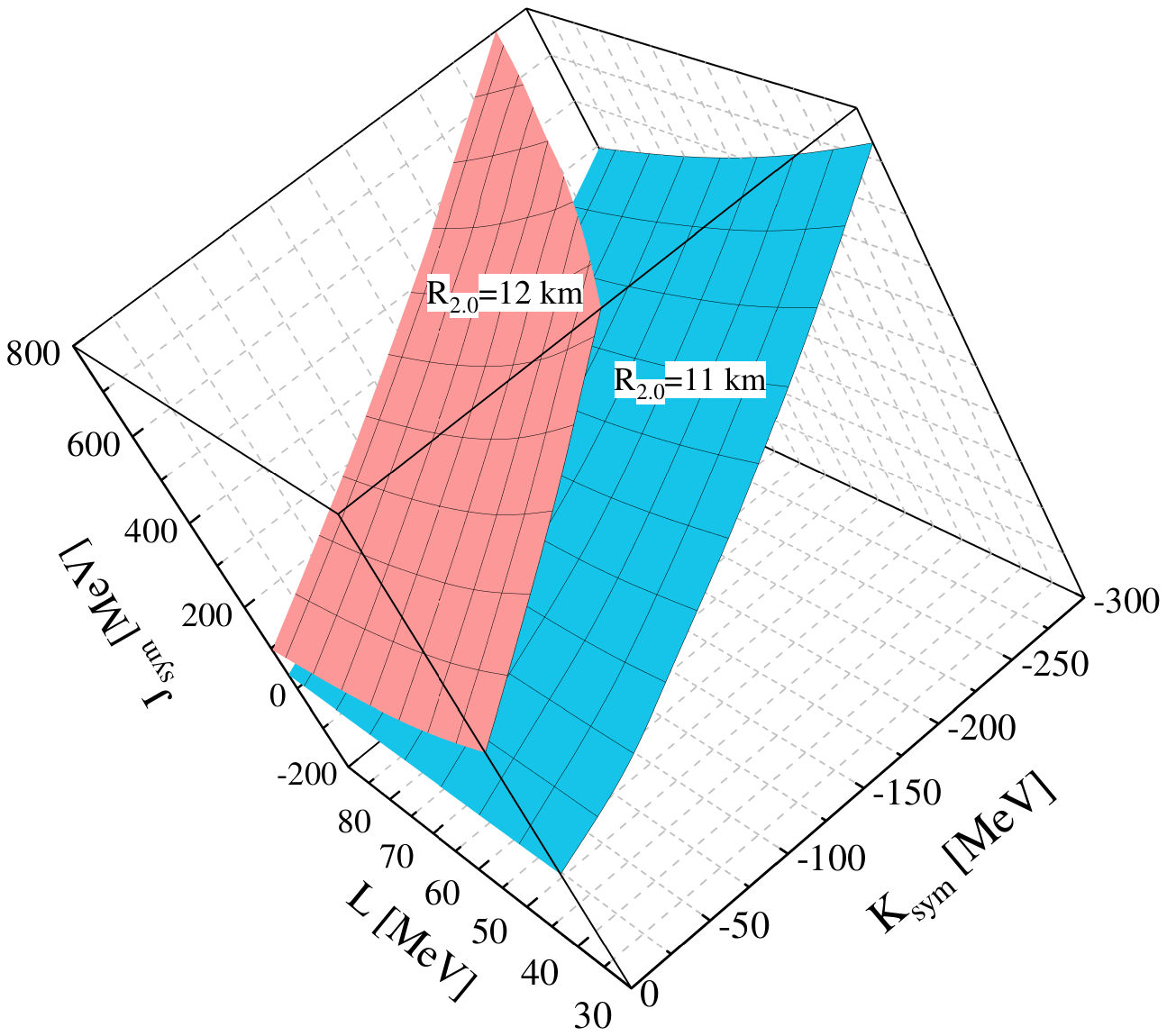}
  }
%   \vspace{-1.3cm}
  \caption{(color online) Constant surfaces of NS radii in the 3-dimensional  parameter space $L-K_{\rm{sym}}-J_{\rm{sym}}$ of high-density nuclear symmetry energy for $R_{1.4}$ (left) and $R_{2.0}$ (Right) with $E_0(\rho_0)=-15.9$ MeV, $K_0=240$ MeV, $E_{\rm sym}(\rho_0)=31.7$ MeV and $J_0=-180$ MeV, respectively. }\label{R14&R20}
\end{center}
\end{figure*}

\begin{figure*}[ht]
\hspace{-10cm}
\begin{center}
 \resizebox{1.1\textwidth}{!}{
  \includegraphics[width=18cm,height=10cm]{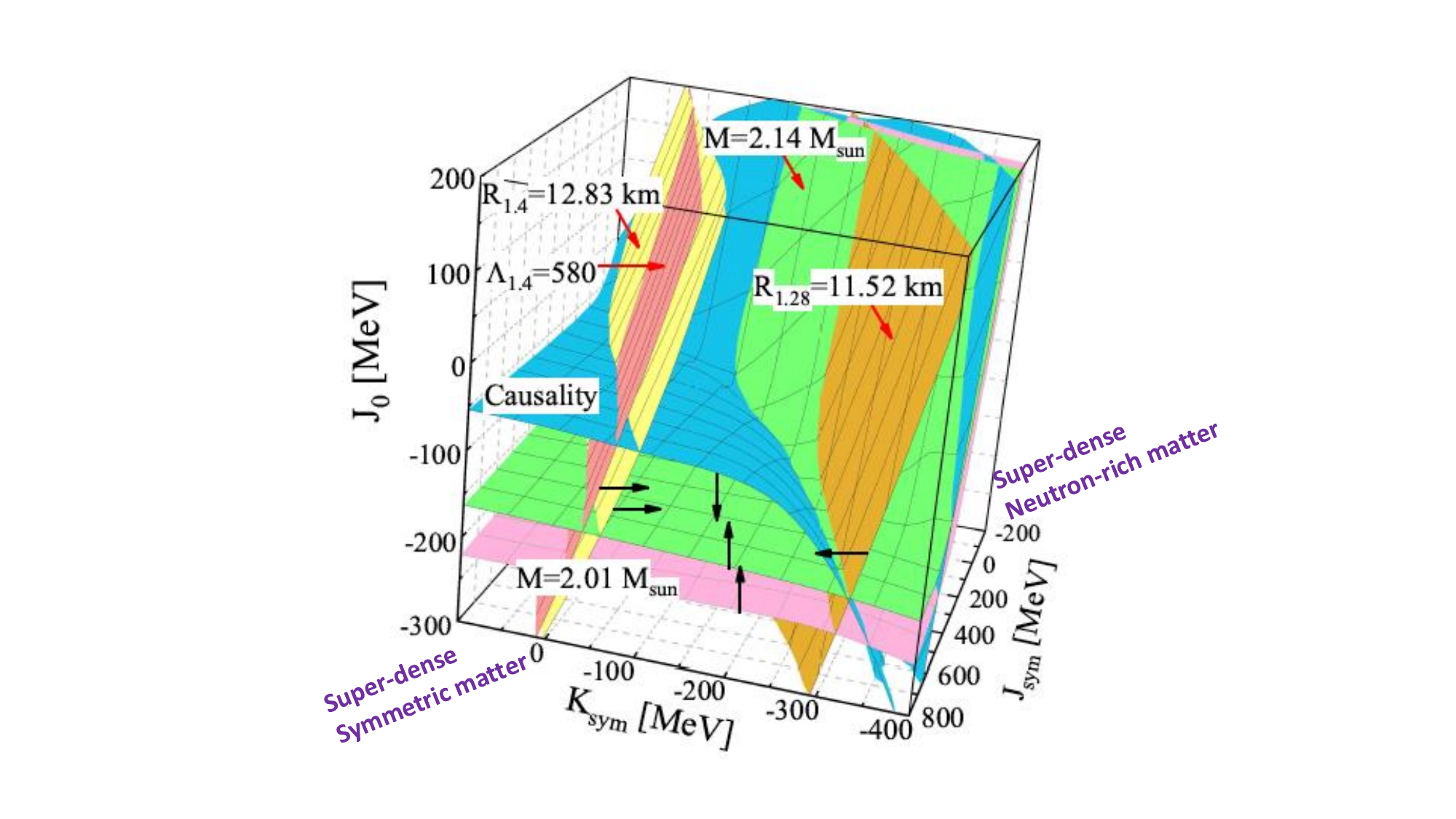}
  }
  \vspace{-1cm}
\caption{The constant surfaces of NS maximum mass M$_{\rm TOV}$=2.01$M_\odot$ (pink) and M$_{\rm TOV}$=2.14$M_\odot$ (green), the upper limit of the tidal polarizability $\Lambda_{1.4} = 580$ (salmon surface) and the corresponding constant radius surface of $R_{1.4} = 12.83$ km (light yellow surface)
for a canonical NS involved in GW170817, $R_{1.28}$ = 11.52 km (orange surface) for a NS of mass 1.28$M_\odot$, and the causality surface (blue surface) on which the speed of sound at NS center is equal to the spped of light in the 3D EOS parameter space of $K_{\rm sym}-J_{\rm sym}-J_0$ with $E_0(\rho_0)=-15.9$ MeV, $K_0=240$ MeV, $E_{\rm sym}(\rho_0)=31.7$ MeV and $L=58.9$ MeV, respectively. The black arrows show the directions supporting the constraints and the red arrows direct to the corresponding surfaces. Modified from Figures originally published in Refs. \cite{zhang2018combined,Zhang:2019fog,Zhang:2021xdt,Zhang-muon} and reviewed in Ref. \cite{Li:2021thg}.}\label{3D}
\end{center}
\end{figure*}

\section{What do we expect to learn at the ultimate precision limit \texorpdfstring{$\Delta R=0$}{Delta R=0} of radius measurement?}\label{Direct}
We investigate this question here by direct inverting NS radii, causality condition and the minimum M$_{\rm TOV}$ in the 3-dimensional (3D) EOS parameter space of dense neutron-rich matter. 
The answer will help us better understand the results of statistical inference of EOS parameters from NS observations.

It has been shown earlier that NS inverse structure problems can be solved numerically by brute force
\cite{zhang2018combined,Zhang:2019fog,Zhang:2020zsc,Zhang:2021xdt} by using the same meta-model EOS discussed above. Because of visualization limitations, for this work, we first show in the left box of Fig. \ref{R14&R20} the surfaces of $R_{1.4}=11, 12$, and 13 km in the $L-K_{\rm{sym}}-J_{\rm{sym}}$ space by fixing the relatively well-constrained saturation parameters at their currently known most probable values, i.e., 
$E_0(\rho_0)=-15.9$ MeV, $K_0=240$ MeV and $E_{\rm sym}(\rho_0)=31.7$ MeV. The skewness $J_0$ of SNM is set to $-180$ MeV based on earlier Bayesian analyses \cite{xie2019bayesian,xie2020bayesian}, and it is known to have little influence on the radii of NSs \cite{zhang2018combined} as we shall demonstrate in Fig. \ref{3D} and discuss in more detail later. These constant-radius surfaces correspond to $\Delta R=0$, thus the ultimate limit of precise radius measurements. Examining the results shown in Fig. \ \ref{R14&R20}, we notice the following points most relevant for our understanding of the Bayesian results:
\begin{enumerate}
    \item All three constant-radius surfaces for $R_{1.4}$ run through the whole range of $J_{\rm{sym}}$ considered, indicating that precise radius measurements of canonical NSs do not provide strong constraints on this parameter characterizing $E_{\rm{sym}}(\rho)$ above about $3\rho_0$. 
\item With a large radius, e.g., $R_{1.4}=13$ km, both $L$ and $K_{\rm{sym}}$ are at the large-positive corner of the $L-K_{\rm{sym}}$ plane. The surface is rather vertical, indicating that its projection onto the $L-K_{\rm{sym}}$ plane covers only a small range in the $K_{\rm{sym}}$ direction, and there is an approximately linear anti-correlation between $L$ and $K_{\rm{sym}}$. On the other hand, with a smaller radius, e.g., $R_{1.4}=11$ km, $L$ can be in its whole prior range from 30 to 90 MeV and $K_{\rm{sym}}$ covers a big range from about 0 to -400 MeV as the surface becomes more inclined in the negative $K_{\rm{sym}}$ direction. Moreover, the correlation between $L$ and $K_{\rm{sym}}$ becomes more complicated.
Furthermore, with the same separation of 1 km in radius, the separation in the 3D EOS parameter space between $R_{1.4}=13$ km and 12 km is much larger than that between $R_{1.4}=11$ km and 12 km, indicating a strong non-linear correspondence between $R_{1.4}$ and the EOS parameters especially the $K_{\rm{sym}}$. 
\item Based on the above observations, one can expect that in Bayesian analyses using NS radius data as the precision $\Delta R$ decreases from about 1 km to 0.1 km gradually, the peaks of the PDFs of $L$ and $K_{\rm{sym}}$ will shift towards their lower boundaries. While the EOS parameters are generated uniformly in their prior ranges in each step of the MCMC process, with observational data of $R_{1.4}$=12 km for example, there are many more combinations of EOS parameters on the right (lower R) than left (higher R) side to produce $R_{1.4}$ values within $12\pm \Delta R$ km. 
\item As the precision $\Delta R$ reaches its ultimate limit of zero, the Bayesian analyses should reveal the constant-radius surface. Every point on the surface represents a unique EOS. Namely, there are an infinite number of nuclear EOSs given a precise radius measurement. All the latter itself can constrain is the correlations among the EOS parameters. In the examples studied here, it is the correlations among $L-K_{\rm{sym}}-J_{\rm{sym}}$. Moreover, more precise radius measurements are expected to strengthen the correlations revealed. Of course, a 3D picture obtained by fixing other EOS parameters at their currently known most probably values is not a complete picture of reality. That is why we have to turn to the statistical inference using multivariate Bayesian analyses. Nevertheless, as we shall demonstrate, the above statements are still true when the uncertainties of all six EOS parameters are considered. Furthermore, as indicated in earlier studies \cite{zhang2018combined,Zhang:2019fog,Zhang:2020zsc,Zhang:2021xdt}, additional observables and/or physics filters/conditions may help break some of the correlations and thus narrow down the uncertainties of the EOS parameters.

\item Shown in the right box of Fig. \ref{R14&R20} are the constant-radius surfaces of $R_{2.0}=11$ km and 12 km, respectively. Since to obtain the mass 2.0M$_{\odot}$, a rather stiff high-density SNM EOS is required. In this example, using the same $J_0=-180$ MeV as for the calculation of constant surfaces of $R_{1.4}$, it is impossible to obtain a large radius of $R_{2.0}=13$ km. The radius $R_{2.0}$ of 2.0 M$_{\odot}$ NS indeed depends more on the high-density symmetry energy parameter $J_{\rm{sym}}$. The surface $R_{2.0}=12$ km requires significantly larger $J_{\rm{sym}}$ values compared to the $R_{2.0}=11$ km surface.  Comparing the 11 km and 12 km surfaces for $R_{1.4}$ and $R_{2.0}$, respectively, it is seen that changing $R_{2.0}$ by 1 km does not require as much change in $K_{\rm{sym}}$ as for $R_{1.4}$. For both values of $R_{2.0}$ considered, the whole prior range of $L$ is acceptable. Thus, one expects the high-precision measurements of $R_{2.0}$ may put more constrains on $J_{\rm{sym}}$ but less on $K_{\rm{sym}}$ and $L$ compared to the measurements of $R_{1.4}$.
\end{enumerate}

The focus of this work is to study how the precision of NS radius measurements affect the inference of high-density EOS parameters through the $P_\mathrm{R}$. Naturally, effects of the latter are intertwined 
with those from enforcing the causality, minimum M$_{\rm TOV}$ and other physics requirements through the 
$P_{\rm{filter}}$ and $P_{\rm{mass,max}}$ factors in calculating the likelihood function of Eq. (\ref{Likelihood}). Thus, completely disentangling the impact from radius measurement and additional conditions is challenging. Nevertheless, much information can still be obtained in both direct inversion and Bayesian inference. In fact, in Section 4 of Ref. \cite{xie2019bayesian} within the same Bayesian+meta-EOS model framework, two of us investigated in detail the individual roles of the model ingredients and the NS radius data on the PDFs of EOS parameters by varying one factor or model ingredient at a time using the same radius data. Among many interesting findings relevant to the present work, we notice particularly
\begin{enumerate}
    \item 
With or without enforcing the causality condition the resulting PDFs of the high-density SNM EOS parameter $J_0$ are dramatically different in shape (Fig.7 of Ref. \cite{xie2019bayesian}). While the PDF of $J_{\rm sym}$ also show significant change quantitatively, all other EOS parameters have little changes. 
    \item 
Comparing calculations without requiring a minimum M$_{\rm TOV}$ that all EOSs have to support with those 
using a sharp cut-off for M$_{\rm TOV}$ at 1.97, 2.01 and 2.17 M$_{\odot}$, respectively, all Gaussian shaped PDFs of $J_0$ are found to center around different means with various standard deviations (Fig.10 of Ref. \cite{xie2019bayesian}). Similar to the above case, while the PDF of $J_{\rm sym}$ also shows significant changes quantitatively, the PDFs of all other EOS parameters have little changes.
\item The mass measurements of the most massive NSs (the targets have been changing) discovered so far also have uncertainties. Comparing calculations assuming the M$_{\rm TOV}$ has a Gaussian distribution centered at 2.01 M$_{\odot}$ with $\sigma_{\rm M}=0.04$ M$_{\odot}$ with another one centered at 2.17 M$_{\odot}$ with $\sigma_{\rm M}= 0.11$ M$_{\odot}$, the PDFs of both $J_0$ and $J_{\mathrm{sym}}$ shift towards higher values as the center mass increases from 2.01 to 2.17 M$_{\odot}$ while the variation of $\sigma_{\rm M}$ has little effect (Fig.11 of Ref. \cite{xie2019bayesian}). Overall, the effects of M$_{\rm TOV}$ on the PDFs of both $J_0$ and $J_{\mathrm{sym}}$ are similar whether the requirement is implemented as a sharp cut-off or through a Gaussian function.
\end{enumerate}

\begin{figure*}[ht]
%\vspace{-0.6cm}
\begin{center}
 \resizebox{1.\textwidth}{!}{
  \includegraphics[width=14cm,height=10cm]{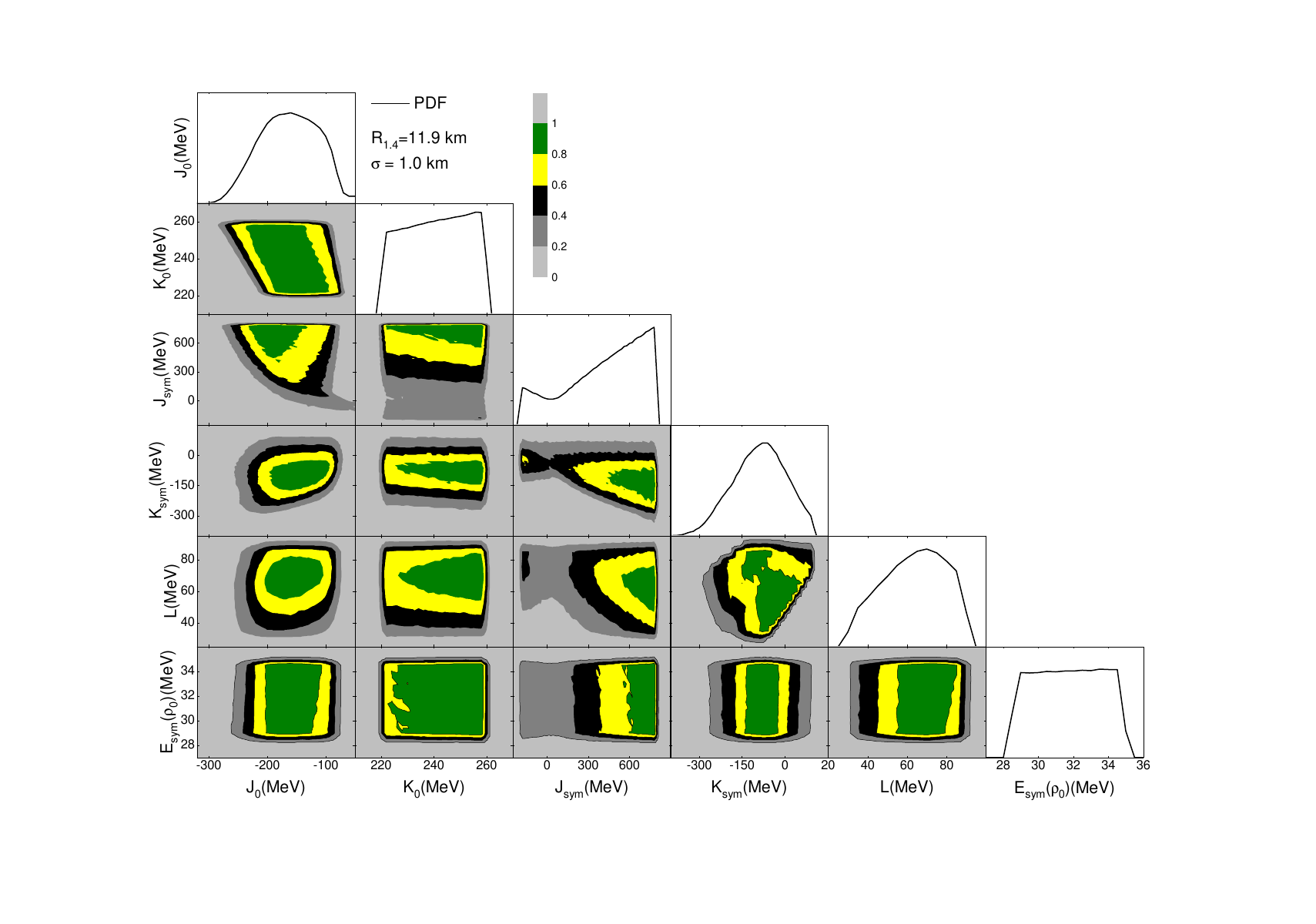}
  }
\vspace{-2.cm}
  \caption{(color online) Corner-plot: 1D posterior PDFs of EOS parameters and 2D posterior heat maps with $R_{1.4}=11.9$ km and a precision of $\Delta R=1.0$ km, respectively.}\label{s10C}
\end{center}
\end{figure*}

\begin{figure*}[ht]
%\vspace{-0.6cm}
\begin{center}
 \resizebox{1.\textwidth}{!}{
  \includegraphics[width=14cm,height=10cm]{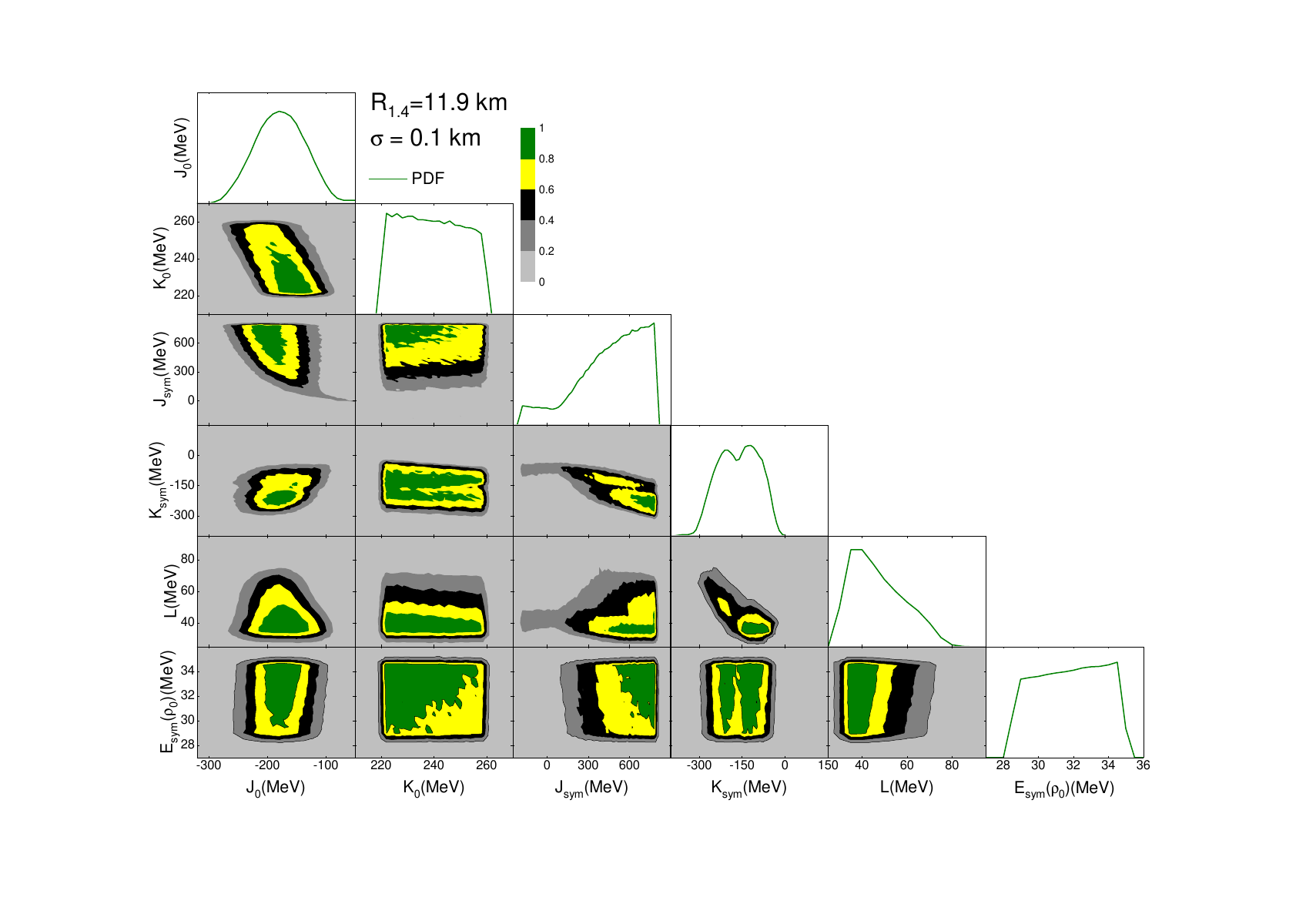}
  }
  \vspace{-2.cm}
  \caption{(color online) Corner-plot: 1D posterior PDFs of EOS parameters and 2D posterior heat maps with $R_{1.4}=11.9$ km and a precision of $\Delta R=0.1$ km, respectively.}\label{s01C}
\end{center}
\end{figure*}

To help better understand the above findings and visualize the intertwining effects of causality, minimum M$_{\rm TOV}$ and NS radius measurements at suprasaturation densities, we examine in Fig. \ref{3D} several constant surfaces in the 3D high-density EOS parameter space of $K_{\rm sym}-J_{\rm sym}-J_0$ with $E_0(\rho_0)=-15.9$ MeV, $K_0=240$ MeV, $E_{\rm sym}(\rho_0)=31.7$ MeV and $L=58.9$ MeV, respectively. 
Defined as the surface on which the speed of sound at NS center is equal to the speed of light \cite{Zhang:2018bwq}, shown in blue is the causality surface. It is seen that towards the front-left corner where the $E_{\rm sym}(\rho)$ is very stiff (with either very large $K_{\rm{sym}}$ and/or $J_{\rm{sym}}$ the $\delta$ approaches zero (SNM), the causality surface becomes very flat indicating some small effects of $J_0$ and almost no effect from $E_{\rm sym}(\rho)$. Ultimately, the speed of sound at NS centers becomes EOS independent reaching a constant determined by the General Relativity (GR) nature of strong-field gravity \cite{Cai23a,Cai23b,Cai23c}. At the back-right corner where the $E_{\rm sym}(\rho)$ is very soft (with either very large but negative $K_{\rm{sym}}$ and/or $J_{\rm{sym}}$ the $\delta$ approaches 1.0 (pure neutron matter), however, the high-density behavior of $E_{\rm sym}(\rho)$ affects significantly the location of the causality surface. The two surfaces corresponding to 
M$_{\rm TOV}$=2.01$M_\odot$ (pink) and M$_{\rm TOV}$=2.14$M_\odot$ (green) behave very similarly. In particular, at the SNM limit they are completely determined by $J_0$. For dense neutron-rich matter, $K_{\rm{sym}}$ and/or $J_{\rm{sym}}$ play important roles in determining the M$_{\rm TOV}$. In particular, towards the limit of pure neutron matter (back-right corner), when the $E_{\rm sym}$ becomes very soft at suprasaturation densities, the $J_0$ has to be higher such that the total pressure in the core of NSs is still high enough to support the same M$_{\rm TOV}$. The above observations about the causality and M$_{\rm TOV}$ surfaces all indicate consistently that they both constrain most tightly the $J_0$ parameter but less tightly the $K_{\rm{sym}}$ and $J_{\rm{sym}}$. 

It is very interesting to see that the upper limit of the tidal polarizability $\Lambda_{1.4} = 580$ (salmon surface) and the corresponding constant radius surface of $R_{1.4} = 12.83$ km (light yellow surface) for a canonical NS involved in GW170817 as well as the $R_{1.28}$ = 11.52 km (orange surface) for the 68\% lower boundary from one analysis of the PSR J0030+0451 observation \cite{riley2019nicer} are all rather vertical, indicating that they are almost independent of $J_0$. As mentioned earlier and stressed in our earlier publications, it means that more precise radius measurements of canonical NSs will not constrain the stiffness of high-density SNM, but the $E_{\rm sym}$ at low-intermediate densities characterized by mostly $L$ and $K_{\rm{sym}}$ and influenced somewhat by the uncertainties of $J_{\rm{sym}}$.  

The space pointed by the black arrows is the allowed high-density EOS parameter space surrounded from the left and right by the NS radii (or tidal polarizability), while from the top and bottom by the causality and minimum M$_{\rm TOV}$ requirement, respectively. The cross lines between some of them (e.g., the causality and M$_{\rm TOV}$=2.01$M_\odot$ surfaces) set the upper or lower boundaries for nuclear symmetry energy and/or SNM EOS. As indicated by Fig. \ref{3D}, even in the 3D space, the various constraints are highly intertwined. In the full 6D EOS parameter space considered in this work, we have to rely on the power of Bayesian analyses to help disentangle statistically the individual roles of various constraints and the radius measurements. 

The correlations themselves among the EOS parameters are inherent in the EOS model (e.g, adjacent coefficients in parameterizing $E_0(\rho)$ and $E_{\rm sym}(\rho)$ are anti-correlated when their constraints from nuclear experiments are applied), TOV equations, and the additional conditions (causality, dynamical stability and the requirement on the minimum M$_{\rm TOV}$). As illustrated by Fig. \ref{3D}, the radius measurement of a given NS introduces additional constraints/correlations among the EOS parameters. Increasing the precision $\Delta R$ of radius measurements is expected to tighten the correlations among the EOS parameters, resulting in a decrease in the statistical uncertainty on the posterior. In particular, the pairwise correlations between EOS parameters inherent in their 2D joint posterior distribution functions (2D PDFs) are expected to be pronounced when the $\Delta R$ decreases.

\section{What new physics can we learn about the EOS of supradense neutron-rich matter from high-precision NS radius measurements?}\label{Results}
In presenting our results, as examples, we first show the standard corner-plots of 1D and 2D posterior distribution functions of the 6 EOS parameters inferred from the radius data with $R_{1.4}=11.9$ km and a precision of $\Delta R=1.0$ km and $0.1$ km in Fig. \ref{s10C} and Fig. \ref{s01C}, respectively. 
Such plots frequently reported in the literature, see, e.g. Ref. \cite{Pang:2023dqj} and/or our previous publications \cite{xie2019bayesian,xie2020bayesian,xie2021bayesian,Xie:2024mxu}, are very informative. The major characteristics of the 1D PDFs, namely, the maximum {\it a posteriori} (MaP) value, the highest posterior density (HPD) 68\% confidence intervals calculated using the approach given in Ref.~\cite{N-Turkkan}, as well as the means and standard deviations of the EOS parameters are listed in Table \ref{MP14} and \ref{Mean14}, respectively. In the case of using $\Delta R=1.0$ km, 
the results are consistent with those in our previous publications focusing on investigating different questions \cite{xie2019bayesian,xie2020bayesian,xie2021bayesian,Xie:2024mxu}. In the case of using $\Delta R=0.1$ km, one qualitatively new feature is the double peak of the PDF of $K_{\rm sym}$ as we shall investigate in great detail. One common feature which deserves some special emphasis is the PDF of $J_{\rm sym}$: it peaks at the upper and lower boundaries of its prior range. As indicted earlier in this paper and our previous publication \cite{xie2020bayesian}, $J_{\rm sym}$ carries information about the poorly known high-density behavior of nuclear symmetry energy. As demonstrated quantitatively and discussed in detail in Section 3 of Ref. \cite{xie2020bayesian}, the PDF of $J_{\rm sym}$ is unstable against the variations of its prior range, indicating that the NS radius data used do not strongly constrain $J_{\rm sym}$. We shall study quantitatively in Section \ref{JKsym} how this uncertainty may affect the structure of PDF($K_{\rm sym}$) especially the strength and location of its double peaks due to the strong anti-correlation between $K_{\rm sym}$ and $J_{\rm sym}$ as well as that between $K_{\rm sym}$ and $L$ as indicated by the 2D PDFs shown already in Fig. \ref{s10C} and Fig. \ref{s01C}. 

Since the main purpose of this work is to investigate how the precision of NS radius measurements affects the 1D and 2D PDFs of high-density EOS parameters, it is necessary for us to compare our results from using different $\Delta R$ values on a single plot for the 1D and 2D PDFs separately. In the following, we will thus focus on comparing the 1D and 2D PDFs of $J_0$, $J_{\rm sym}$, $K_{\rm sym}$ and $L$ on separate plots instead of the traditional corner-plots. For most parts, we skip the discussions about the results related to $K_0$ and $E_{\rm sym}(\rho_0)$ as they are relatively well determined and have little effect on NS radii and/or tidal deformations. 

\begin{figure*}[ht]
%\vspace{-0.6cm}
\begin{center}
 \resizebox{0.9\textwidth}{!}{
  \includegraphics[width=10cm,height=10cm]{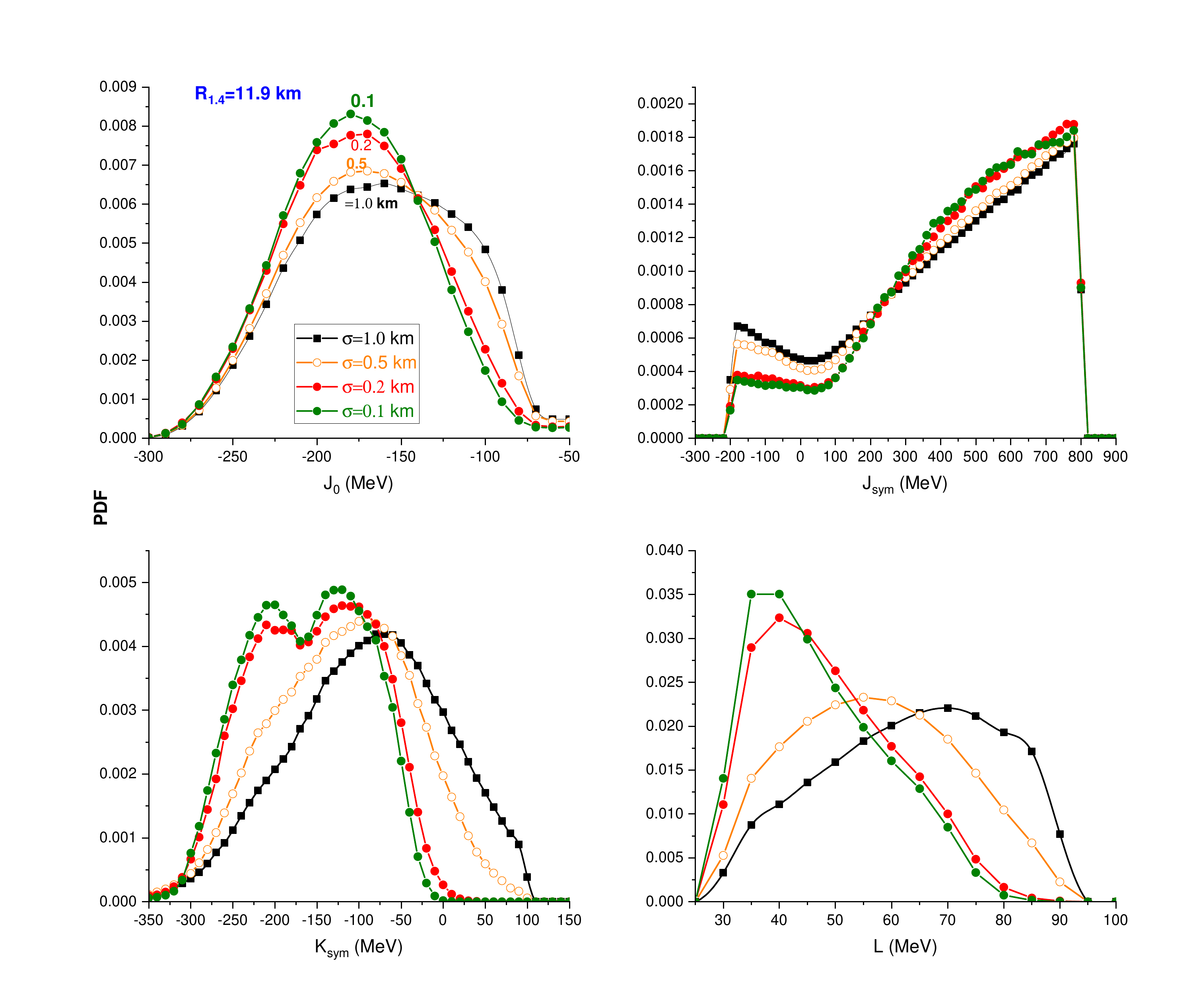}
  }
%    \vspace{-1.3cm}
  \caption{(color online) Posterior PDFs of EOS parameters with $R_{1.4}=11.9$ km and a precision of $\Delta R=1.0, 0.5, 0.2$, and $0.1$ km, respectively.}\label{R14PDF}
\end{center}
\end{figure*}

\begin{figure*}[ht]
%\vspace{-0.6cm}
\begin{center}
 \resizebox{0.9\textwidth}{!}{
  \includegraphics[width=16cm,height=5cm]{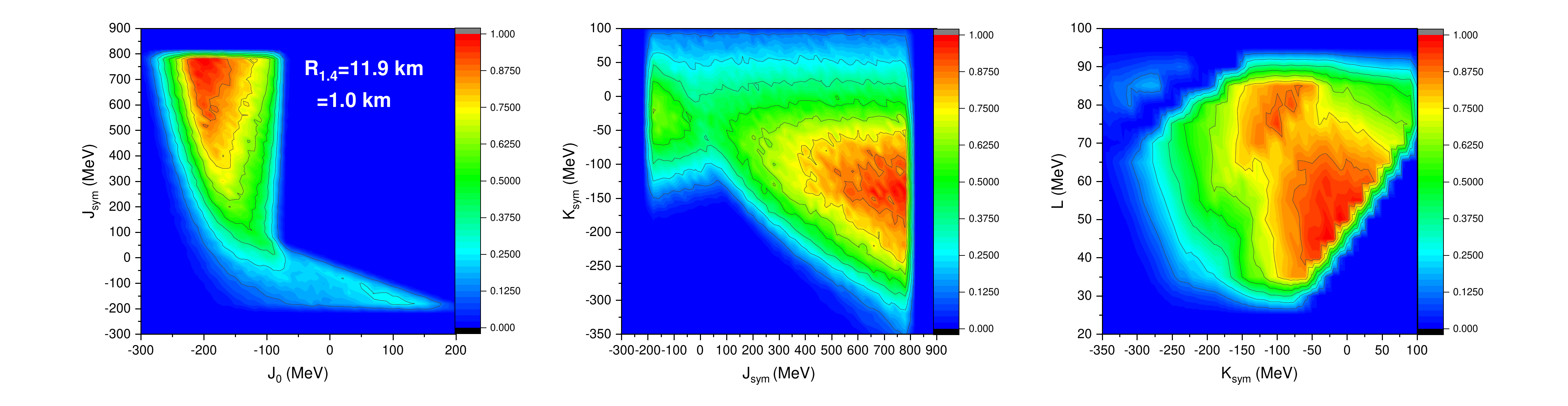}
  }
 \resizebox{0.9\textwidth}{!}{
  \includegraphics[width=16cm,height=5cm]{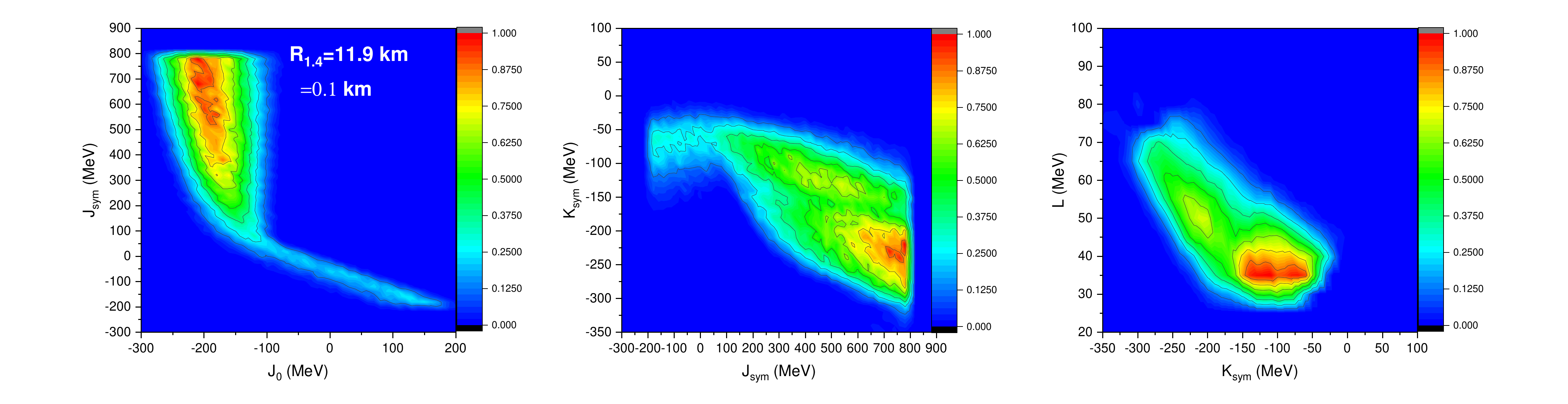}
  }
   \resizebox{0.9\textwidth}{!}{
  \includegraphics[width=16cm,height=5cm]{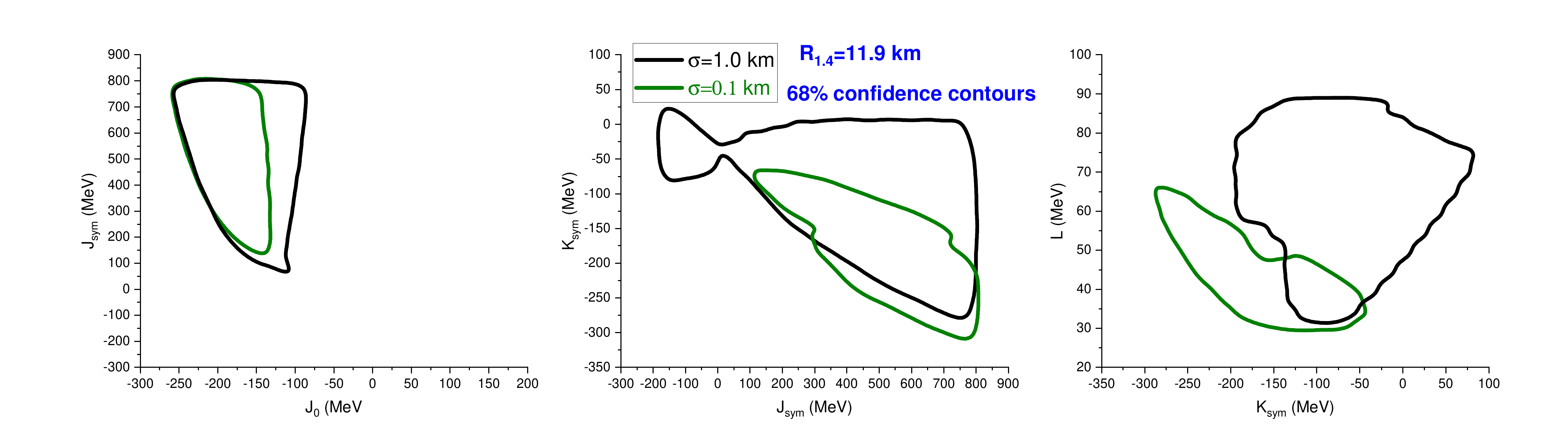}
  }
 %  \vspace{-1.3cm}
  \caption{(color online) Posterior 2D probability distribution functions among the four high-density EOS parameters using R$_{1.4}$=11.9 km and a precision of $\Delta$R=1.0 km (upper panels) and 0.1 km (middle panels), respectively. The bottom panels compare their 68\% confidence boundaries.}\label{R14COR}
\end{center}
\end{figure*}

\subsection{What can we learn from comparing the 1D and 2D posterior PDFs of EOS parameters inferred from the data \texorpdfstring{$R_{1.4}=11.9$}{R1.4=11.9} km with precision \texorpdfstring{$\Delta R=1.0, 0.5, 0.2$}{Delta R=1.0, 0.5, 0.4, 0.3, 0.2}, and \texorpdfstring{$0.1$}{0.1} km?}
Shown in Fig. \ref{R14PDF} is a comparison of the 1D posterior PDFs of $J_0$, $J_{\rm{sym}}$, $K_{\rm{sym}}$, and $L$, using the radius data of $R_{1.4}=11.9$ km with a precision of $\Delta R=1.0, 0.5, 0.2$, and $0.1$ km, respectively. The corresponding MaP values and 68\% HPD confidence intervals as well as 
the means and standard deviations of the four parameters are compared in Table \ref{MP14} and Table \ref{Mean14}, respectively. We notice that in our analyses we used a bin size of 5 MeV for $L$, but 10 MeV is used for the other three EOS parameters shown here. Several interesting observations can be made: 
\begin{enumerate}
    \item Both the MaP values and means of $J_0$, $K_{\rm{sym}}$ and $L$ gradually decrease, especially for the last two, as $\Delta R$ decreases. 
    They become approximately stabilized within the bin sizes used after 
    $\Delta R\leq 0.2$ km. This is consistent with our expectations based on the results shown in Fig. \ref{R14&R20}. It means that the currently available MaP values and means of these parameters, especially $K_{\rm{sym}}$ and $L$, from various Bayesian analyses by many groups in the community of the available data with NS radii measured mostly with $\Delta R\geq 1.0$ km are unreliable. High precision measurements with $\Delta R\leq 0.2$ km will be necessary to extract stable MaP values and means of $J_0$, $J_{\rm{sym}}$, $K_{\rm{sym}}$ and $L$. 
  \item On the other hand, the mean of skewness $J_{\rm{sym}}$ of symmetry energy becomes larger as the precision of measuring the radius improves. This is mainly due to its anti-correlation with both $J_0$ and $K_{\rm{sym}}$ as we shall discuss. It is known from previous Bayesain analyses \cite{xie2019bayesian,xie2021bayesian} of NS radius data, 
  the radius $R_{1.4}$ does not effectively constrain $J_{\rm{sym}}$ as indicated by the peak of its PDF at its upper boundary. The sharp drop there is due to the cut-off from its prior PDF used. Nevertheless, it is interesting to see that as $\Delta R$ decreases, because the pairwise anti-correlations of $J_{\rm{sym}}-J_0$ and $J_{\rm{sym}}-K_{\rm{sym}}$ become pronounced, the mean value of $J_{\rm{sym}}$ increases to counterbalance the decreases of $J_0$ and $K_{\rm{sym}}$. 
  \item Very interestingly, a distinguished two-peak structure is revealed in the PDF of $K_{\rm{sym}}$ when the precision of NS radius measurement is better than about 0.2 km. Generally speaking, it is not surprising that fine structures can be seen when the probes used have better resolutions. As we shall discuss next, the two-peak structure of PDF($K_{\rm{sym}}$) is completely due to the strong anti-correlations of $K_{\rm{sym}}-J_{\rm{sym}}$ and $K_{\rm{sym}}-L$ that are only visible 
  when the precision $\Delta R$ is better than about 0.2 km.
\end{enumerate}
\begin{table}[htbp]
\centering
\caption{The MaP values and their corresponding 68\% HPD credible intervals of the four EOS parameters inferred from Bayesian analyses using {\bf R$_{1.4}$=11.9 km} with precision $\Delta$R from 1.0 to 0.1 km. }\label{MP14}
\begin{tabular}{lccccccc}
  \hline\hline
  Parameters (MeV) & $\Delta$R=1.0 &  $\Delta$R=0.5 &$\Delta$R=0.2 &$\Delta$R=0.1 \\
  \hline\hline\\
%  \vspace{0.2cm}
 $J_0:$ &$-160_{-50}^{+60}$& $-170_{-50}^{+60}$ &$-170_{-50}^{+40}$& $-180_{-40}^{+50}$\\
 $J_{\mathrm{sym}}:$ &$780_{-480}^{+0}$& $780_{460}^{+0}$ &$780_{-420}^{+0}$& $780_{-420}^{+0}$ \\
 $K_{\mathrm{sym}}:$ &$-70_{-100}^{+80}$& $-100_{-100}^{+70}$ &$-120_{-100}^{+50}$& $-130_{-100}^{+50}$ \\
 $L:$ &$70_{-40}^{+5}$& $55_{-25}^{+15}$ &$40_{-10}^{+15}$& $35_{-5}^{+20}$\\
  \hline
 \end{tabular}
\end{table}
\begin{table}[htbp]
\centering
\caption{Means and standard deviations of the four EOS parameters (MeV) inferred from Bayesian analyses using {\bf R$_{1.4}$=11.9 km} with precision $\Delta$R from 1.0 to 0.1 km. }\label{Mean14}
\begin{tabular}{lccccccc}
  \hline\hline
  Pars. &$\Delta$R=1.0 &$\Delta$R=0.5 &$\Delta$R=0.2 &$\Delta$R=0.1 \\
  \hline\hline\\
%  \vspace{0.2cm}
 $J_0:$&$-149\pm 71$&$-149\pm 71$&$-161\pm 71$&$-164\pm 70$\\
  $J_{\rm sym}:$&~~$417\pm 277$&$431\pm 269$&$461\pm 248$&$463\pm 243$\\
  $K_{\rm sym}:$&$-89\pm 91$&$-119\pm 84$&$-154\pm 72$&$-162\pm 69$\\
   $L:$&$63\pm 16$&$57\pm 15$&$49\pm 12$&$47\pm 12$\\
  \hline
 \end{tabular}
\end{table}

Shown in Fig. \ref{R14COR} are the 2D posterior PDFs of the four high-density EOS parameters using R$_{1.4}$=11.9 km with a precision of $\Delta$R=1.0 km (upper panels) and 0.1 km (middle panels), respectively. The bottom panels compare their 68\% HPD confidence boundaries to be more quantitative. 
The 2D PDFs convey the full co-varying probability landscape including 
information about the pairwise correlation strength between the two parameters.
In parameterizing the 
SNM EOS $E_0(\rho)$ and symmetry energy $E_{\rm{sym}}(\rho)$, any two adjacent parameters are mathematically expected to be anti-correlated. 
While the correlations among parameters of $E_0(\rho)$ and those of $E_{\rm{sym}}(\rho)$ are mostly due to the physical requirements, e.g., $\beta-$equilibrium, charge neutrality, the final EOS has to be stiff enough to support NSs at least as massive as 1.97M$_{\odot}$ and causality condition as we discussed in the previous section. As shown in Eq. (\ref{pressure1}), the total pressure has contributions from both SNM EOS and nuclear symmetry energy. For instance,
the contributions from $J_0$ and $J_{\rm{sym}}$ are complementary, their PDFs inferred from NS data are thus expected to be anti-correlated, albeit weakly, especially for canonical NSs. 

Indeed, as shown in Fig. \ref{R14COR}, $J_{\rm{sym}}-J_0$ and $J_{\rm{sym}}-K_{\rm{sym}}$ are clearly anti-correlated. Most interestingly, it is seen that the strengths of all correlations depend strongly on the precision $\Delta R$. In particular, comparing the 2D PDFs obtained with $\Delta R=$1.0 km and 0.1 km, the strong anti-correlations of $K_{\rm{sym}}-J_{\rm{sym}}$ around ($K_{\rm{sym}}=-250$ MeV, $J_{\rm{sym}}=800$ MeV) and $K_{\rm{sym}}-L$ around ($K_{\rm{sym}}=-120$ MeV, $L=35$ MeV), especially the latter, is only clearly visible with $\Delta R=0.1$ km. Thus, in the case of $\Delta R=0.1$ km, corresponding to the peaks of PDFs around $J_{\rm{sym}}=800$ MeV and $L=35$ MeV, there are two peaks in the PDF($K_{\rm{sym}}$) around $K_{\rm{sym}}=-250$ MeV and $-120$ MeV, respectively, as shown in Fig. \ref{R14PDF}. This clearly shows the resolving power of high-precision NS radius measurements. To be more quantitative, in the bottom panels we have compared the 68\% HPD boundaries of the 2D PDFs obtained with $\Delta R=1.0$ km and 0.1 km, respectively. It is seen that the correlations are more pronounced as $\Delta R$ decreases, especially for $K_{\rm{sym}}-J_{\rm{sym}}$ and $K_{\rm{sym}}-L$ as $K_{\rm{sym}}$ is the most important parameter determining $R_{1.4}$ as mentioned earlier, with the high-precision measurement of $\Delta R=0.1$ km.

\begin{figure}[ht]
%\vspace{-1cm}
\begin{center}
 \resizebox{0.7\textwidth}{!}{
  \includegraphics[width=12cm,height=9cm]{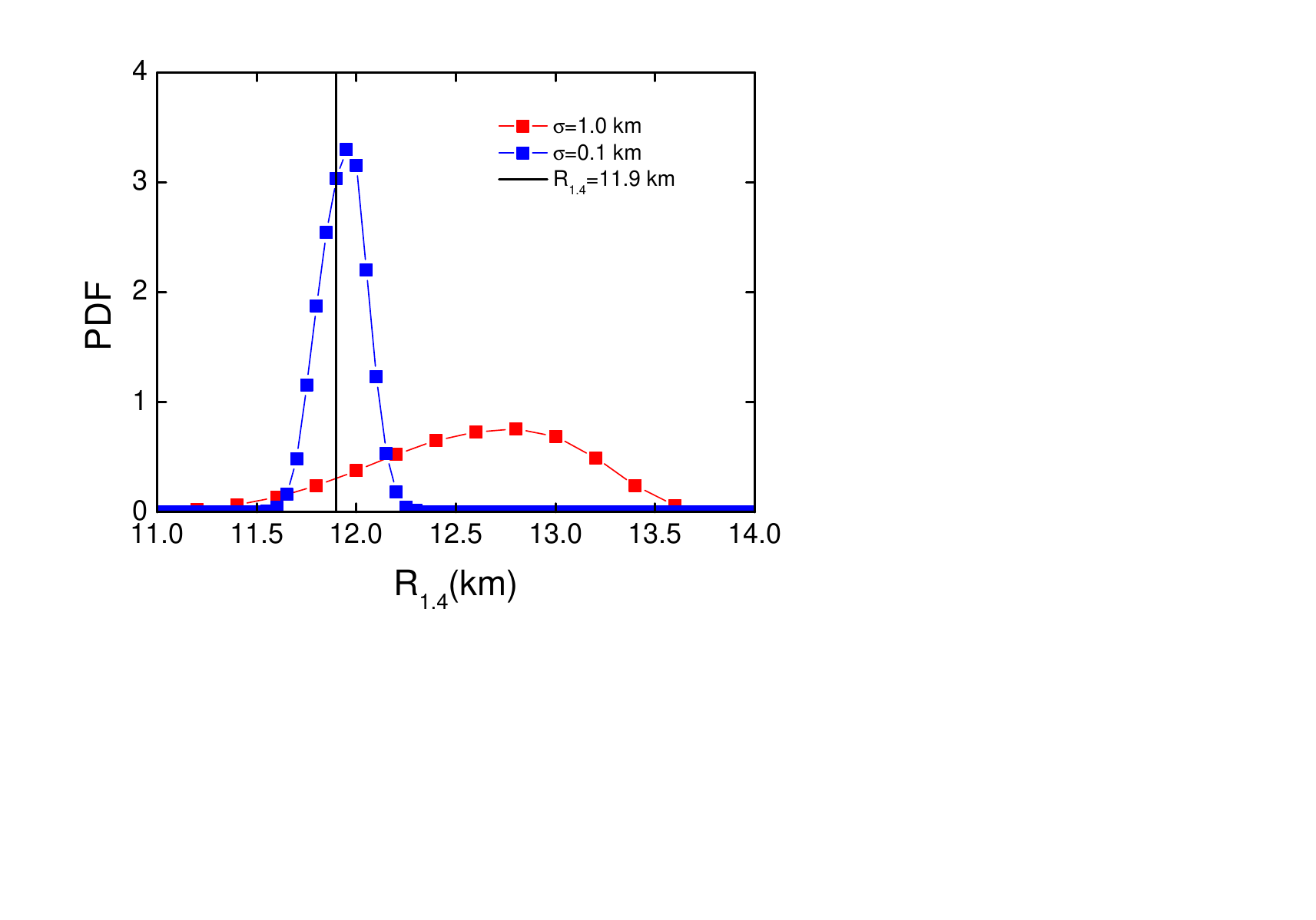}
  }
   \vspace{-3.5cm}
  \caption{(color online) Posterior PDFs of $R_{1.4}$ obtained with the input data $R_{1.4}=11.9$ km and a precision of $\Delta R=1.0$ km and $0.1$ km, respectively.}\label{R0110PDF}
\end{center}
\end{figure}

\begin{figure*}[ht]
%\vspace{-0.6cm}
\begin{center}
 \resizebox{1.\textwidth}{!}{
  \includegraphics[width=10cm,height=8cm]{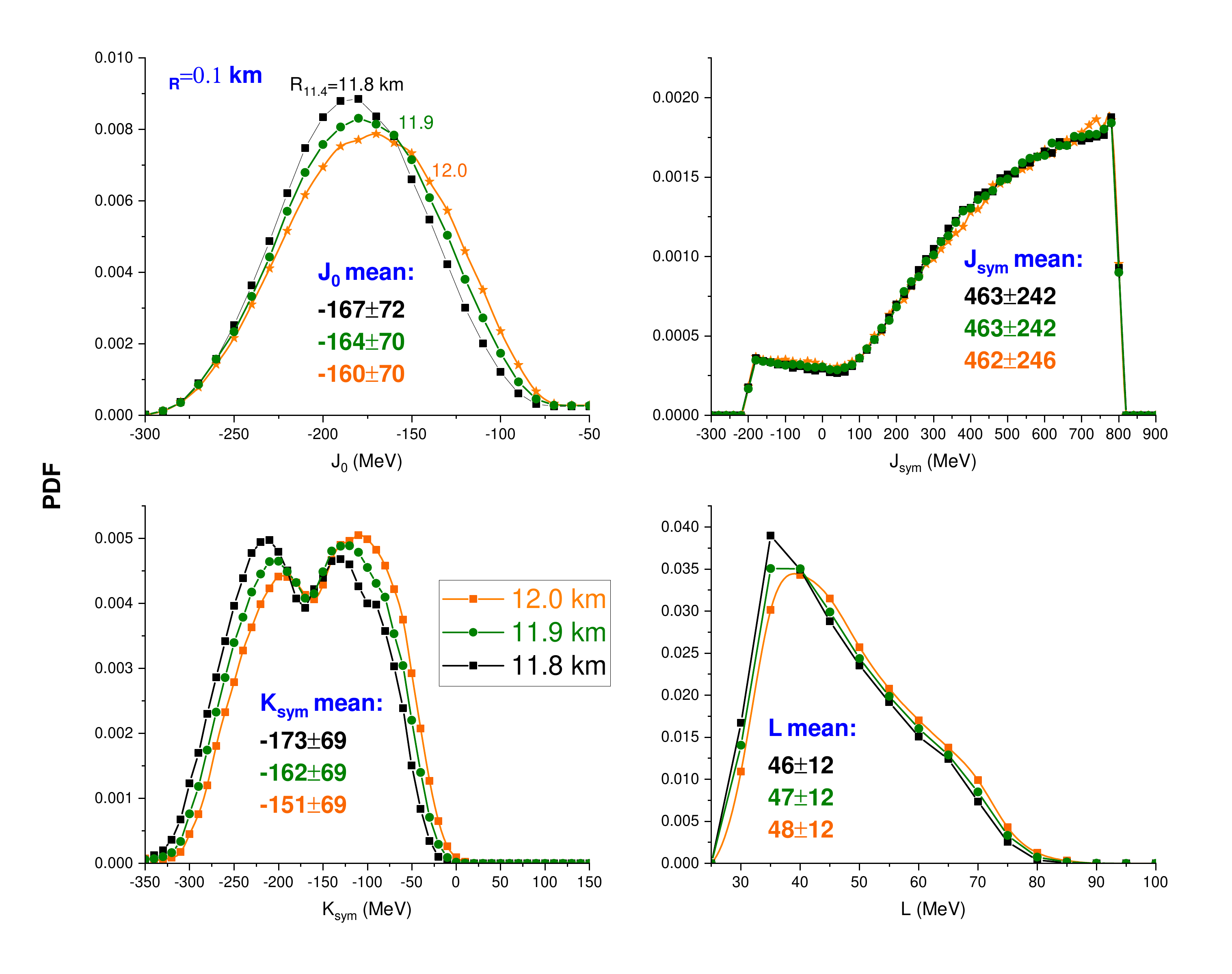}
  }
  \vspace{-1cm}
  \caption{(color online) Posterior 1D PDFs of the four EOS parameters with $R_{1.4}=11.8$, 11.9 and 12.0 km all with a precision of $\Delta R=0.1$ km, respectively.}\label{R18120}
\end{center}
\end{figure*}

To further understand the results presented above, shown in Fig. \ref{R0110PDF} are the posterior PDF of $R_{1.4}$ from the Bayesian analysis using the NS radius data of $R_{1.4}$=11.9 km with the precision $\Delta R$=1.0 km and $\Delta R$=0.1 km as indicated. As listed in Table \ref{Mean14}, when $R_{1.4}$=11.9 km with different precision from 1.0 to 0.1 km are used as input data in the Bayesian analysis, the necessary NS EOS effectively becomes softer as shown by the decreasing mean values of $J_0$, $L$, and $K_{\rm{sym}}$. This implies that the posterior NS radius predicted is smaller using the precision $\Delta R$=0.1 km than that using $\Delta R$=1.0 km. Several factors can make this happen. As we mentioned earlier, while the EOS parameters are initialized randomly and uniformly within their prior range, the predicted NS radii are not uniform or symmetric about some central values as the underlying physics is highly nonlinear. Also, although the likelihood function of observable $R$ is symmetric with respect to the input mean value of $R_{1.4}$ data, the physics requirements/filters in the MCMC process break the symmetry. Consequently, the accepted predictions of $R_{1.4}$ during the MCMC process are not necessarily symmetric about the mean value of $R_{1.4}$ data. This is consistent with the results of direct inversions shown in Fig. \ref{R14&R20} and the associated expectations discussed in Section \ref{Direct}. Indeed, as clearly indicated by the posterior PDF of the accepted $R_{1.4}$ in Fig. \ref{R0110PDF}, finally a larger (smaller) posterior radius $R_{1.4}$ is preferred when $\Delta R$=1.0 (0.1) km is used with the same input $R_{1.4}$=11.9 km in the calculations. Interestingly, as the precision improves, the MaP values of the posterior $R_{1.4}$ moves towards its input value of 11.9 km (dark vertical line). Even with $\Delta R=0.1$ km, however, it is still about 0.1 km away. This finding demonstrates once again the importance of high-precision radius measurements.

\subsection{What more can we learn from comparing the results using a radius of \texorpdfstring{$R_{1.4}=11.8, 11.9$}{R1.4=11.8, 11.9} and \texorpdfstring{$12.0$}{12.0} km all with a precision of \texorpdfstring{$\Delta R=0.1$}{Delta R=0.1} km?}

\begin{figure*}[ht]
%\vspace{-0.6cm}
\begin{center}
 \resizebox{0.95\textwidth}{!}{
  \includegraphics[width=16cm,height=5cm]{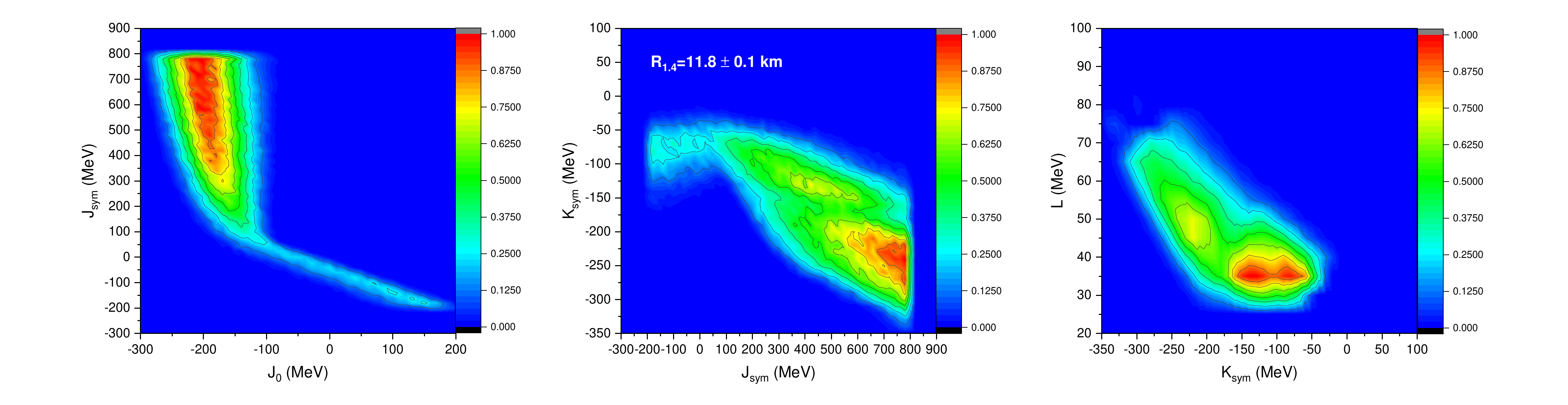}
  }
 \resizebox{0.95\textwidth}{!}{
  \includegraphics[width=16cm,height=5cm]{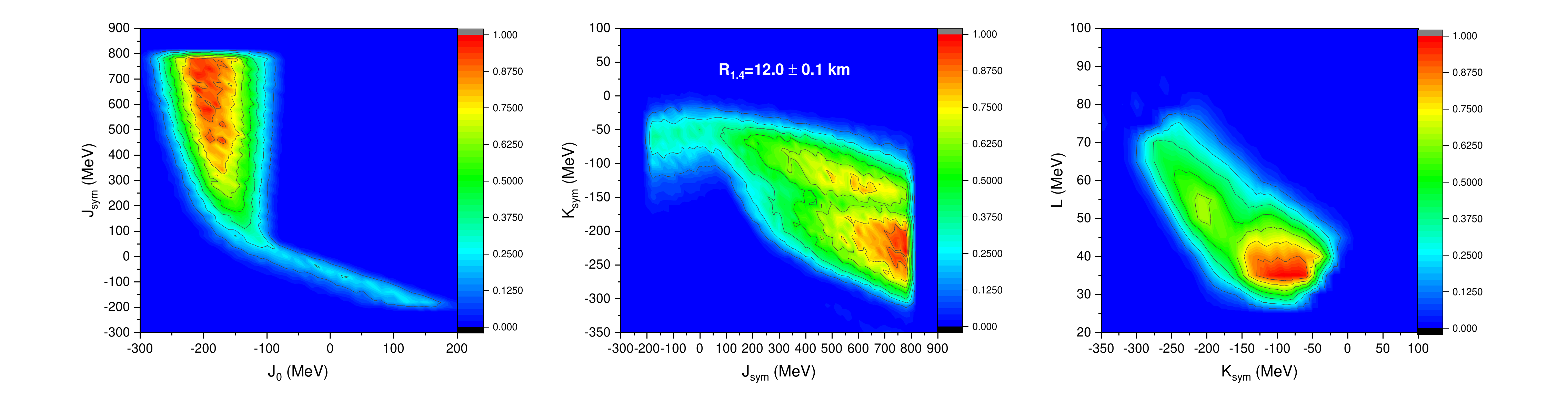}
  }
 %  \vspace{-1.3cm}
  \caption{(color online) Posterior 2D probability distribution functions among the four high-density EOS parameters using R$_{1.4}$=11.8 km (upper panels) and 12.0 km (lower panels) and a precision of $\Delta$R=0.1 km, respectively.}\label{R1812COR}
\end{center}
\end{figure*}
In the studies so far, we have adopted the mean radius $R_{1.4}=11.9$ km based on the analysis of GW17087 by the LIGO/VIRGO Collaboration \cite{abbott2017gw170817}. It is necessary to study if the two-peak structure of PDF($K_{\rm{sym}}$) persists if $R$ itself is changed by about the same amount as the precision $\Delta R$. We thus compare in Fig. \ref{R18120} the PDFs of the four key EOS parameters obtained by using \texorpdfstring{$R_{1.4}=11.8, 11.9$}{R1.4=11.8, 11.9} and \texorpdfstring{$12.0$}{12.0} km all with a precision \texorpdfstring{$\Delta R=0.1$}{Delta R=0.1} km. The mean values and the standard deviations of the four EOS parameters are also shown in the plot. Firstly, it is seen that the two-peak structure of PDF($K_{\rm{sym}}$) persists in all three cases. Secondly, as the radius increases from 11.8 km to 12.0 km, the mean value of $J_0$ increases by about 4\%, that of $J_{\rm sym}$ has little change, while those of $K_{\rm sym}$ and $L$ increase by about 15\% and 2\%, respectively, to make the EOS more stiff as required by the slightly larger radius. Again, it indicates that $K_{\rm sym}$ is most important for determining the radii of canonical NSs.

With the increasing radius $R_{1.4}$, the first peak of PDF($K_{\rm{sym}}$) around $K_{\rm{sym}}=-125$ MeV gradually shifts towards larger $K_{\rm{sym}}$ values and the peak of PDF($L$) shifts towards higher $L$ values as the resulting symmetry energy has to be stiff enough to make the radius larger. 
The underlying physics causing the two peaks are still due to the $K_{\rm{sym}}-J_{\rm{sym}}$ and $K_{\rm{sym}}-L$ correlations as shown in Fig. \ref{R1812COR}. Consistent with the PDFs shown in Fig. \ref{R18120}, there are small shifts in the locations of the highest correlation strength in the 2D PDFs. Comparing the two sets of 2D PDFs with $R_{1.4}=11.8$ km and 12.0 km shown in Fig. \ref{R1812COR}, 
we also notice that with the large radius, the $K_{\rm{sym}}-J_{\rm{sym}}$ correlation around $K_{\rm{sym}}=-220$ MeV and $J_{\rm{sym}}=800$ MeV becomes weaker indicated by the more broad spread in the area while the $K_{\rm{sym}}-L$ correlation becomes stronger around $K_{\rm{sym}}=-100$ MeV and $L=40$ MeV. This leads to the slightly different heights of the two peaks in the 1D PDF($K_{\rm{sym}}$) shown in Fig. \ref{R18120}

\begin{figure*}[ht]
%\vspace{-0.6cm}
\begin{center}
 \resizebox{0.8\textwidth}{!}{
  \includegraphics[width=10cm,height=10cm]{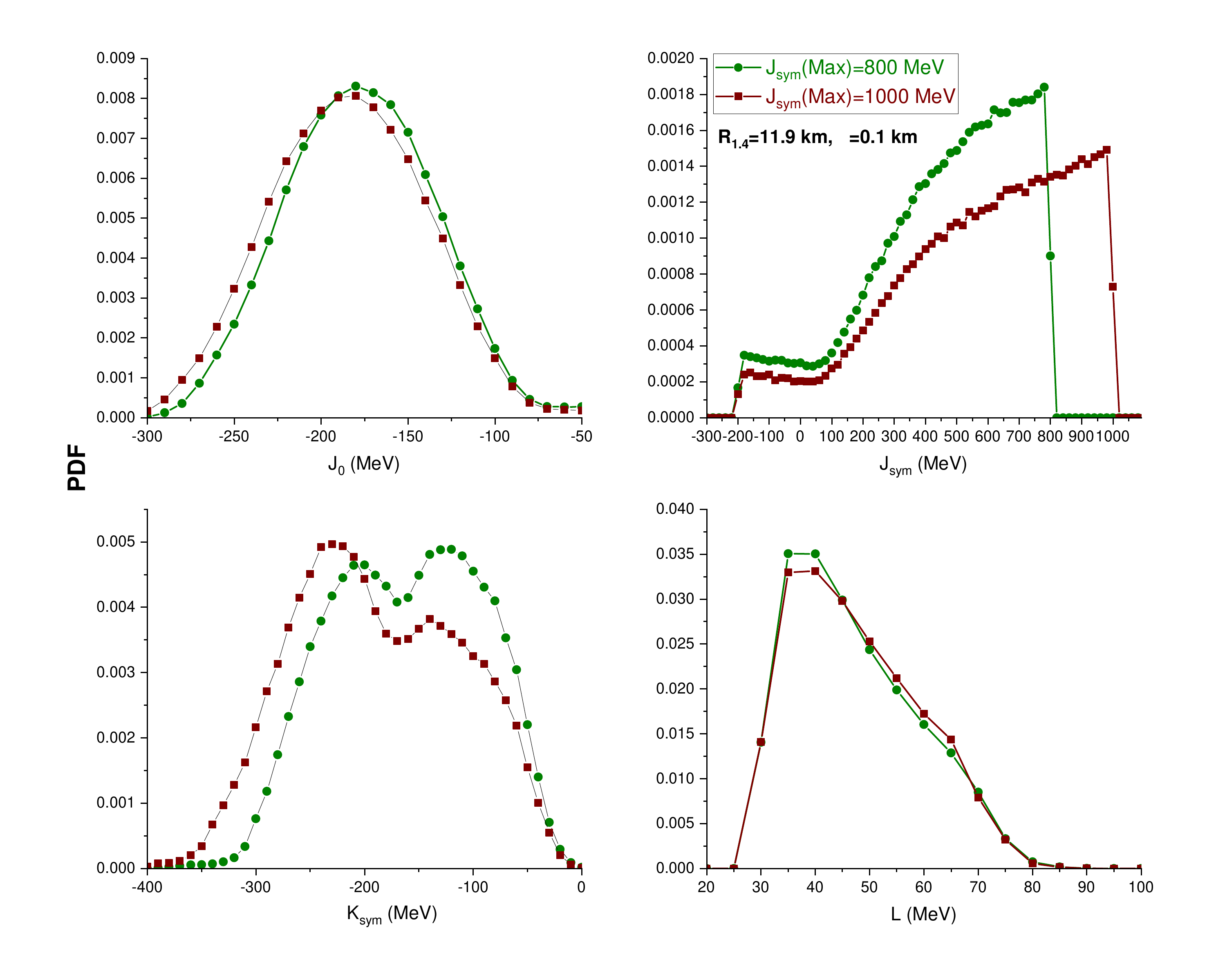}
  }
\vspace{-0.8cm}
  \caption{(color online) Comparisons of the 1D PDFs of high-density EOS parameters with $R_{1.4}=11.9$ km and a precision of $\Delta R=0.1$ km, with the maximum prior value of $J_{\rm{sym}}$ set at $800$ (green) and 1000 MeV (wine), respectively.}\label{JsymE1}
\end{center}
\end{figure*}
\begin{figure*}[ht]
%\vspace{-0.6cm}
\begin{center}
 \resizebox{0.95\textwidth}{!}{
  \includegraphics[width=16cm,height=5cm]{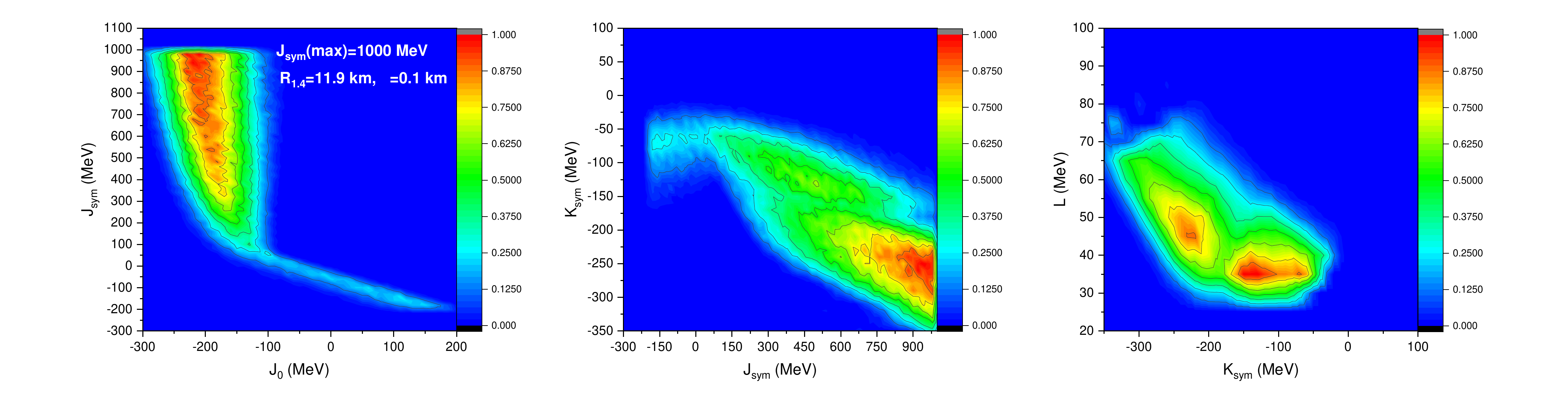}
  }
 %  \vspace{-1.3cm}
  \caption{(color online) Posterior 2D probability distribution functions among the four high-density EOS parameters with $R_{1.4}=11.9$ km and a precision of $\Delta R=0.1$ km, with the maximum prior value of $J_{\rm{sym}}$ set at $1000$.}\label{JsymE2}
\end{center}
\end{figure*}

\subsection{What are the effects of the high-density symmetry energy parameter $J_{\rm sym}$ on the PDF($K_{\rm sym})$?}\label{JKsym}
As mentioned previously, the prior range $-200<J_{\rm sym}<800$ MeV \cite{Dutra:2012mb,Dutra:2014qga,Zhang17,Tews17,Li-PPNP} for $J_{\rm sym}$ is still very large. 
The fundamental cause for this situation or generally our poor knowledge about the high-density behavior of symmetry energy is the still mysterious spin-isospin dependence of 3-body force and/or tensor force as well as the isospin-dependence of short-range nucleon-nucleon correlations induced by them in dense neutron-rich matter, see, e.g. Refs \cite{Li-PPNP,Li-news} for reviews. Current energy density functional theories mostly use some phenomenological approximations for the 3-body force and often neglect the tensor force. Moreover, as discussed earlier, our $J_{\rm sym}$ parameter represents effectively all high-density terms in a Taylor expansion of energy density functionals. Therefore, the upper limit of $J_{\rm sym}$ could be very different from predictions of existing nuclear many-body theories. Unfortunately, there is currently no constraint from laboratory experiments on this parameter. It is thus interesting to examine how the 1D and 2D PDFs of the high-density EOS parameters may vary if we modify the upper limit for the prior range of $J_{\rm sym}$. Most importantly, how does the two-peak structure of 1D PDF($K_{\rm sym})$ change? For these purposes, we performed another calculation with $J_{\rm sym} ({\rm max})=1000$ MeV for the case of $R_{1.4}=11.9$ km and a precision of $\Delta R=0.1$ km. 

Shown in Fig. \ref{JsymE1} are comparisons of the 1D PDFs of high-density EOS parameters with $R_{1.4}=11.9$ km, $\Delta R=0.1$ km and $J_{\rm{sym}}({\rm max})$ set at $800$ (green) and 1000 MeV (wine), respectively. The 2D PDFs for the latter case are shown in Fig. \ref{JsymE2}.
It is seen that as the $J_{\rm{sym}}({\rm max})$ is moved {\it a priori} from 800 to 1000 MeV,
the peak of the posterior PDF($J_{\rm{sym}})$ moves outward to $J_{\rm{sym}}({\rm max})$=1000 MeV accordingly, indicating again that the NS radius data is not restricting this parameter. This finding is consistent with that found earlier in Ref. \cite{xie2020bayesian}. Consequently, the second peak in the 1D PDF($K_{\rm{sym}})$ due to the anti-correlation between $J_{\rm{sym}}$ and $K_{\rm{sym}}$ moves downward to 
$K_{\rm{sym}}\approx -250$ MeV, about $25$ MeV lower than the second peak obtained with $J_{\rm{sym}}({\rm max})$=800 MeV. Indeed, as shown in Fig. \ref{JsymE2}, the anti-correlation between $J_{\rm{sym}}$ and $K_{\rm{sym}}$ is the strongest around $K_{\rm{sym}}\approx -250$ MeV. Similarly, the anti-correlation between $K_{\rm{sym}}$ and $L$ makes the 1D PDF($L$) to move slightly towards larger values. As a result of the competition of the two anti-correlations, the PDF($K_{\rm{sym}})$ ends up still having two peaks but the one around $K_{\rm{sym}}\approx -250$ MeV is significantly more probable. The above comparisons of the two calculations indicate again the importance of high-precision radius measurements for inferring the fine features of EOS parameters.

Despite of the clearly demonstrated physics cause of the two-peak structures in the PDF($K_{\rm sym}$) above, some people may still suspect that it is an artifact of something went wrong, e.g., the autocorrelation in MCMC is too large. It is well known that the latter can lead to ``ridges" in the 1D posterior PDFs \cite{Cowles96,trotta2017bayesian}. It is therefore necessary to check the level of auto-correlations in our Bayesian analyses. In Appendix \ref{auto}, the autocorrelation functions \cite{trotta2017bayesian} of all six EOS parameters are evaluated for a single chain with
$R_{1.4}=11.9$ km and $\Delta R=0.1$ km and 1.0 km, respectively. In particularly, in examining the autocorrelation in 20,000 $K_{\rm sym}$ samples from our Bayesian analyses with $\Delta R=0.1$ km, for comparison we use a sample of the same size randomly generated uniformly with $K_{\rm sym}$ in the same prior range. We found no significant autocorrelation at a 99\% significance level besides the statistical fluctuation that exist also in the random sample.

Moreover, in Appendix \ref{LL}, the jumping maps (spectrum or time series) of the posterior (i.e., the product of the likelihood function and the prior), $L$ and $K_{\rm sym}$ during the MCMC process are examined. Visually, we found no obvious evidence for any ``slow mixing" or ``chain sticking" that can occur when the MCMC sampler gets stuck too long in a particular region of the EOS parameter space. We thus have no reason to assume the double peak in PDF($K_{\rm sym}$) is an artifact in our sampling method. Nevertheless, as emphasized in many places in the literature (e.g., Refs. \cite{Cowles96,trotta2017bayesian}), it is impossible to say with certainty that a finite sampler using an MCMC algorithm is completely free of convergence issues with absolutely no autocorrelation.

\subsection{What more can we learn from the radius \texorpdfstring{$R_{2.0}=11.9$}{R2.0=11.9} km of a massive NS of mass 2.0M\texorpdfstring{$_{\odot}$}{sun} with a varying precision of \texorpdfstring{$\Delta R=1.0, 0.5, 0.2$}{Delta R=1.0, 0.5, 0.2}, and \texorpdfstring{$0.1$}{0.1} km?}

\begin{figure*}[ht]
%\vspace{-0.6cm}
\begin{center}
 \resizebox{0.9\textwidth}{!}{
  \includegraphics[width=10cm,height=10cm]{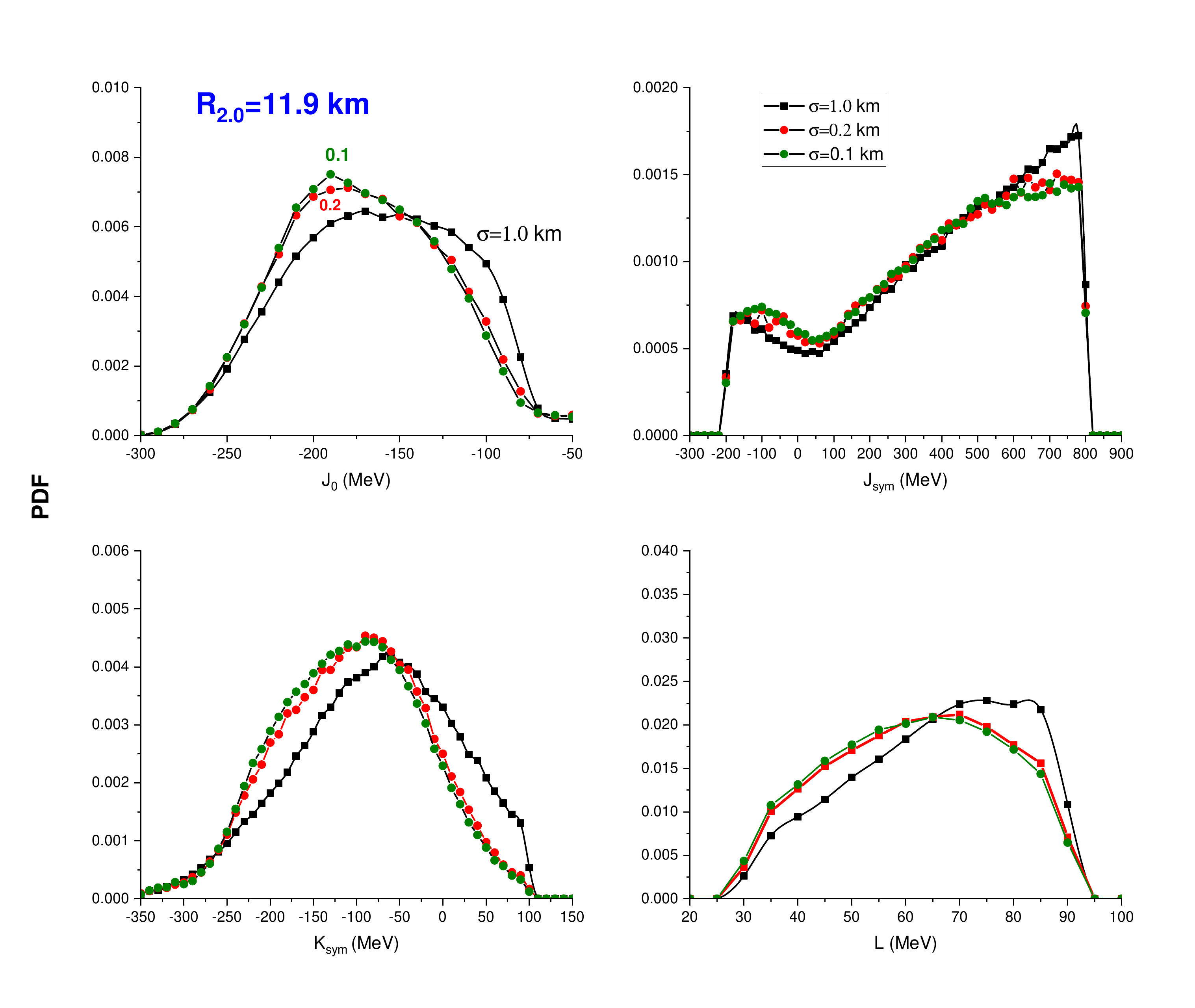}
  }
 \vspace{-1.cm}
  \caption{(color online) Posterior PDFs of the four EOS parameters obtained with $R_{2.0}=11.9$ km and a precision of $\Delta R=1.0$ km, 0.5, 0.2 and $0.1$ km, respectively..}\label{R20PDF}
\end{center}
\end{figure*}

\begin{figure*}[ht]
%\vspace{-0.6cm}
\begin{center}
 \resizebox{0.9\textwidth}{!}{
  \includegraphics[width=16cm,height=5cm]{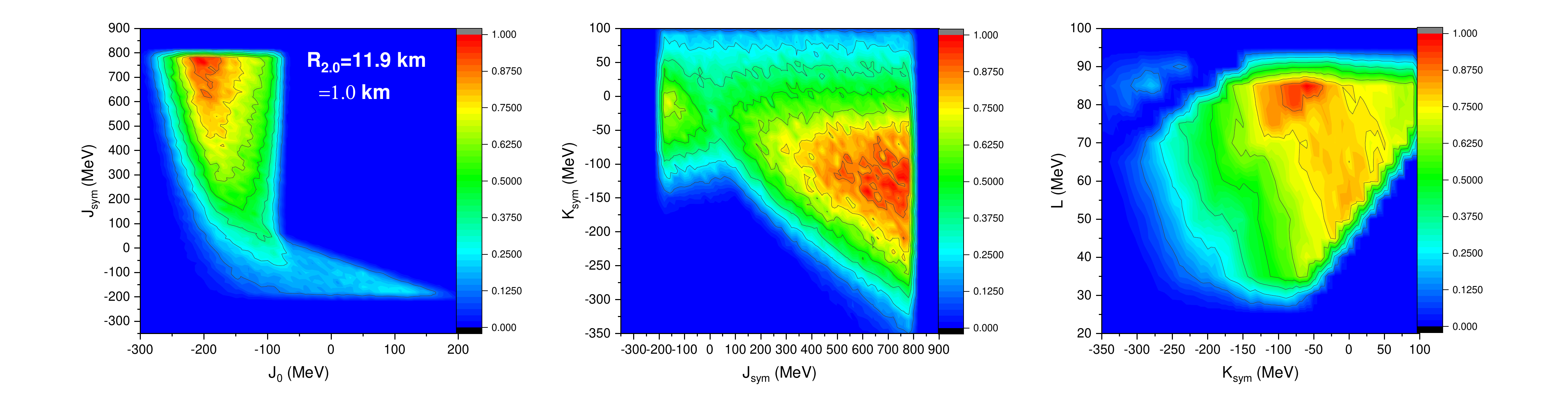}
  }
 \resizebox{0.9\textwidth}{!}{
  \includegraphics[width=16cm,height=5cm]{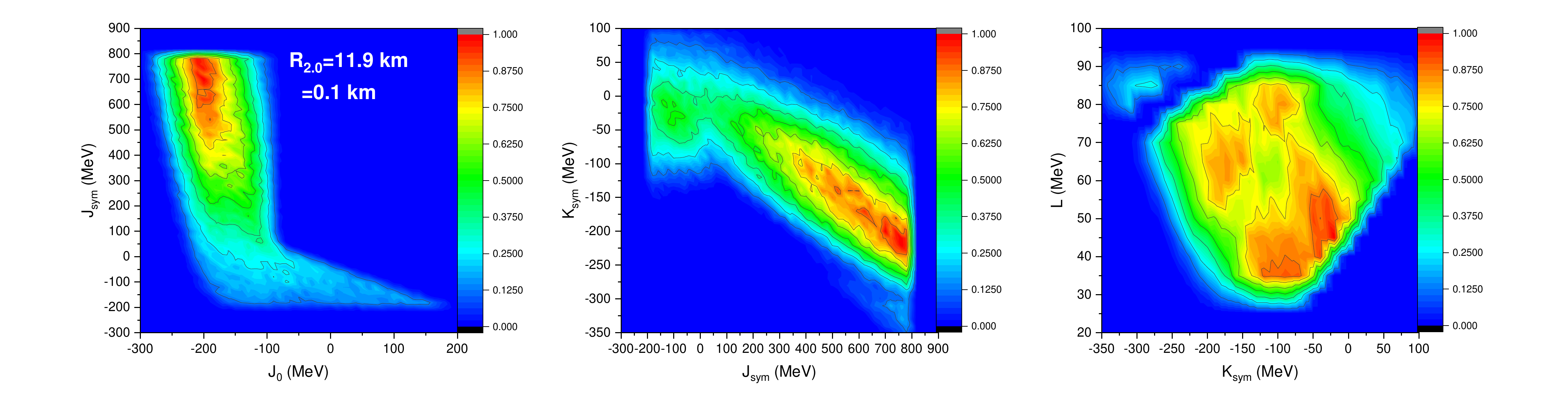}
  }
    \resizebox{0.9\textwidth}{!}{
  \includegraphics[width=16cm,height=5cm]{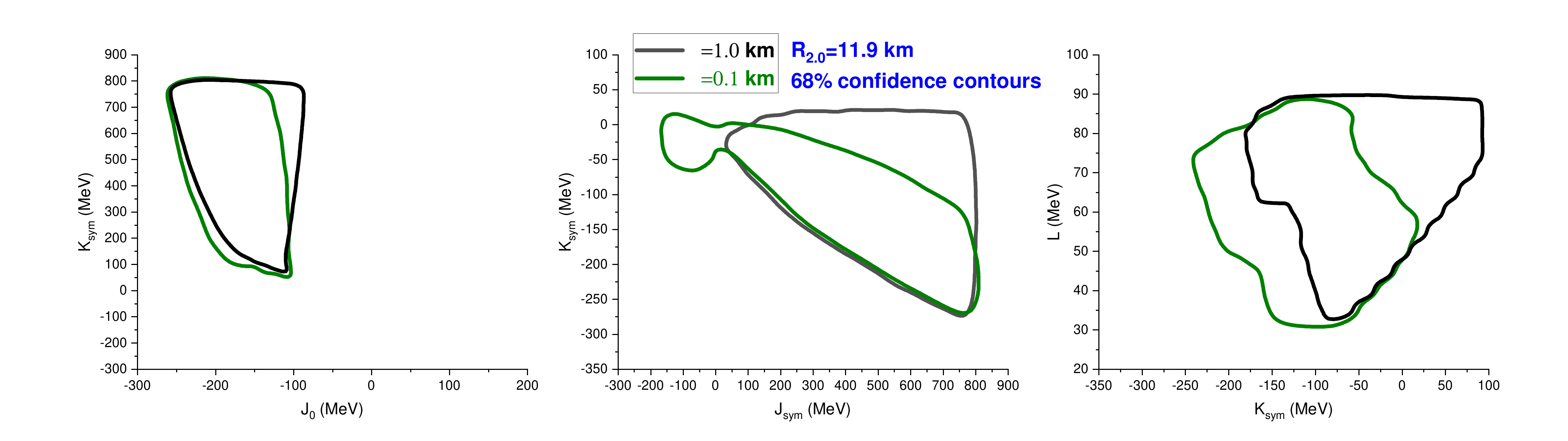}
  }
 %  \vspace{-1.3cm}
  \caption{(color online) 2D posterior probability distribution functions among the four high-density EOS parameters using {\bf R$_{2.0}$=11.9 km} and a precision of $\Delta$R=1.0 km (upper panels) and 0.1 km (middle panels), respectively. The bottom panels compare their 68\% confidence boundaries.}\label{R20COR}
\end{center}
\end{figure*}

\begin{figure}[ht]
%\vspace{-0.6cm}
\begin{center}
 \resizebox{0.45\textwidth}{!}{
  \includegraphics[width=5cm,height=8cm]{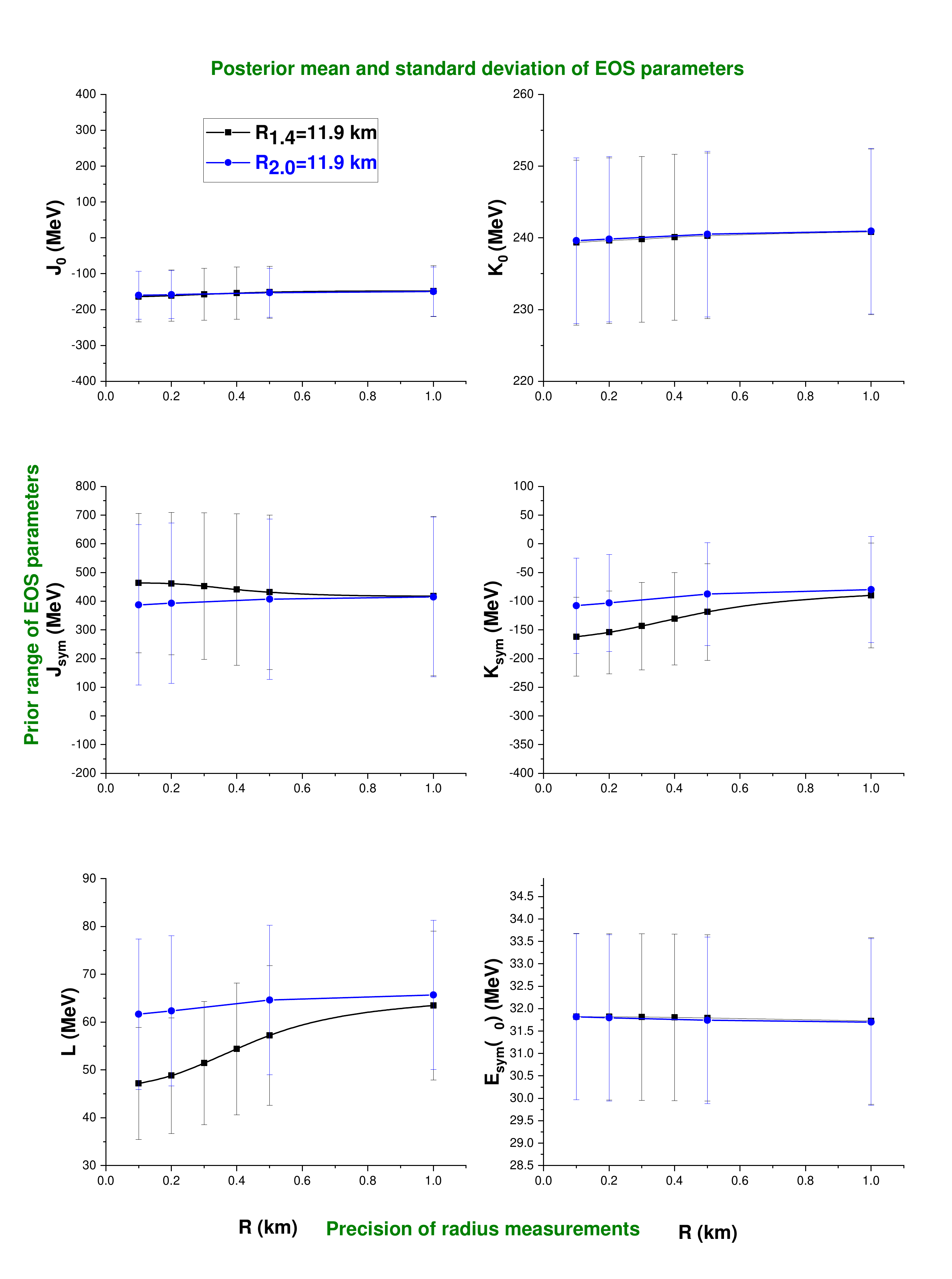}
  }
% \resizebox{0.45\textwidth}{!}{
%  \includegraphics[width=10cm,height=15cm]{Manuscript523/R20-ste.pdf}
%  }
 %  \vspace{-1.3cm}
  \caption{(color online) Comparisons of the mean and standard deviations of all six EOS parameters inferred from using the $R_{1.4}$ and $R_{2.0}$ data with a varying precision from 1.0 to 0.1 km.}\label{SteCom}
\end{center}
\end{figure}
Compared to what we have learned from the above analyses using the radii of canonical NSs, what possibly more physics can we learn from studying the radii of more massive NSs? To answer this question, we performed comprehensive anaylyses using $R_{2.0}=11.9$ km with different precision. Shown in Figs. \ref{R20PDF} and \ref{R20COR} are the 1D and 2D PDFs of the four key EOS parameters. Their MaP values with 68\% HPD boundaries, means and standard deviations are listed in Tables \ref{MP20} and \ref{Mean20}, respectively. It is interesting to note particularly the following points:
\begin{enumerate}
    \item The effects of precision $\Delta R$ in measuring $R_{2.0}$ on the 1D and 2D PDFs of $K_{\rm{sym}}$ and $L$ are not as strong as in the case of measuring $R_{1.4}$. This is not surprising as $R_{2.0}$ depends more on the EOS at higher densities than $R_{1.4}$. Indeed, it is seen that the PDF of $J_{\rm{sym}}$ near its upper limit is more affected by the $\Delta R$, indicating the potential of better constraining this very poorly known parameter characterizing the symmetry energy at densities around $(3-4)\rho_0$. Comparing the PDF of $J_{\rm{sym}}$ in this case to that with $R_{1.4}=11.9$ km,
    its peak near the upper limit of $J_{\rm{sym}}$ is significantly lower. Consequently, the mean value of $J_{\rm{sym}}$ is reduced from $463\pm243$ (with $R_{1.4}=11.9$ km) to $387\pm 279$ (with $R_{2.0}=11.9$ km) at $\Delta R=0.1$ km as listed in Tables \ref{Mean14} and \ref{Mean20}. One reason for this finding is that the required skewness $J_0$ of SNM is slightly higher to make the EOS stiff enough to support a 2.0M$_{\odot}$ NS as also listed in Tables \ref{Mean14} and \ref{Mean20}. Since $J_{\rm{sym}}$ and $J_0$ are anti-correlated as shown in Fig. \ref{R20COR}, the mean value of $J_{\rm{sym}}$ needs to be lower corresponding to a higher $J_0$ with $R_{2.0}=11.9$ km. However, we notice that this effect is expected to be small as we already used the requirement that all EOSs have to be stiff enough to support NSs at least as massive as 1.97M$_{\odot}$. Another possibly more important reason as we shall elaborate more next is the anti-correlations of $K_{\rm{sym}}$ with both $J_{\rm{sym}}$ and $L$. For a NS of mass 2.0M$_{\odot}$ to have the same radius of 11.9 km as a 1.4M$_{\odot}$ NS, the required mean value of $K_{\rm{sym}}$ is higher than that in the case of $R_{1.4}=11.9$ km especially in the high-precision measurements. Thus, the anti-correlation between $K_{\rm{sym}}$ and $J_{\rm{sym}}$ also requires the latter to become smaller in the case of $R_{2.0}=11.9$ km.
    
    \item There is no longer any two-peak structure in the PDF of $K_{\rm{sym}}$ with $R_{2.0}=11.9$ km even at high precision. Its stabilized single peak around $K_{\rm{sym}}\approx -100$ MeV with $\Delta R=0.1$ or 0.2 km is the result of a competition between $K_{\rm{sym}}-L$ and $K_{\rm{sym}}-J_{\rm{sym}}$ anti-correlations. As shown in Fig. \ref{R20COR}, the $K_{\rm{sym}}-L$ anti-correlation is the strongest around ($K_{\rm{sym}}=-40$ MeV, $L=50$ MeV) with $\Delta R=0.1$ km, while the $K_{\rm{sym}}-J_{\rm{sym}}$ anti-correlation peaks around ($K_{\rm{sym}}=-220$ MeV, $J_{\rm{sym}}=800$ MeV). Whether there exists two-peaks or a single peak in the PDF of $K_{\rm{sym}}$ and its location depend on both the strengths of the above two anti-correlations as well as the mean values of $J_{\rm{sym}}$ and $L$. Besides the changes of the mean values of $J_{\rm{sym}}$ mentioned above, it is also seen in Tables \ref{Mean14} and \ref{Mean20} that the mean value of $L$ changes from $47\pm 12$ (with $R_{1.4}=11.9$ km) to $62\pm 16$ MeV (with $R_{2.0}=11.9$ km) with $\Delta R=0.1$ km. These changes of posterior mean values of the EOS parameters are also shown in Fig. \ref{SteCom}. Since a larger $L$ prefers a smaller $K_{\rm{sym}}$ while a reduced $J_{\rm{sym}}$ prefers a larger one both with $R_{2.0}=11.9$ km, compared to the case with $R_{1.4}=11.9$ km, their competition makes the PDF($K_{\rm{sym}})$ to have a single peak around $-100$ MeV as observed instead of two peaks. 
    It is seen from Figs. \ref{R20COR} and \ref{R14COR}, the anti-correlation $K_{\rm{sym}}-L$ is much weaker or more spread out with $R_{2.0}=11.9$ km, as $R_{2.0}$ does not constrain tightly $L$ which characterizes the symmetry energy below about $1.5\rho_0$. It is also seen that improving the precision of measuring $R_{2.0}$ from $\Delta R=$1.0 km to 0.1 km has the strongest effect in narrowing down the $K_{\rm{sym}}$-$J_{\rm{sym}}$ correlation.
    \end{enumerate}
    
Shown in Fig. \ref{SteCom} are comparisons of the mean and standard deviations of all six EOS parameters inferred from using the $R_{1.4}$ and $R_{2.0}$ data with $\Delta R$ from 1.0 to 0.1 km. It is seen that at poor precision, e.g., $\Delta R$=1.0 km, the results from using the $R_{1.4}$ and $R_{2.0}$ data are about the same. As one expects and demonstrated above, the differences gradually appear as the precision gets better especially using the $R_{1.4}$ data. Overall, only the symmetry energy parameters show appreciable changes with respect to their individual prior scales. We also notice that most of the 68\% HPD ranges and/or standard deviations do not vary coherently with MaP or mean values. This is mostly because of the cut-off introduced by the prior ranges of some of the EOS parameters. For example, the uncertainties/ranges of $K_0$ and $E_{\rm{sym}}(\rho_0)$ are entirely set by our prior knowledge from nuclear physics. For $J_{\rm{sym}}$, because the NS radius data provide little constraints, its uncertainty range essentially remains the same as its prior range. Then, due to its strong anti-correlations with $K_{\rm{sym}}$ and $J_0$ and a weak correlation with $L$, the 68\% HPD ranges and/or standard deviations of all EOS parameters do not necessarily decrease obviously with the decreasing $\Delta R$.
    
It is seen in Fig. \ref{SteCom} that $L$ and $K_{\rm{sym}}$ are lower while the $J_{\rm{sym}}$ is higher to obtain $R_{1.4}=11.9$ km than $R_{2.0}=11.9$ km at high precision. This has important implications on the density profiles of proton fraction $x_p$ at $\beta-$equilibrium, and subsequently whether and where the fast cooling through the direct URCA process (requiring $x_p\geq$ 11\%) \cite{Lat91,Klahn:2006ir} can happen or not in the two NSs considered. Lower values of $L$ and $K_{\rm{sym}}$ indicate that the symmetry energy at low-intermediate densities $(1-2)\rho_0$ is lower/softer, while a higher $J_{\rm{sym}}$ implies that the symmetry energy around $(3-4)\rho_0$ is higher/stiffer in a canonical NS with $R_{1.4}=11.9$ km compared to a NS of mass 2.0M$_{\odot}$ with the same radius. It is well known that to minimize the total energy of isospin asymmetric nuclear matter because of the $E_{\rm{sym}}(\rho)\delta^2$ term in its EOS, wherever/whenever the symmetry energy is higher the isospin asymmetry $\delta$ there is lower. The resulting density profile $\delta(\rho)$ at $\beta$-equilibrium exhibits the so-called isospin fractionation phenomenon ($E_{\rm sym}(\rho_1)\delta_1\approx E_{\rm sym}(\rho_2)\delta_2$) for two regions of density $\rho_1$ and $\rho_2$ in $\beta$-equilibrium \cite{Pawel,Zhang-EPJA23}), see, e.g., Ref. \cite{LCK} for a review. The results shown in Fig. \ref{SteCom} obtained under the assumption that $R_{2.0}=R_{1.4}=11.9$ km imply that the massive NS of 2.0M$_{\odot}$ is more neutron-rich at high densities ($(3-4)\rho_0$) but neutron-poor at low-intermediate densities ($(1-2)\rho_0$) compared to a canonical NS. As indicated, however, this information can only be inferred from high-precision radius measurements. In this sense, the latter will help better understand the cooling mechanisms of protoneutron stars and their NS mass dependence \cite{Boukari:2024wrg}.

\begin{table}[htbp]
\centering
\caption{The maximum {\it a posteriori} values and their corresponding 68\% credible intervals of the four EOS parameters inferred from Bayesian analyses using {\bf R$_{2.0}$=11.9 km} with precision $\Delta$R from 1.0 to 0.1 km. }\label{MP20}
\begin{tabular}{lccccccc}
  \hline\hline
  Parameters (MeV) & $\Delta$R=1.0 &  $\Delta$R=0.5 &$\Delta$R=0.2 &$\Delta$R=0.1 \\
  \hline\hline\\
%  \vspace{0.2cm}
 $J_0:$ &$-170_{-40}^{+70}$& $-170_{-50}^{+60}$ &$-180_{-40}^{+60}$& $-190_{-30}^{+70}$\\
 $J_{\mathrm{sym}}:$ &$780_{-500}^{+0}$& $780_{460}^{+0}$ &$780_{-520}^{+0}$& $780_{-520}^{+0}$ \\
 $K_{\mathrm{sym}}:$ &$-60_{-100}^{+90}$& $-70_{-100}^{+80}$ &$-90_{-100}^{+70}$& $-90_{-100}^{+70}$ \\
 $L:$ &$75_{-40}^{+0}$& $70_{-40}^{+5}$ &$70_{-40}^{+0}$& $65_{-35}^{+5}$\\
  \hline
 \end{tabular}
\end{table}

\begin{table}[htbp]
\centering
\caption{Means and variances of the four EOS parameters (MeV) inferred from Bayesian analyses using {\bf R$_{2.0}$=11.9 km} with precision $\Delta$R from 1.0 to 0.1 km. }\label{Mean20}
\begin{tabular}{lccccccc}
  \hline\hline
  Pars. & $\Delta$R=1.0 &  $\Delta$R=0.5 &$\Delta$R=0.2 &$\Delta$R=0.1 \\
  \hline\hline\\
%  \vspace{0.2cm}
 $J_0:$&$-150\pm 69$&$-153\pm 68$& $-158\pm 67$ &$-160\pm 67$& \\
 $J_{\mathrm{sym}}:$ &$414\pm 278$& $407\pm 279$ &$393\pm 279$& $387\pm 279$ \\
 $K_{\mathrm{sym}}:$&$-80\pm 92$ &$-88\pm 90$&$-103\pm 84$ &$-108\pm 83$ \\
 $L:$&$66\pm 16$ & $65\pm 16$&$62\pm 16$& $62\pm 16$\\
  \hline
 \end{tabular}
\end{table}

\subsection{What more can we learn from assuming \texorpdfstring{$R_{1.4}=R_{1.6}=R_{1.8}=R_{2.0}=11.9$}{R1.4=R1.6=R1.8=R2.0=11.9} km with varying precision?}
\begin{figure*}[ht]
%\vspace{-0.6cm}
\begin{center}
 \resizebox{0.7\textwidth}{!}{
  \includegraphics[width=15cm,height=10cm]{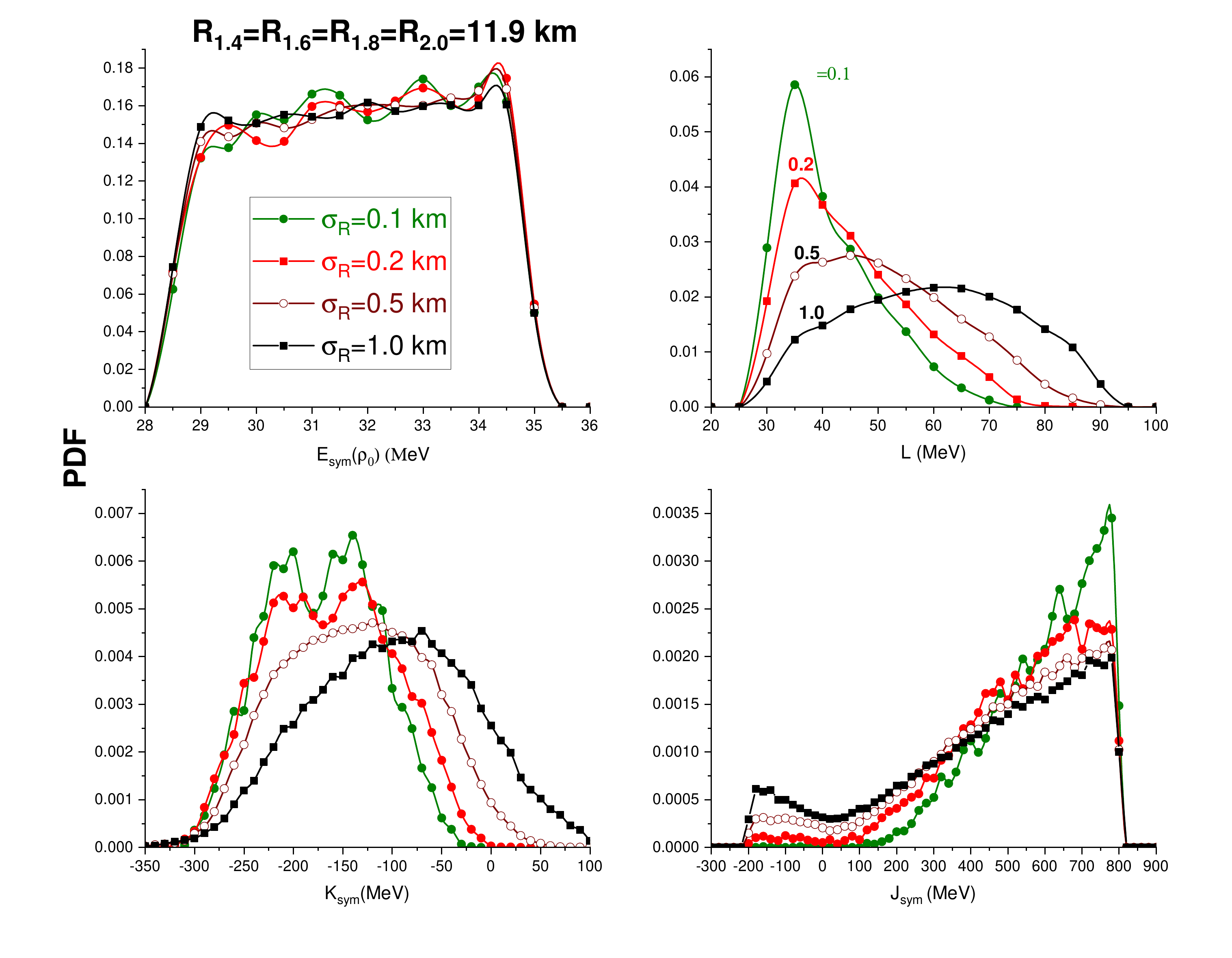}
  }
  \caption{(color online) Posterior PDFs of the four parameters characterizing the density dependence of nuclear symmetry energy assuming \texorpdfstring{$R_{1.4}=R_{1.6}=R_{1.8}=R_{2.0}=11.9$}{R1.4=R1.6=R1.8=R2.0=11.9} with a varying precision indicated.}\label{CR-Esym}
\end{center}
\end{figure*}

\begin{figure*}[ht]
%\vspace{-0.6cm}
\begin{center}
 \resizebox{0.7\textwidth}{!}{
  \includegraphics[width=15cm,height=10cm]{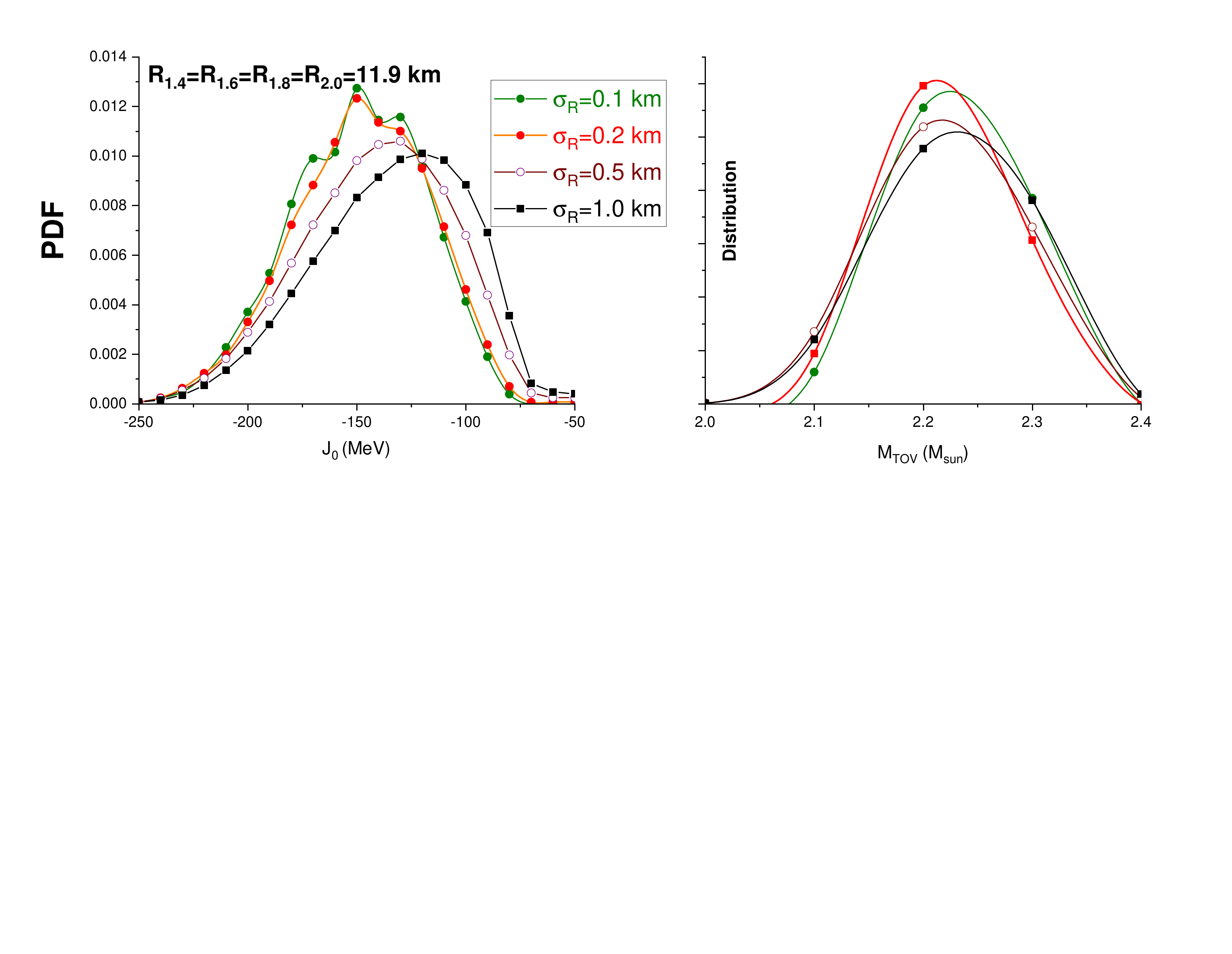}
  }
   \vspace{-4.3cm}
  \caption{(color online) Posterior PDFs of the skewness parameter $J_0$ of SNM EOS (left) and the maximum mass M$_{\rm TOV}$ (right) assuming \texorpdfstring{$R_{1.4}=R_{1.6}=R_{1.8}=R_{2.0}=11.9$}{R1.4=R1.6=R1.8=R2.0=11.9} km with a varying precision indicated.}\label{CR-J0}
\end{center}
\end{figure*}
With the current precision of about $\Delta R\geq 1.0$ km in measuring the radii of NSs, most of the existing data indicate empirically 
$R_{1.4}\approx R_{1.6}\approx R_{1.8}\approx R_{2.0}\approx 11.9$ km, see e.g., Ref. \cite{MR-Russia} for a recent summary. As we have shown above, with $\Delta R=1.0$ km, one can not distinguish the PDFs and correlations of EOS parameters inferred from $R_{1.4}=11.9$ or $R_{2.0}=11.9$ km. However, as the precision improves, their differences are clearly visible. 
In particular, at high precision, the PDF($K_{\rm{sym}}$) has a distinctive two-peak structure for 1.4M$_{\odot}$ but not for 2.0M$_{\odot}$ NS. While we have not studied at what NS mass this two-peak feature will start disappearing because of the extremely expensive computing costs, we present here results of one calculation using combined data (the radius likelihood functions of the four NSs of different masses are multiplied together) assuming \texorpdfstring{$R_{1.4}=R_{1.6}=R_{1.8}=R_{2.0}=11.9$}{R1.4=R1.6=R1.8=R2.0=11.9} km. One generally expects the NS central density to increase with its mass. Thus, it is interesting to know which EOS parameters are required to change most when the four NSs of different masses are forced to have the same radius.

\begin{figure*}[ht]
%\vspace{-0.6cm}
\begin{center}
 \resizebox{0.7\textwidth}{!}{
  \includegraphics[width=10cm,height=10cm]{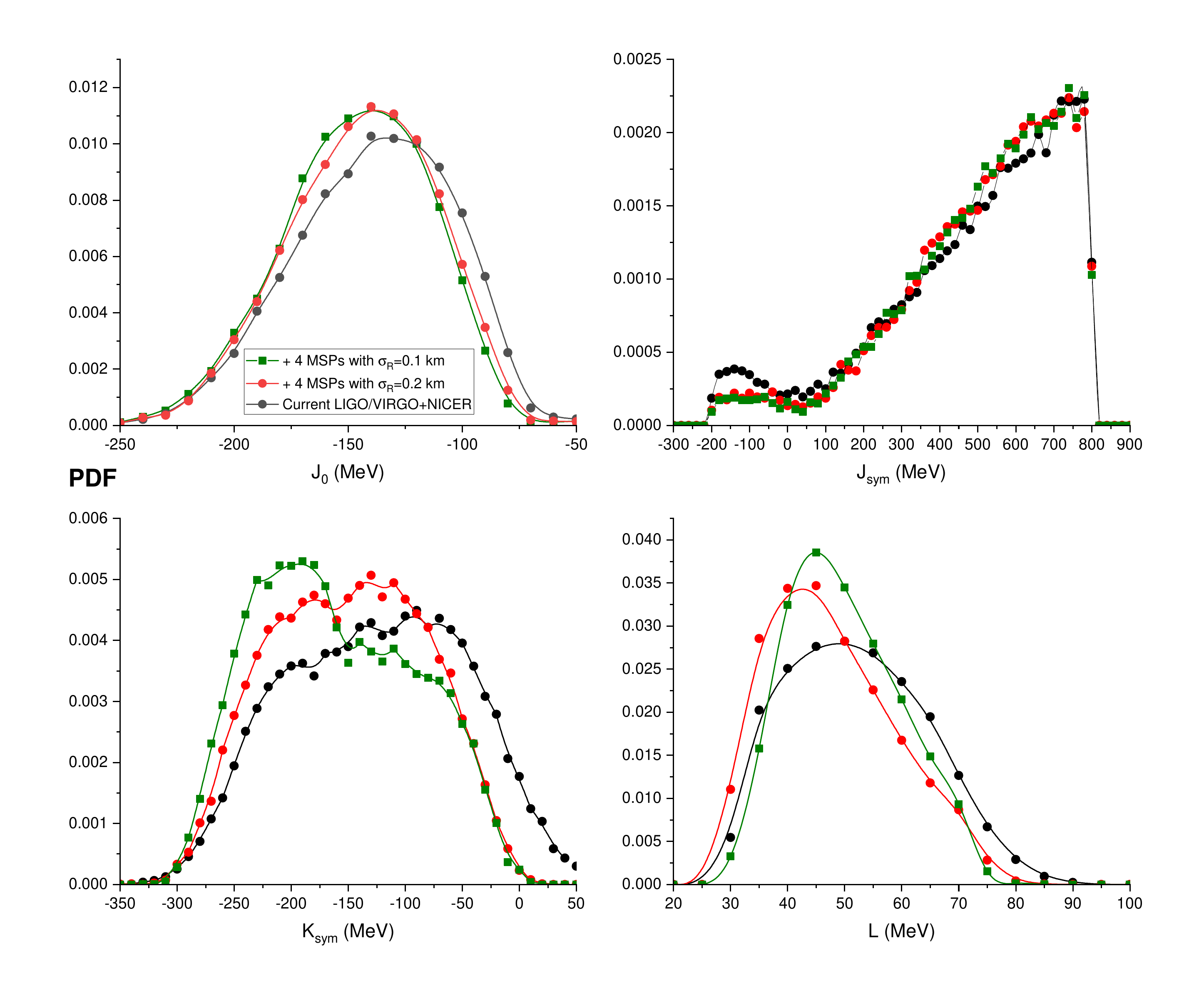}
  }
%   \vspace{-1.3cm}
  \caption{(color online) Posterior PDFs of the four key EOS parameters obtained from including the mass and radius data of the four MSPs listed in Table \ref{tab-data} with the expected precision of  \texorpdfstring{$\Delta R=0.2$}{Delta R=0.2} km and \texorpdfstring{$0.1$}{0.1} km, respectively, with respect to the results obtained from using only the radius data from GW170817 by LIGO/VIRGO, as well as those for PSR J0030+0451 and PSR J0740+6620 from NICER listed also in Table \ref{tab-data}.}\label{4MSP}
\end{center}
\end{figure*}

Shown in Fig. \ref{CR-Esym} are the posterior PDFs of the four parameters characterizing the density dependencies of nuclear symmetry energy. It is seen that the PDFs of $L$, $K_{\rm{sym}}$, and $J_{\rm{sym}}$ all show significant dependencies on $\Delta R$. Most features observed in analyzing only the $R_{1.4}=11.9$ km data are again apparent.
For the PDF of $E_{\rm{sym}}(\rho_0)$, while the fluctuations become more obvious with higher precision as one expects, overall the high-precision NS radius measurement does not provide any new information about $E_{\rm{sym}}(\rho_0)$ compared to our prior knowledge. The two-peak structure of the PDF($K_{\rm{sym}}$) remains and the PDF($J_{\rm{sym}}$) appears to show stronger sensitivity to $\Delta R$ compared to the results from studying separately the $R_{1.4}=11.9$ km and $R_{2.0}=11.9$ km cases earlier. This observation is understandable. As noticed before, the peak of PDF($K_{\rm{sym}}$) around $K_{\rm{sym}}\approx-220$ MeV is due to the anti-correlation between $K_{\rm{sym}}$ and $J_{\rm{sym}}$. While its second peak around $K_{\rm{sym}}\approx -120$ MeV is due to the anti-correlation between $K_{\rm{sym}}$ and $L$ that is most visible for canonical NSs with high precision. Thus, the combined analyses assuming \texorpdfstring{$R_{1.4}=R_{1.6}=R_{1.8}=R_{2.0}=11.9$}{R1.4=R1.6=R1.8=R2.0=11.9} km at high precision maintain both peaks in the PDF($K_{\rm{sym}}$). Moreover, 
with more massive NSs involved in the analyses, the $J_{\rm{sym}}$ plays a more significant role. Consequently, its PDF($J_{\rm{sym}}$) appears more sensitive to $\Delta R$.

Shown in Fig. \ref{CR-J0} are the posterior PDFs of the skewness parameter $J_0$ of SNM EOS (left) and the maximum mass M$_{\rm TOV}$ (right) assuming \texorpdfstring{$R_{1.4}=R_{1.6}=R_{1.8}=R_{2.0}=11.9$}{R1.4=R1.6=R1.8=R2.0=11.9} km with varying precision.
The PDF$(J_0)$ is consistent with that observed earlier from analyzing the $R_{1.4}=11.9$ km and $R_{2.0}=11.9$ km cases. The maximum mass M$_{\rm TOV}$ of NSs on the mass-radius sequence given a NS EOS is determined mostly by the $J_0$ while the $J_{\rm{sym}}$ plays a less important role, see e.g., Refs. \cite{Zhang:2018bwq,LiEPJA} for reviews. Because of the anti-correlation between $J_{\rm{sym}}$ and $J_0$ as shown earlier, the PDF of M$_{\rm TOV}$ peaked around M$_{\rm TOV}\approx 2.25$ M$_{\odot}$ does not show much dependence on the precision $\Delta R$. This is not surprizing as the latter affects directly only the radii of NSs. Its secondary effect on M$_{\rm TOV}$ is thus small. It is interesting to point out here that the 
PDF(M$_{\rm TOV}$) obtained here is in perfect agreement with the very recent finding of M$_{\rm TOV}=2.25^{+0.08}_{-0.07}$ M$_{\odot}$ from a comprehensive analysis using two approaches: (1) modeling the mass function of 136 NSs with a sharp cut-off at 2.28 M$_{\odot}$, and (2) Bayesian inference of the EOS from the available combined multi-messenger data of NSs considering constraints from the chiral effective field theory at low densities and the perturbative QCD at very high densities \cite{Fan:2023spm}.

\subsection{What more can we learn from the radii of the four MSPs that are about 1.0 km apart with \texorpdfstring{$\Delta R=1.0, 0.5, 0.2$}{Delta R=1.0, 0.5, 0.2} and \texorpdfstring{$0.1$}{0.1} km, with respect to what we knew from analyzing the radii of GW170817, PSR J0030+0451 and PSR J0740+6620?}
In the new era of multimessenger astronomy, combined data of multiple obserbables obtained using different tools over possibly multiple sources are often used in Bayesian analyses. Based on what we have learned so far from the above analyses, different precision of different messengers from NSs of different masses may be a concern to be taken care of. Nevertheless, we emphasize that such practice may be important for studying the local derivative $dM/dR$ of the mass-radius sequence. 

Listed in the last 4 rows in Table \ref{tab-data} are four MSPs with precisely measured masses between about 1.2 to 2 M$_{\odot}$. Their presently known radii are about 1.0 km apart at the corners of a rotated {\Large Z} curve as shown in Fig. \ref{MR1plot}. Their radii are not well constrained according to Refs. \cite{Wat,Nice,Webb:2019tkw,Guo:2021bqa,NANOGrav:2017wvv}.
The differences between their radii are at the same level as the current best precision $\Delta R\approx 1.0$ km. Since these four MSPs cover approximately the same mass range as the case we studied in the pevious subsection assuming \texorpdfstring{$R_{1.4}=R_{1.6}=R_{1.8}=R_{2.0}=11.9$}{R1.4=R1.6=R1.8=R2.0=11.9} km with a varying precision, a comparative study here with high precision simultaneously for all four MSPs is at least educational. 
We will also compare the results from this study with the ones (used as a reference below) obtained from using only the data from GW170817 by LIGO/VIRGO, as well as the PSR J0030+0451 and PSR J0740+6620 data from NICER listed in Table \ref{tab-data}. 

Our results are shown in Fig. \ref{4MSP}. It is seen that the PDF($J_0$) shifts towards a slightly lower MaP value of $J_0$ around $-150$ MeV with both $\Delta R=0.1$ and 0.2 km compared to the reference from using the combined data from LIGO/VIRGO+NICER. This is consistent with our findings in the previous subsections. The PDF($J_{\rm sym}$) does not seem to show any difference mostly due to the mixing of the radius data of four MSPs on the {\Large Z} curve. Similarly, the situation holds for $L$, although one can still distinguish its PDF obtained with different precision. On the other hand, since $K_{\rm sym}$ is the parameter most sensitive to the variation of NS radii, its PDF shows some structures. Again, these structures at high precision are due to the anti-correlations of $K_{\rm sym}$ with $J_{\rm sym}$ and $L$. The relative heights of the two-peaks (or a peak and a shoulder) depend sensitively on the precision $\Delta R$ as the latter affects strongly different contributions to the total likelihood function from the four MSPs when the EOSs are sampled during the MCMC process. Obviously, because of the mixing of NS radius data about 1.0 km apart, the effects of the precision are not as clear as in the previous case when all four NSs are assumed to have the same radius.

\section{Summary and Conclusions}\label{summary}
In summary, future high-precision X-ray and gravitational wave observatories are designed to measure NS radii with unprecedented precision. Some recent analyses and simulations have forecasted that a network consisting of one Cosmic Explorer and the Einstein Telescope can measure the radii of NSs in the mass range of $(1.0-2.0)$ M$_{\odot}$ with an accuracy $\sigma_R\equiv \Delta R$ less than 0.1 km that is more than ten times better than current constraints from LIGO-Virgo-KAGRA and NICER. One outstanding scientific goal of these high-precision radius measurements is to constrain more precisely the nuclear EOS especially at suprasaturation densities.  However, other than the stiffness generally spoken of in the literature, it is presently unclear what particular aspects of the EOS and to what precision they will be better constrained by the proposed high-precision NS radius measurements. 

Within a Bayesian framework using a meta-model EOS for NSs consisting of $npe\mu$ matter at $\beta-$equilibrium, we inferred systematically the 1D and 2D PDFs of the EOS parameters using several imagined data sets. We found that some interesting fine-features of the EOS of superdense neutron-rich matter can be extracted from the expected high-precision NS radius data. 
In particular, we found that (1) the precision $\Delta R$ of measuring NS radii has little effect on inferring the SNM stiffness (in terms of its incompressibility $K_0$ and skewness $J_0$ measuring the stiffness of SNM around $\rho_0$ and $(3-4)\rho_0$, respectively), (2) NS radius measurements with higher precision reveal more accurately not only the PDFs of but also pairwise correlations among parameters characterizing the high-density behavior of nuclear symmetry energy $E_{\rm sym}(\rho)$, (3) the 1D PDF of curvature $K_{\rm sym}$ has two peaks corresponding to its strong anti-correlations with the slope $L$ and skewness $J_{\rm sym}$ of $E_{\rm sym}(\rho)$ when $\Delta R$ in measuring the radii of canonical NSs is less than about 0.2 km, and the locations of the two peaks are sensitive to the maximum value of $J_{\rm{sym}}$ reflecting the stiffness of $E_{\rm{sym}}(\rho)$ above about 3$\rho_0$,
(4) high-precision radius measurements for canonical NSs is more useful than massive ones for constraining the EOS of nucleonic matter at supersaturation densities around $2\rho_0$. Overall, important new physics regarding the EOS of supradense neutron-rich nuclear matter can be obtained with the planned NS radius measurements with unprecedented precision.

There are certainly caveats in our work. For example, as stated clearly earlier, we used conservatively the so-called minimum model for NSs to reduce the number of model assumptions, parameters and their uncertainties. There is no doubt that high-precision NS radius measurements may help reveal other new physics when more degrees of freedom, particles and/or phases are considered in modeling NSs. Nevertheless, results from our present work provide a useful foundation for further exploring new physics with the expected high-precision NS radius data. Having used a lot of imagined NS radius data in our present analyses, we are enthusiastically waiting for the real NS radius data with precision better than 0.1 km.

\begin{figure*}[ht]
%\vspace{-0.6cm}
\begin{center}
 \resizebox{0.49\textwidth}{!}{
  \includegraphics[width=10cm,height=9cm]{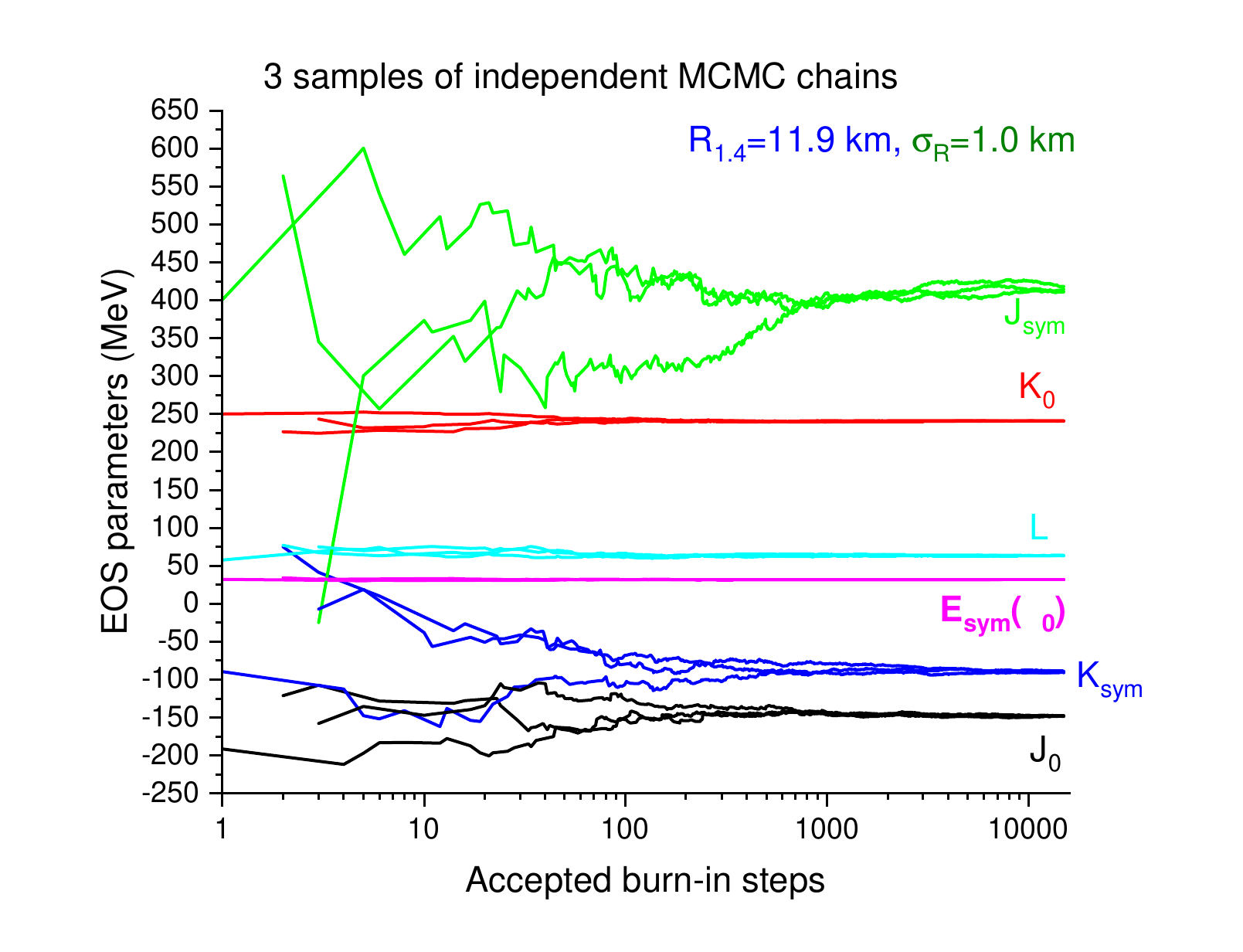}
  }
 \resizebox{0.49\textwidth}{!}{
  \includegraphics[width=10cm,height=9cm]{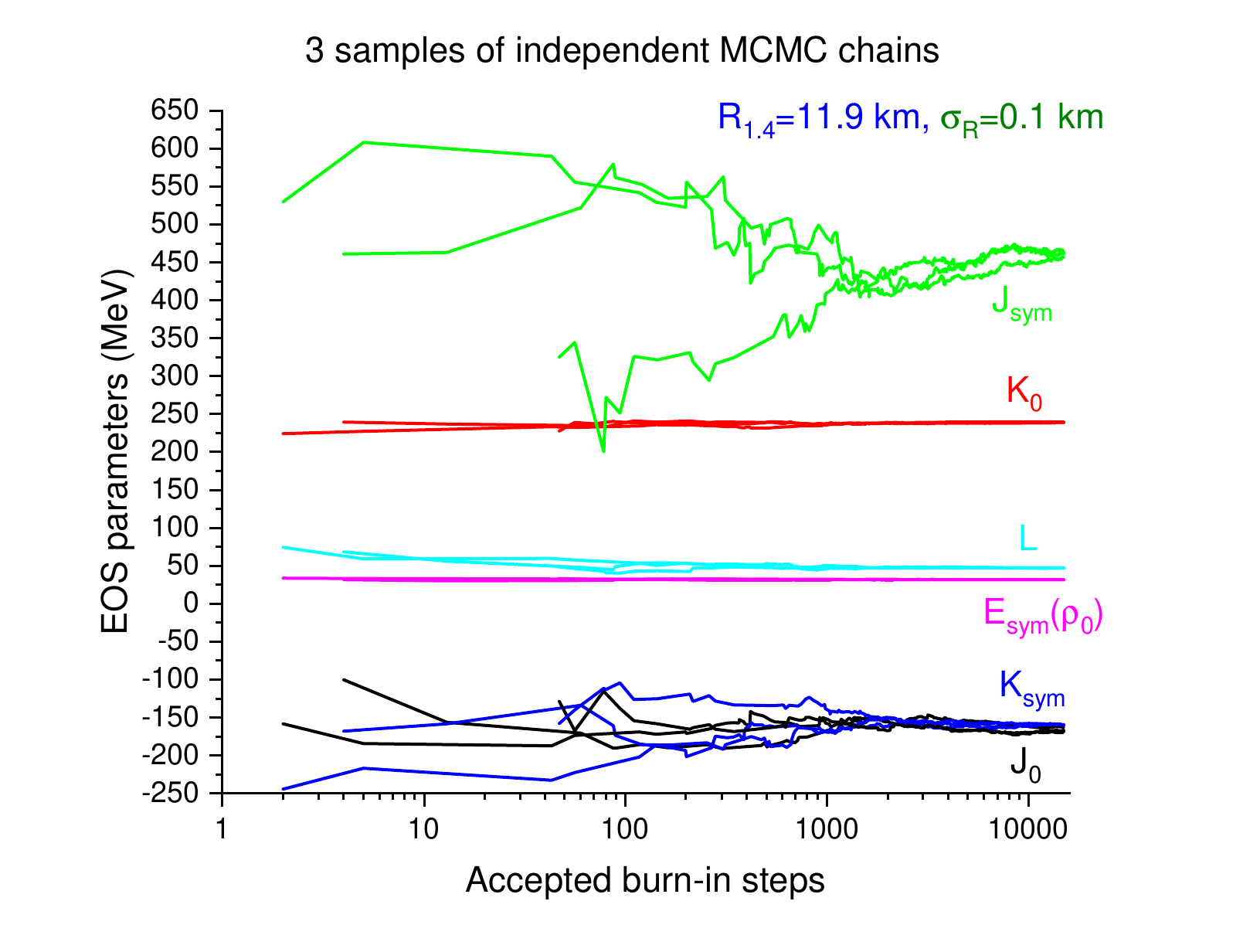}
  }
%   \vspace{-1.3cm}
  \caption{(color online) Examples of the tracing plots (running averages) of the 6 EOS parameters during the burn-in stage of the MCMC process from 3 independent walkers using $R_{1.4}=11.9$ km with $\Delta R=1.0$ km (left) and 0.1 km (right), respectively.}\label{burnin}
\end{center}
\end{figure*}
\begin{figure*}[ht]
%\vspace{-0.6cm}
\begin{center}
 \resizebox{0.49\textwidth}{!}{
  \includegraphics[width=14cm,height=9cm]{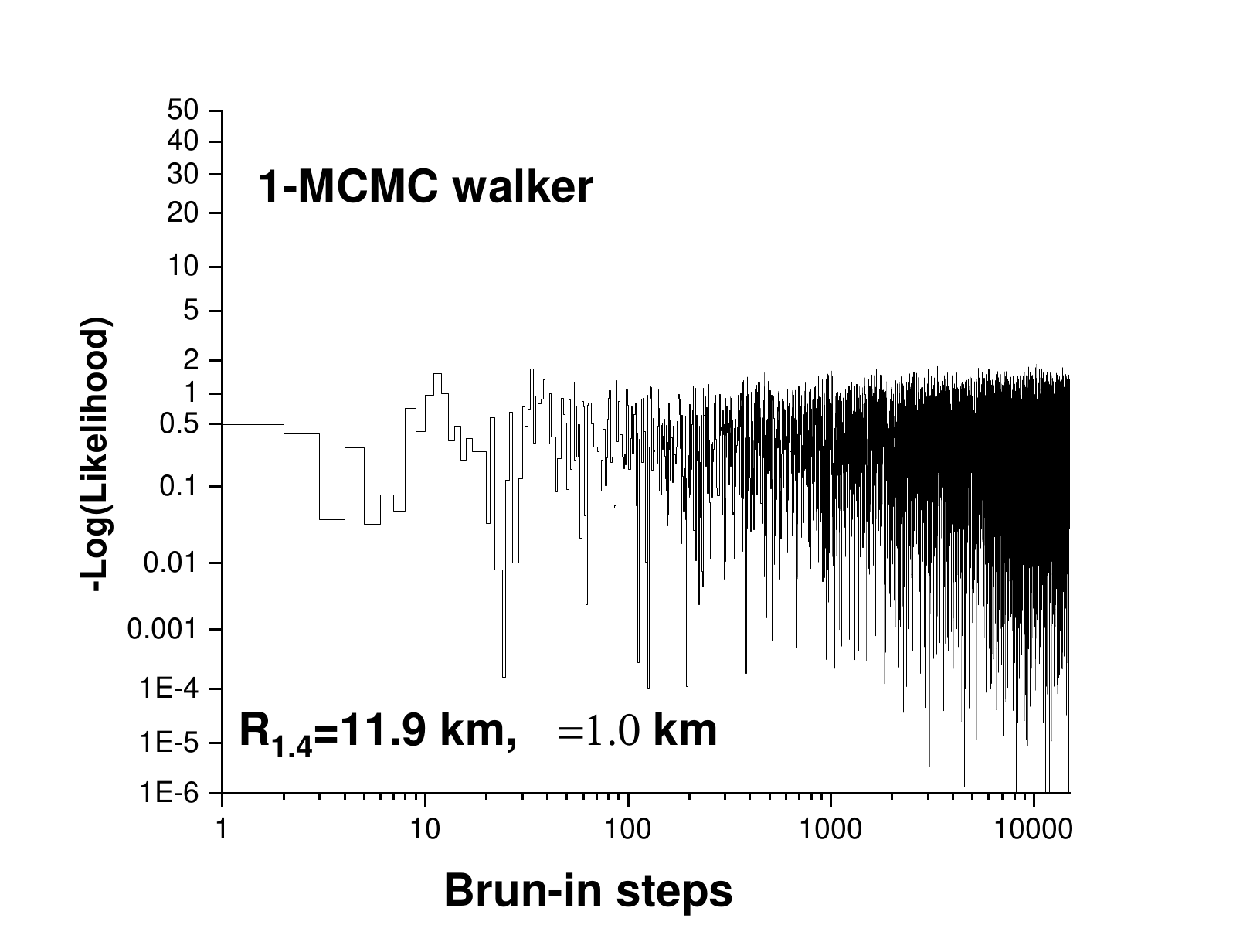}
  }
 \resizebox{0.49\textwidth}{!}{
  \includegraphics[width=14cm,height=9cm]{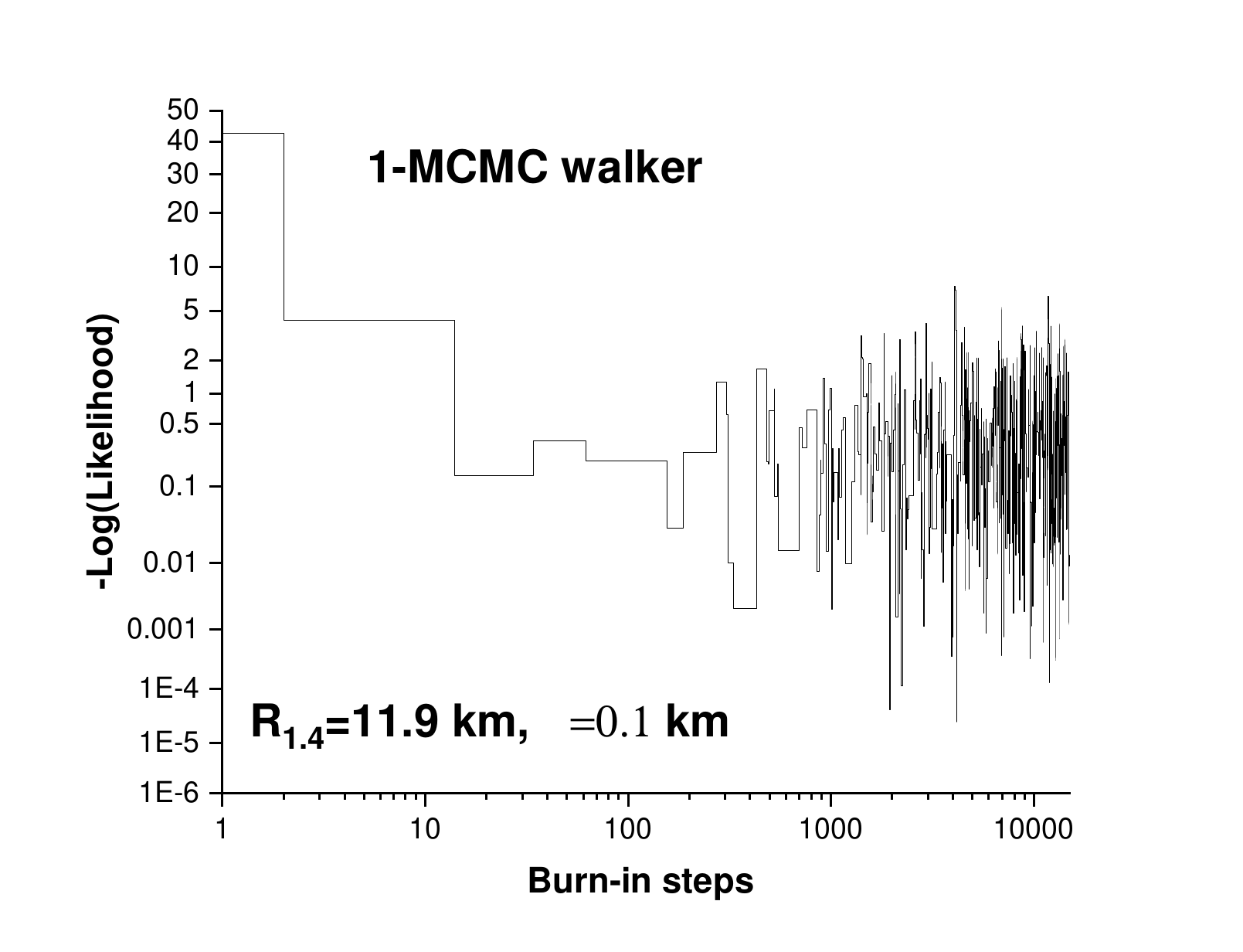}
  }
%   \vspace{-1.3cm}
  \caption{(color online)The $-{\rm Log}({\rm Likelihood)}$ as a function of the number of accepted MCMC steps (in Log scale) during the burn-in stage for a single walker using $R_{1.4}=11.9$ km with $\Delta R=1.0$ km (left) and 0.1 km (right), respectively.}\label{burnin2}
\end{center}
\end{figure*}

\section{Appendices}
It is well known that in Bayesian analyses there are many technical issues besides the science or engineering questions that one wishes to address without much biases/problems from the technical ones, see, e.g., Ref. \cite{trotta2017bayesian}. The community has developed many tools to diagnose the issues, e.g., convergence and autocorrelations, and {\it many} methods to mitigate their effects. For example, 13 convergence diagnostics were reviewed in Ref. \cite{Cowles96}. To our best knowledge, various indicators 
of convergence and autocorrelations are not really independent but could be complementary. While we did not test our codes/models exhaustively with all existing tools, we conducted testings with several widely used ones to gain confidence about our work in several situations. In the following, we provide some more technical details and examples testing convergence and autocorrelations that are most closely related to the physics issues we investigated in the main text. 

\appendix
\section{MCMC burn-in steps}
In our Bayesian analyses, the Metropolis-Hastings algorithm is utilized in the MCMC process to generate posterior PDFs in the EOS multi-parameters space. By integrating all other parameters via the marginal estimation approach, one can obtain not only the 1D PDFs for individual parameters but also the 2D PDFs containing inherently information about two-parameter correlations. 

During the initial phase, known as the burn-in period, samples are discarded to ensure that the posterior PDFs represent equilibrium distributions, as recommended by Trotta (2017)\cite{trotta2017bayesian}. Shown in Fig. \ref{burnin} are examples of the running averages of the six EOS parameters during the burn-in stage of the MCMC process from 3 independent walkers/runners. As examples, here we used only $R_{1.4}=11.9$ km with $\Delta R=1.0$ km (left) and 0.1 km (right), respectively. It is seen that they quickly converge (indicating the constraining power of the observables and filters used), especially in the case of $\Delta R=1.0$ km. We also notice that $K_0$ and $E_{\rm sym}(\rho_0)$ do not change much as they have little influence on the radii of NSs compared to the supersaturation density parameters $L$, $K_{\rm{sym}}$, $J_{\rm{sym}}$ and $J_0$. In the present work, we thus focus on the latter four parameters. It is interesting to also note that the stabilized mean values of $L$ and $K_{\rm{sym}}$ decrease appreciably while the mean of $J_{\rm{sym}}$ increases when the $\Delta R$ changes from 1.0 km to 0.1 km. The other three parameters show no obvious change. 

Shown in Fig. \ref{burnin2} are the $-{\rm Log}({\rm Likelihood})$ as a function of the number of accepted MCMC steps (in Log scale) during the burn-in stage for a single walker using $R_{1.4}=11.9$ km with $\Delta R=1.0$ km (left) and 0.1 km (right), respectively. Generally speaking, the walker can quickly jump from one extreme spot to another with very different likelihoods without getting stuck for too long at a particular area of the EOS space. As expected, a lot more steps are accepted with $\Delta R=1.0$ km (left) than 0.1 km (right). 

Both the running averages and the $-{\rm Log}({\rm Likelihood})$ plots show strong indications of efficiently converging MCMC chains. It is seen that 15,000 burn-in steps are more than enough for all runners to reach the equilibrium stage. All results presented in this work are obtained from using at least 600,000 subsequent steps by each runner for calculating the posterior PDFs. Since gradually fewer steps are accepted in the MCMC process with higher precision $\Delta R$, 24 to 48 runners are used with different radius precision. We do not use any quantitative exit criteria on any particular quantity, but requiring all 1D PDFs to become stable within statistical errors. Specifically, by adding or reducing 8 runners in accumulating the posterior EOS data we examine if the 1D PDFs have any obvious changes to more runners or longer chains are necessary.
We also note that the average acceptance rate in our MCMC process for $R_{1.4}=11.9$ km with $\Delta R=1.0$ km and 0.1 km is about 28\% and 6\%, respectively. 

%The Bayes factor from the ratio of evidences in the above 2 cases is found to be about 4, meaning that the data with larger errors can accommodate more EOS models as one expects. For the purposes of this work, since we could not find any physically new information from studying the Bayes factors compared to what we have presented already, Bayes factors are not discussed anywhere in the main text.

\begin{figure*}[ht]
%\vspace{-0.6cm}
\begin{center}
 \resizebox{0.7\textwidth}{!}{
  \includegraphics[width=16cm,height=6cm]{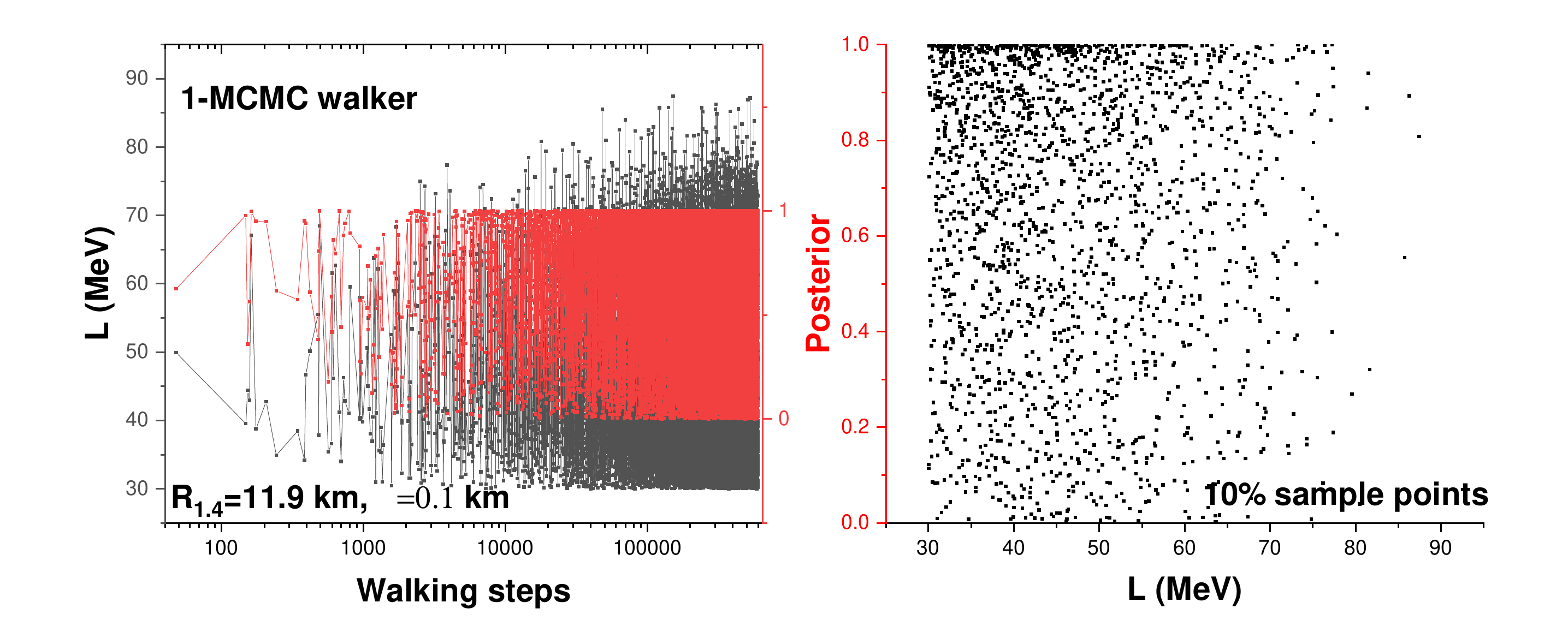}
  }
 \resizebox{0.7\textwidth}{!}{
  \includegraphics[width=16cm,height=6cm]{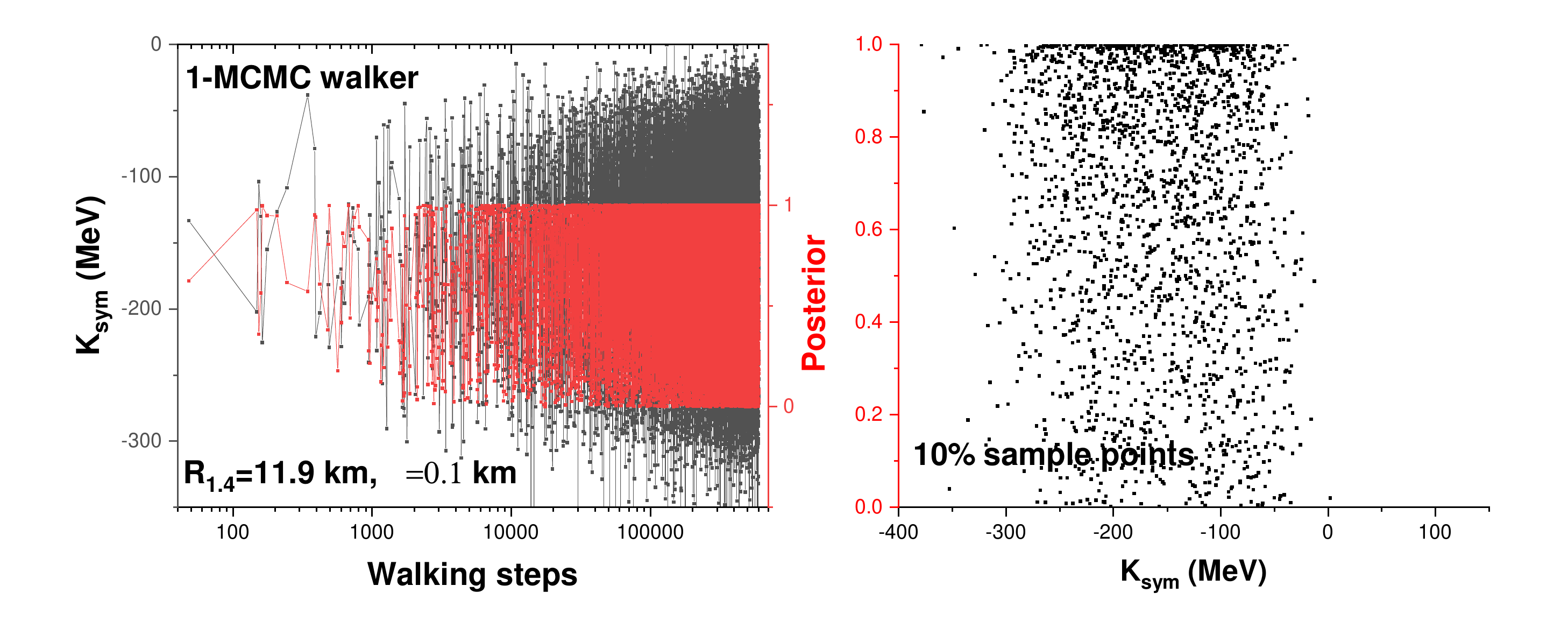}
  }
%   \vspace{-1.3cm}
  \caption{(color online) Left: The jumping maps (time series) of posterior $P$ (red), $L$ and $K_{\rm sym}$ (black) for a single MCMC walker with $R_{1.4}=11.9$ km and $\Delta R=0.1$ km. Right: scatter plots of $P$ versus $L$ and $P$ versus $K_{\rm sym}$, respectively, from 10\% accepted samples shown in the left windows.
  }\label{PLK}
\end{center}
\end{figure*}

\begin{figure}[ht]
%\vspace{-0.6cm}
\begin{center}
 \resizebox{0.5\textwidth}{!}{
  \includegraphics[width=7cm,height=8cm]{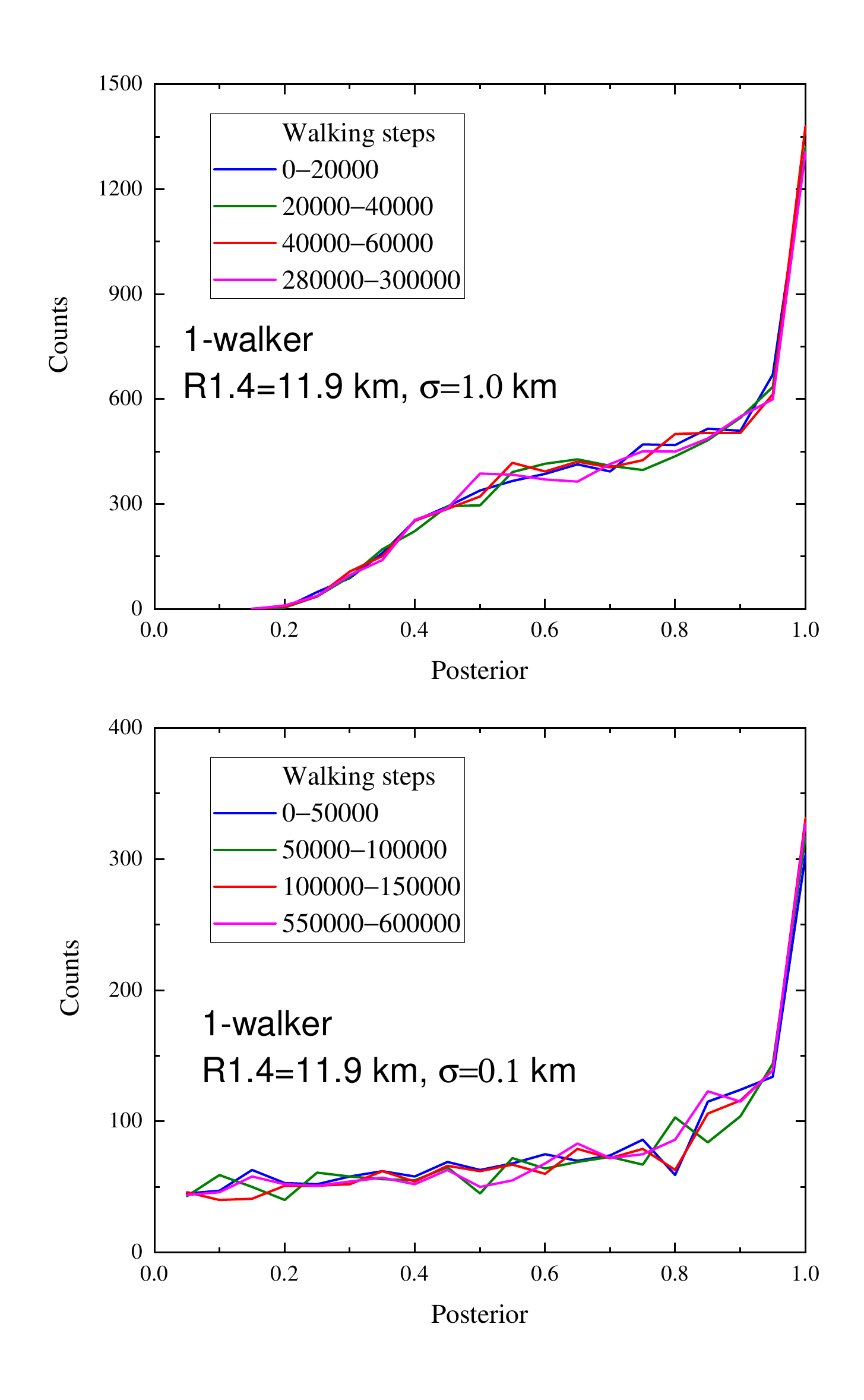}
  }
 \vspace{-1cm}
  \caption{(color online) Distribution of the posterior P in equal intervals of walking steps for a single MCMC walker with $R_{1.4}=11.9$ km with $\Delta R=1.0$ km (upper) and $\Delta R=0.1$ km (lower), respectively.}\label{NBZ-P}
\end{center}
\end{figure}
\begin{figure*}[ht]
\vspace{-10cm}
\begin{center}
 \resizebox{1.\textwidth}{!}{
  \includegraphics[width=8cm,height=12cm]{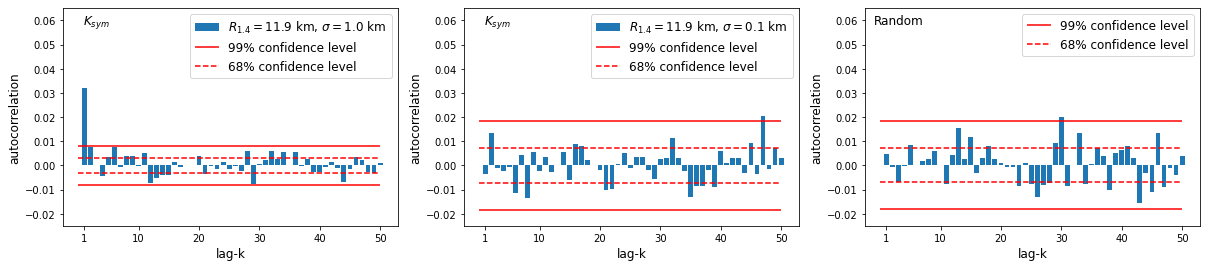}
  }
  \vspace{-11cm}
  \caption{(color online) In blue, autocorrelation as a function of lag $k$ for a single walker with $R_{1.4} = 11.9$ km, $\sigma = 1.0$ km, $R_{1.4} = 11.9$ km, $\sigma = 0.1$ km, and $20,000$ random numbers generated uniformly between $-400$ and $100$. The red dashed (solid) lines represent the approximate $68\%$ ($99\%$) significance bounds.}\label{ksymAutocorr}
\end{center}
\end{figure*}
\section{Further testing MCMC convergence}\label{LL}
To further test the convergence of MCMC chains and check if there is any obvious "sticking" in sampling the PDFs of $K_{\rm{sym}}$ and $L$ that are most important for determining NS radii, shown in the left of Fig. \ref{PLK} are the 
jumping maps (time series) of posterior $P$ (i.e., the product of the likelihood function and the prior) (red), $L$ and $K_{\rm sym}$ (black) for a single MCMC walker with $R_{1.4}=11.9$ km and $\Delta R=0.1$ km. The right panels are the scatter plots of $P$ versus $L$ and $P$ versus $K_{\rm sym}$, respectively, from 10\% of the accepted samples shown in the left windows. For this single-walker after 15,000 burn-in steps, the mean values of both $L$ and $P$ versus $K_{\rm sym}$ stay almost constants, while their variances took more steps to reach their stationary values. With the 24-48 walkers we used in the final analyses of various cases, even the variances are constants. We see no obvious evidence for any ``slow mixing" or ``chain sticking" that can occur when the MCMC sampler gets stuck too long in a particular region of the EOS parameter space. It is interesting to see that in earlier steps shown in Log scale, when the $L$ or $K_{\rm sym}$ is continuously going to a particular direction (e.g., their lower or upper limit set by their priors), the corresponding $P$ in those steps moves oppositely to bring the walker back to stay around their stationary values.  
The scatter plots of $P-L$ and $P-K_{\rm sym}$ from the 10\% accepted samples are consistent with the posterior 1D PDFs shown in the main text. 

To measure more quantitatively the degree of convergence, the jumping maps of the posterior $P$ shown in Fig. \ref{PLK} can be further analyzed. Shown in Fig. \ref{NBZ-P} are the distributions (in terms of the number of accepted steps) of the posterior P in equal intervals of steps for a single MCMC walker with $R_{1.4}=11.9$ km with $\Delta R=1.0$ km (upper) and $\Delta R=0.1$ km (lower), respectively. In both cases considered, the distributions of $P$ remain the same in all step intervals. Of course, with a much smaller $\Delta R$, statistical fluctuations are larger. Thus more walkers/steps are used with $\Delta R=0.1$ km.

\begin{figure*}[ht]
%\vspace{-2.cm}
\begin{center}
 \resizebox{0.8\textwidth}{!}{
  \includegraphics[width=12cm,height=18cm]{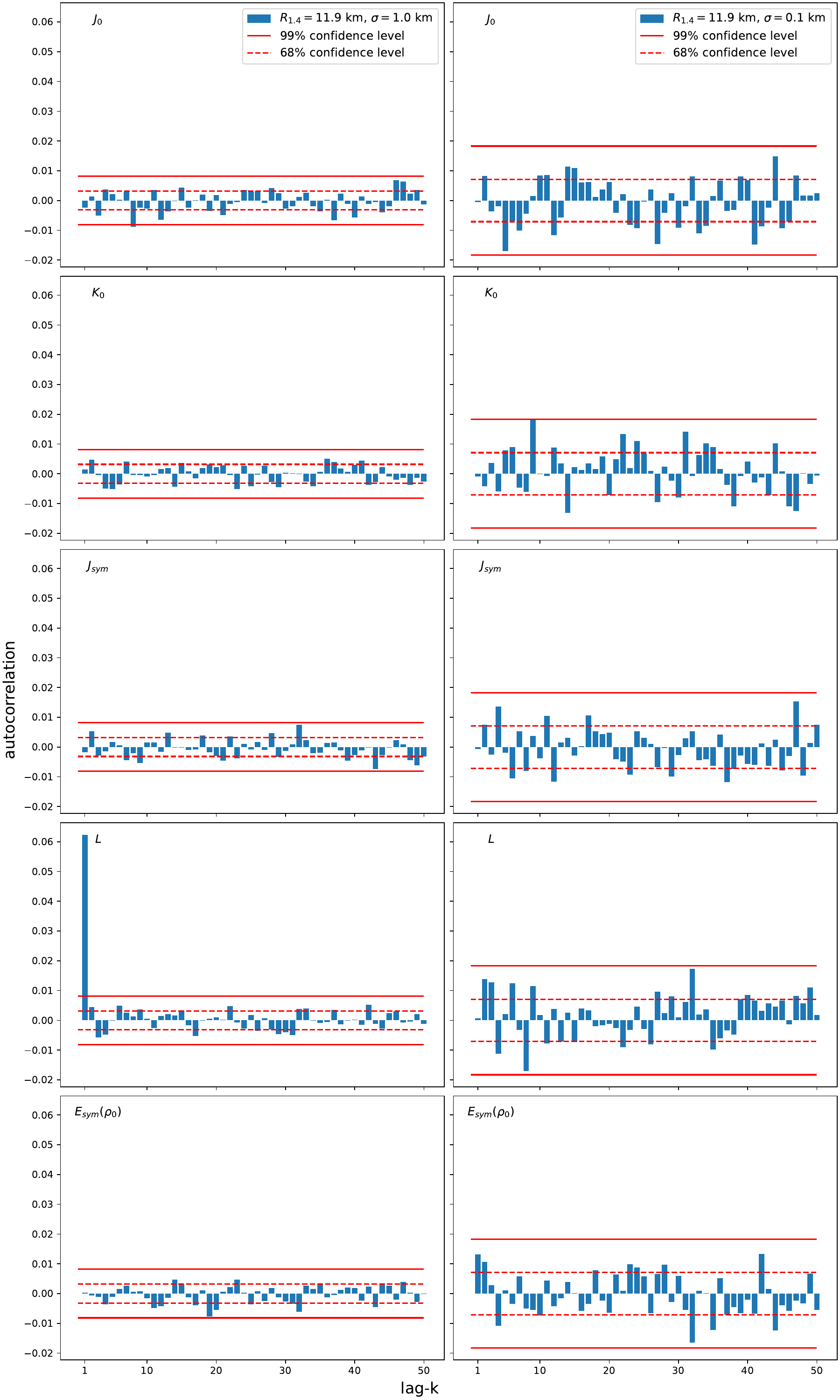}
  }
%  \vspace{-2cm}
  \caption{(color online) Examples of the tracing plots (running averages) of the 6 EOS parameters during the burn-in stage of the MCMC process from 3 independent runners.}\label{paraAutocorr}
\end{center}
\end{figure*}

\section{Evaluating MCMC auto-correlations}\label{auto}
In any Markov Chain, it is expected that there be some amount of correlation between samples \cite{trotta2017bayesian,Hogg_2018,madras1988}. If this autocorrelation is too high, the Markov Chain will be slow to explore the entire target distribution, but will stall in a certain location. This is inefficient and reduces the effective sample size. While these issues are mitigated by the use of multiple chains running for a long time after significant burn-in, we report it here to demonstrate that our sampling is not significantly biased by autocorrelation.

While the visual diagnostics of convergence and autocorrelation are useful, more quantitative measures are necessary. For this purpose, we examine here the autocorrelation function ($ACF$). It is a measure of the correlation between samples $k$ steps apart, referred to as lag. ACF at lag $k$ is estimated from a sample by \cite{trotta2017bayesian}:
\begin{equation}
    ACF(k) = \frac{\Sigma_{i=0}^{N-k}(\theta_i - \bar\theta)(\theta_{i+k}-\bar\theta)}{\Sigma_{i=0}^{N-k}(\theta_i - \bar\theta)^2}
\end{equation}
Where $N$ is the number of samples and $\bar\theta$ is the mean of the parameter. The bounds for the confidence level can be approximated by $\pm z_{1-\alpha/2} / \sqrt{N}$ \cite{Brockwell2016}, i.e., the z-score corresponding to the significance level $\alpha$ of a two-tailed test, divided by the square root of the number of samples. We report here values for $z_{1-\alpha/2}$ of $1$ and $2.576$, which correspond to the z-scores at 68\% and 99\% confidence levels, respectively, or equivalently $\alpha = 0.32$, and $0.01$ significance levels. So long as most of the $ACF(k)$ values fall within these bounds, we can conclude there is not a significant amount of autocorrelation.

In Figs. \ref{ksymAutocorr} and \ref{paraAutocorr}, we report the $ACF(k)$ for $k=1$ to $50$ from a single walker with $R_{1.4} = 11.9$ km, $\sigma = 1.0$ km and $\sigma = 0.1$ km. We do not report $k=0$, because this is always one, since every sample is perfectly correlated with itself. As can be seen in the plots, for $k$ greater than zero, the autocorrelation drops significantly below one.

We will focus on $K_{sym}$, shown in Fig. \ref{ksymAutocorr}, because it has the double peak in its PDF that we will show is not due to a sampling bias, but the other parameters are similar. We first note that the bounds for the confidence levels are smaller for the case of $\sigma = 1.0$ km. This is because the bounds are proportional to $1/\sqrt{N}$ samples and a single chain with $\sigma = 1.0$ km had approximately five times as many accepted steps as that with $\sigma = 0.1$ km. Also shown in Fig. \ref{ksymAutocorr}, is the autocorrelation of $20,000$ random numbers generated uniformly between $-400$ and $100$ as comparison. In all three graphs, most values do indeed fall within the $99\%$ bound as desired, indicating no significant autocorrelation.

The only significant break is lag $k = 1$ for $\sigma = 1.0$ km. To completely remove this we could thin our samples by throwing out every other one. This is not always recommended if it is not too computationally expensive to simply run and store longer or multiple chains \cite{roy2019mcmcDiagnostics}. These other solutions will eventually reach the same effective sample size, as if we had more independent (i.e. less autocorrelated) samples, as mentioned earlier.

We conclude, therefore, that the level of autocorrelation in our Markov Chains is modest, and the small amount present is mitigated by our use of many long chains. It is thus reasonable to assume the double peak in $K_{sym}$ is not due to autocorrelations in our sampling method.
\\
\newpage
\noindent{\bf Acknowledgement:} BAL and XG were supported in part by the U.S. Department of Energy, Office of Science, under Award Number DE-SC0013702, the CUSTIPEN (China-U.S. Theory Institute for Physics with Exotic Nuclei) under the US Department of Energy Grant No. DE-SC0009971, and the Office of Vice President for Research and Economic Development at Texas A\&M University-Commerce. WJX was supported in part by the Shanxi Provincial Foundation for Returned Overseas Scholars under Grant No 20220037, the Natural Science Foundation of Shanxi Province under Grant No 20210302123085, the Open Project of Guangxi Key Laboratory of Nuclear Physics and Nuclear Technology, No. NLK2023-03 and the Central Government Guidance Funds for Local Scientific and Technological Development, China (No. Guike ZY22096024). NBZ is supported in part by the National Natural Science Foundation of China under Grant No. 12375120, the Zhishan Young Scholar of Southeast University under Grant No. 2242024RCB0013 and the Start-up Research Fund of Southeast University under Grant No. RF1028623060.

%\newpage
\bibliographystyle{nst}
%\bibliographystyle{elsarticle-num}
%\bibliography{references}
%\bibliographystyle{elsarticle-num}

\clearpage
\end{document}